\documentclass[
amsmath,amssymb,
prd,nofootinbib,floatfix,12pt,
]{revtex4}

\usepackage{amsmath,amssymb,amsfonts}
\usepackage[dvips]{color}
\usepackage{xcolor}
\usepackage{dsfont}
\usepackage{mathrsfs}
\usepackage{float}
\usepackage{hyperref}
\usepackage{graphicx}
\usepackage{graphics}
\usepackage{setspace}
\usepackage{lmodern}
\usepackage{tikz}
\usepackage{cancel}
\usepackage{placeins}

\makeatletter
\newlength\xvec@height%
\newlength\xvec@depth%
\newlength\xvec@width%
\newcommand{\xvec}[2][]{%
  \ifmmode%
    \settoheight{\xvec@height}{$#2$}%
    \settodepth{\xvec@depth}{$#2$}%
    \settowidth{\xvec@width}{$#2$}%
  \else%
    \settoheight{\xvec@height}{#2}%
    \settodepth{\xvec@depth}{#2}%
    \settowidth{\xvec@width}{#2}%
  \fi%
  \def\xvec@arg{#1}%
  \def\xvec@dd{:}%
  \def\xvec@d{.}%
  \raisebox{.2ex}{\raisebox{\xvec@height}{\rlap{%
    \kern.05em
    \begin{tikzpicture}[scale=1]
    \pgfsetroundcap
    \draw (.05em,0)--(\xvec@width-.05em,0);
    \draw (\xvec@width-.05em,0)--(\xvec@width-.15em, .075em);
    \draw (\xvec@width-.05em,0)--(\xvec@width-.15em,-.075em);
    \ifx\xvec@arg\xvec@d%
      \fill(\xvec@width*.45,.5ex) circle (.5pt);%
    \else\ifx\xvec@arg\xvec@dd%
      \fill(\xvec@width*.30,.5ex) circle (.5pt);%
      \fill(\xvec@width*.65,.5ex) circle (.5pt);%
    \fi\fi%
    \end{tikzpicture}%
  }}}%
  #2%
}
\makeatother

\hypersetup{colorlinks,citecolor=blue,urlcolor=blue, linkcolor=black}
\newcommand\beq{\begin{eqnarray}}
\newcommand\eeq{\end{eqnarray}}

\allowdisplaybreaks
\begin{document}
\renewcommand{\theequation}{\arabic{section}.\arabic{equation}}
\renewcommand{\thefigure}{\arabic{section}.\arabic{figure}}
\renewcommand{\thetable}{\arabic{section}.\arabic{table}}

\title{\large
Isosinglet vectorlike leptons at $e^+e^-$ colliders}

\author{Prudhvi N. Bhattiprolu$^{\dag}$,
        Stephen P. Martin$^{\ddag}$,
        Aaron Pierce$^{\dag}$
}

\affiliation{\it \baselineskip=18pt
  $^{\dag}${Leinweber Center for Theoretical Physics,\\ University of Michigan, Ann Arbor, MI 48109, USA}\\
  $^{\ddag}${Northern Illinois University, DeKalb IL 60115, USA}
}

\begin{abstract}\normalsize 
 We study weak isosinglet vectorlike leptons that decay through a small mixing with the tau lepton, for which the discovery and exclusion reaches of the Large Hadron Collider and future proposed hadron colliders are limited.  We show how an $e^+ e^-$ collider may act as a discovery machine for these $\tau^{\prime}$ particles, demonstrate that the $\tau^{\prime}$ mass peak can be reconstructed in a variety of distinct signal regions, and explain how the $\tau^{\prime}$ branching ratios may be measured.
\end{abstract}

\maketitle

\tableofcontents

\setcounter{footnote}{1}
\setcounter{figure}{0}
\setcounter{table}{0}
\newpage

\section{Introduction\label{sec:introduction}}
\setcounter{equation}{0}
\setcounter{figure}{0}
\setcounter{table}{0}
\setcounter{footnote}{1}

It is sometimes argued that hadron colliders have the best reach for  particle discovery beyond the Standard Model (SM), while electron-positron colliders are best suited for precision studies and indirect searches. The larger cross-sections for strongly interacting particles and the ease of attaining higher energies at hadron colliders makes this statement plausible, but for particles with only electroweak interactions it need not be the case. The present paper is concerned with an example where the exclusion reach of the Large Hadron Collider (LHC)  and future proposed hadron colliders is limited: a weak isosinglet vectorlike lepton, which we refer to as $\tau^{\prime}$, that decays through a small mixing with the tau lepton.  We will argue that an $e^+e^-$ collider 
\cite{CLIC:2018fvx,CLICdp:2018cto,Roloff:2018dqu,Bambade:2019fyw,LCCPhysicsWorkingGroup:2019fvj,ILCInternationalDevelopmentTeam:2022izu,Bai:2021rdg,FCC:2018evy,Bernardi:2022hny,CEPCStudyGroup:2018ghi,CEPCPhysicsStudyGroup:2022uwl} may act as a discovery machine for the $\tau^{\prime}$, even for low beam energies.

Vectorlike fermions are hypothetical particles whose couplings to the SM gauge bosons are the same for the left- and right-handed components. There are many new physics models--motivated by solutions to the hierarchy problem and other puzzles--that incorporate vectorlike fermions. Examples in the context of supersymmetry are found in refs.~\cite{Moroi:1991mg,Moroi:1992zk,Babu:2004xg,Babu:2008ge,Martin:2009bg,Graham:2009gy,Endo:2011xq,Martin:2012dg,Endo:2012cc,Fischler:2013tva,Bhattiprolu:2021rrj}, and a more general survey can be found in
Ref.~\cite{Ellis:2014dza}. The non-chiral couplings permit mass terms that do not require electroweak symmetry breaking. This means vectorlike fermions decouple efficiently from low-energy observables, and indirect constraints from oblique electroweak corrections do not provide a strong constraint \cite{delAguila:2008pw,Martin:2012dg}. Mixing with the known Standard Model fermions can produce constraints from flavor violation, including lepton number violation, but the size of this mixing is \emph{a priori} undetermined, and such constraints are easily evaded while still allowing for decays that are prompt on the length scales relevant for collider detectors. 

In the following Section, we outline the basic phenomenology of the $\tau^{\prime}$, discussing its production at $e^{+} e^{-}$ colliders as well as its decay pattern.  The $\tau^{\prime}$ decays contain a $\nu_{\tau}$, or else a $\tau$, which will itself have neutrino(s) in its decay.  These neutrinos necessitate a strategy to effectively reconstruct the $\tau^{\prime}$ mass peak.  In Sec.~\ref{sec:peakreco}  we discuss such a strategy for a variety of channels of interest. In Sec.~\ref{sec:Results} we present the results of this strategy for a $\sqrt{s} = 500$ GeV $e^{+} e^{-}$ collider with an integrated luminosity of 4 ab$^{-1}$, and demonstrate that the reconstruction of an observable mass peak will be possible up to nearly the kinematic limit.  We also discuss how one might effectively determine the branching ratios of the $\tau^{\prime}$, allowing a robust test of its identity.  We choose this intermediate center of mass energy to present our most detailed results, but also show relevant results for other $\sqrt{s}$.  At lower energies we discuss $\sqrt{s}= 250$ GeV with 2 ab$^{-1}$, relevant for a Higgs factory, as well as $\sqrt{s}= 380$ GeV with 1.5 ab$^{-1}$, a top factory.   The latter has been discussed as an initial option for the compact linear collider (CLIC) \cite{CLIC:2018fvx,CLICdp:2018cto,Roloff:2018dqu}, and slightly lower energies have been proposed for top factories at an International Linear Collider (ILC) \cite{Bambade:2019fyw,LCCPhysicsWorkingGroup:2019fvj,
ILCInternationalDevelopmentTeam:2022izu,
Bai:2021rdg} or a Future Circular Collider (FCC-ee) \cite{FCC:2018evy,Bernardi:2022hny} or a Circular Electron Positron Collider (CEPC) \cite{CEPCStudyGroup:2018ghi,
CEPCPhysicsStudyGroup:2022uwl}.  We also discuss machines that would be further into the future, including a linear collider at $\sqrt{s} = 1$ TeV with 5 ab$^{-1}$ as well as CLIC options of $\sqrt{s} = 1.5$ TeV with 2.5 ab$^{-1}$ and  $\sqrt{s} = 3$ TeV  with 5 ab$^{-1}$. Our benchmark choices for beam energies and luminosities are inspired by Table 1-1 of the Snowmass 2021 Energy Frontier report found in Ref.~\cite{Narain:2022qud}.

Other $e^+ e^-$ collider studies of singlet vectorlike leptons that decay through mixing with the $\tau$ can be found in Refs.~\cite{Shang:2021mgn,Yang:2021dtc,Li:2022hzl,Shang:2023rfv,Bose:2022obr}. Our results below concur that discovery should be easily accomplished for masses up to very close to the kinematic limit at $e^+e^-$ colliders.
However, unlike previous studies, our study is done in the context of mass peak reconstruction and provides a detailed account of different final-state contributions to a comprehensive set of distinct signal regions, which should enable branching ratio determinations.
This approach not only identifies an excess but also clarifies the nature of this excess.
We also discuss the non-trivial efficiencies and the mass reconstruction widths for various $\tau^\prime$ masses and $\sqrt{s}$.
We clearly demonstrate that the feasibility of reconstructing mass peaks varies considerably with $\sqrt{s}$ and the choice of collider, even for masses well below the threshold.
Given the wide variety of colliders under consideration at present, understanding the differences in discovery potential between them is of paramount interest.

While our study pertains to a specific model that presents significant challenges for the LHC and future proposed hadron colliders,
other recent studies of pair-production of vectorlike leptons at colliders, but with different assumptions for decay modes, can be found in
\cite{Dermisek:2014qca,Kawamura:2019rth,Freitas:2020ttd,OsmanAcar:2021plv,Kawamura:2021ygg,Bernreuther:2023uxh,Kawamura:2023zuo}. 
In particular, hadron collider search strategies for vectorlike leptons that decay through mixing with the electron or muon, instead of the tau as in the present paper, have been given in \cite{Dermisek:2014qca}, with experimental LHC limits in \cite{ATLAS:2015qoy}.

\section{Production and decay of isosinglet vectorlike leptons\label{sec:singletvll}}
\setcounter{equation}{0}
\setcounter{figure}{0}
\setcounter{table}{0}
\setcounter{footnote}{1}

The isosinglet $\tau^{\prime}$ can be written as two left-handed fermions that transform under the gauge group $SU(3)_c \times SU(2)_L \times U(1)_Y$ in the representations
\beq
\tau'_L, \tau_R^{\prime \dagger} \>\,\sim\>\, ({\bf 1}, {\bf 1}, -1) \>+\> ({\bf 1}, {\bf 1}, +1),
\label{eq:tauprimeReps}
\eeq
which should be contrasted to the tau lepton of the Standard Model, 
\beq
\tau_L, \tau_R^\dagger \>\,\sim\>\, ({\bf 1}, {\bf 2}, -1/2) \>+\> ({\bf 1}, {\bf 1}, +1).
\label{eq:tauReps}
\eeq
We assume the mass of the $\tau$ is determined (as usual in the Standard Model) by a Yukawa coupling $y_\tau$ to the Higgs field, and the mass of the $\tau'$ is predominantly due to the large mass parameter $M$. Mixing between the $\tau$ and $\tau'$ is determined by a small Yukawa coupling $\epsilon$ which can be treated as a perturbation. Thus, the mass matrix for $\tau, \tau^\prime$ is given by \cite{Kumar:2015tna}  
\beq
\begin{pmatrix} 
\label{eq:matrix}
y_\tau v & 0 
\\
\epsilon v & M
\end{pmatrix}
,
\eeq  
where $v \approx 174$ GeV is the SM Higgs vacuum expectation value, with mass eigenvalues
\beq
m_{\tau} &=& y_\tau v \left ( 1 - \frac{\epsilon^2 v^2}{2 M^2} + \ldots \right ), 
\\
M_{\tau'} &=& M \left ( 1 + \frac{\epsilon^2 v^2}{2 M^2} + \ldots \right ).
\eeq
To a good approximation these are simply $y_\tau v$ and $M$, respectively.

\begin{figure}[!t]
\begin{minipage}[]{0.5\linewidth}
  \includegraphics[width=9cm]{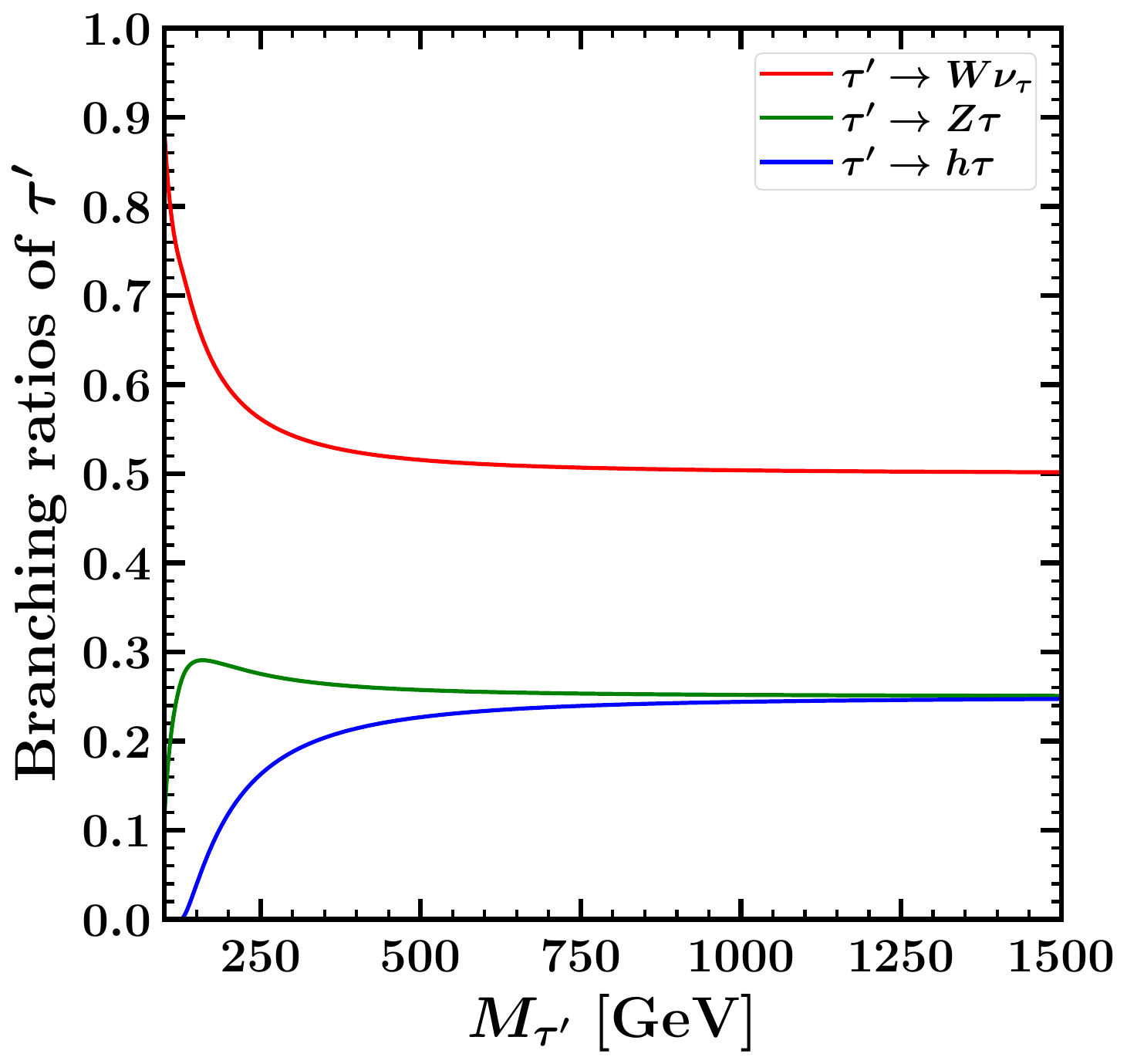}
\end{minipage}
\hspace{0.05\linewidth}
\begin{minipage}[]{0.425\linewidth}  
\caption{Branching ratios for weak isosinglet vectorlike leptons $\tau'$ 
as a function of $M_{\tau'}$. Lepton number violation is assumed negligible. 
\label{fig:BRs}}
\end{minipage}
\end{figure}

\begin{figure}[!t]
\begin{minipage}[]{0.5\linewidth}
  \includegraphics[width=9cm]{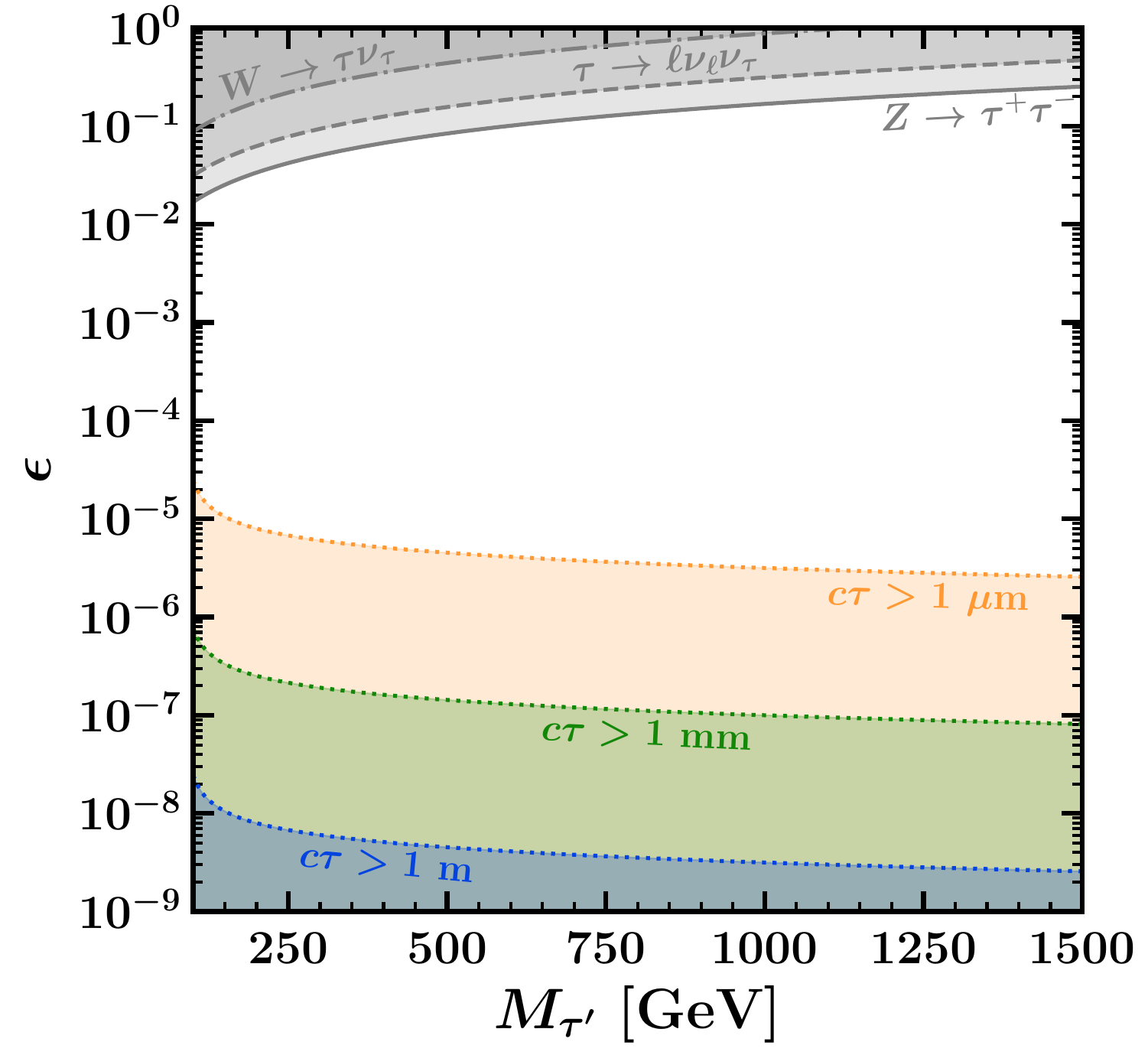}
\end{minipage}
\hspace{0.05\linewidth}
\begin{minipage}[]{0.425\linewidth}  
\caption{Upper bounds at 95\% CL on the Yukawa coupling $\epsilon$ (see Eq.~(\ref{eq:matrix})) as a function of $M_{\tau^\prime}$ are shown as gray shaded regions with solid, dashed, and dash-dotted borders, as labeled. Regions where $\tau^\prime$ has a decay length $c\tau >$ (1 $\mu$m, 1 mm, 1 m) are shown as (orange, green, blue) shaded regions, respectively, with dotted borders. 
\label{fig:MixingYukawa}}
\end{minipage}
\end{figure}

The unitary matrices that diagonalize the $\tau$-$\tau^\prime$ mass matrix in Eq.~(\ref{eq:matrix}) can be parameterized by a mixing angle $\theta_L$ related to the Yukawa coupling $\epsilon$ by 
\beq
\sin \theta_L &=& \frac{\epsilon v}{M_{\tau^\prime}},
\label{eq:SinThetaMix}
\eeq
while dropping terms of order $m_\tau^2/M_{\tau^\prime}^2$.
For the singlet vectorlike lepton model, to the lowest order in the mixing Yukawa coupling $\epsilon$, the decay rates are
\beq
\Gamma (\tau' \rightarrow W \nu_\tau) &=& \frac{\epsilon^2}{32 \pi} M_{\tau'} (1 + 2 r_W) (1 - r_W)^2,
\label{eq:BR1}
\\
\Gamma (\tau' \rightarrow Z \tau) &=& \frac{\epsilon^2}{64 \pi} M_{\tau'} (1 + 2 r_Z) (1 - r_Z)^2,
\label{eq:BR2}
\\
\Gamma (\tau' \rightarrow h \tau) &=& \frac{\epsilon^2}{64 \pi} M_{\tau'} (1 - r_h)^2,
\label{eq:BR3}
\eeq
where $r_X = M_X^2/M_{\tau'}^2$ for $X = W, Z, h$. The factors of $1 + 2 r_W$ and $1 + 2 r_Z$ in the vector boson decay modes are due to the transverse $(2 r_X)$ and the longitudinal (1) parts, with the latter associated with the would be Nambu-Goldstone modes that give the vector bosons their masses. It follows that in the limit of large $M_{\tau'}$ (or, equivalently, small $r_X$), the branching fractions approach the Goldstone equivalence limit,
\beq
{\rm BR}(\tau' \rightarrow W \nu_\tau) : {\rm BR}(\tau' \rightarrow Z \tau) : {\rm BR}(\tau' \rightarrow h \tau)
&=&
0.5 : 0.25 : 0.25, 
\eeq
as illustrated in Figure \ref{fig:BRs}. In this paper, we assume that the $\tau^\prime$ decays promptly on collider scales. The mixing Yukawa coupling $\epsilon$ required for $\tau^\prime$ to have a decay length of (1 $\mu$m, 1 mm, 1 m) as a function of $M_{\tau^\prime}$ is shown as the (orange, green, blue) dotted line, respectively, in Figure~\ref{fig:MixingYukawa}. For longer decay lengths, search strategies treating the $\tau'$ as a long-lived charged particle would be possible.\footnote{If $\tau^\prime$ is stable over detector lengths, then it can be inferred that $M_{\tau^\prime} \gtrsim 750$ GeV based on the
$-dE/dx$ and time of flight measurements in searches for long lived charginos at the LHC \cite{ATLAS:2019gqq,Bhattiprolu:2019vdu}.}

The mass mixing of $\tau^\prime$ with the SM $\tau$ modifies the the couplings of the SM $\tau_L$ to the $Z$ and $W$ bosons. They become $-\frac{g}{c_W} \left( s_W^2 - \frac{1}{2} \cos^2\theta_L \right)$ and $-\frac{g}{\sqrt{2}} \cos\theta_L$, respectively, where $s_W$ ($c_W$) is the (co)sine of the weak mixing angle and $g$ is the $SU(2)_L$ gauge coupling. These modifications can lead to deviations in the partial widths of $\tau \rightarrow \ell \bar{\nu}_\ell \nu_\tau$,
$Z \rightarrow \tau^+ \tau^-$, and 
$W \rightarrow \tau \nu_\tau$, etc.,
which can be used to infer constraints on the mixing angle $\theta_L$.
In the following, we obtain upper bounds on $\sin\theta_L$ from the following observables:
$\Gamma(\tau \rightarrow \ell \nu_\ell \nu_\tau)$,
$\Gamma(Z \rightarrow \tau^+ \tau^-)/\Gamma(Z \rightarrow \ell^+ \ell^-)$, and
$\Gamma(W \rightarrow \tau \nu_\tau)/\Gamma(W \rightarrow \ell \nu_\ell)$,
where $\ell = e, \mu$. Using the experimental values from Ref.~\cite{ParticleDataGroup:2022pth} for each of the observables,
we find the $\tau$-$\tau^\prime$ mixing angle is constrained to be
\beq
\sin \theta_L
&\lesssim&
\begin{cases}
    2.9 \times 10^{-2} \text{ from } \Gamma(Z \rightarrow \tau^+ \tau^-)/\Gamma(Z \rightarrow \mu^+ \mu^-),\\
    5.5 \times 10^{-2} \text{ from } \Gamma(\tau \rightarrow \mu \nu_\mu \nu_\tau),\\
    1.5 \times 10^{-1} \text{ from } \Gamma(W \rightarrow \tau \nu_\tau)/\Gamma(W \rightarrow e \nu_e),
\end{cases}
\label{eq:MixingBound}
\eeq
at 95\% CL.
The strongest constraints on $\sin \theta_L$ are from $Z \rightarrow \tau^+ \tau^-$ and $\tau \rightarrow \ell \nu_\ell \nu_\tau$ decays.  As expected, the constraint from $W \rightarrow \tau \nu$ decays is weaker. We have also verified that the constraints from $h \rightarrow \tau^+ \tau^-$ decays and oblique electroweak precision observables are also less stringent.  The bounds on $\sin {\theta_L}$ can be translated into bounds on $\epsilon$ as a function of $M_{\tau^\prime}$ using Eq.~(\ref{eq:SinThetaMix}).  These are shown in Figure~\ref{fig:MixingYukawa}.

Direct searches at colliders constrain $M_{\tau^{\prime}}$.
The exclusion $M_{\tau^\prime} < 101.2$ GeV at 95\% CL can be inferred from the non-observation of new charged heavy leptons at the Large Electron Positron (LEP) collider \cite{L3:2001xsz}. 
LHC searches for weak isosinglet $\tau^\prime$ by the CMS collaboration have excluded \cite{CMS:2022nty}:
\beq
125\text{ GeV} < M_{\tau^\prime} < 150\text{ GeV} \quad \text{(95\% CL exclusion)}
.
\eeq
Substantial improvement of these limits at the LHC, even with high luminosity, appears challenging\footnote{Note that the situation is very different for weak isodoublet vectorlike leptons, for which LHC experimental limits \cite{CMS:2019hsm,ATLAS:2023sbu} 
are much stronger. See \cite{Kumar:2015tna,Bhattiprolu:2019vdu} for theoretical discussions.} owing to small production cross-sections and large branching fractions to $W \nu_\tau$ \cite{Kumar:2015tna}. Exclusion and discovery prospects for isosinglet vectorlike leptons also seem limited at future $pp$ colliders \cite{Bhattiprolu:2019vdu}. This opens a window for $e^{+} e^{-}$ colliders to act as discovery machines.

\begin{figure}[!t]
 \begin{center}
   \includegraphics[width=15cm]{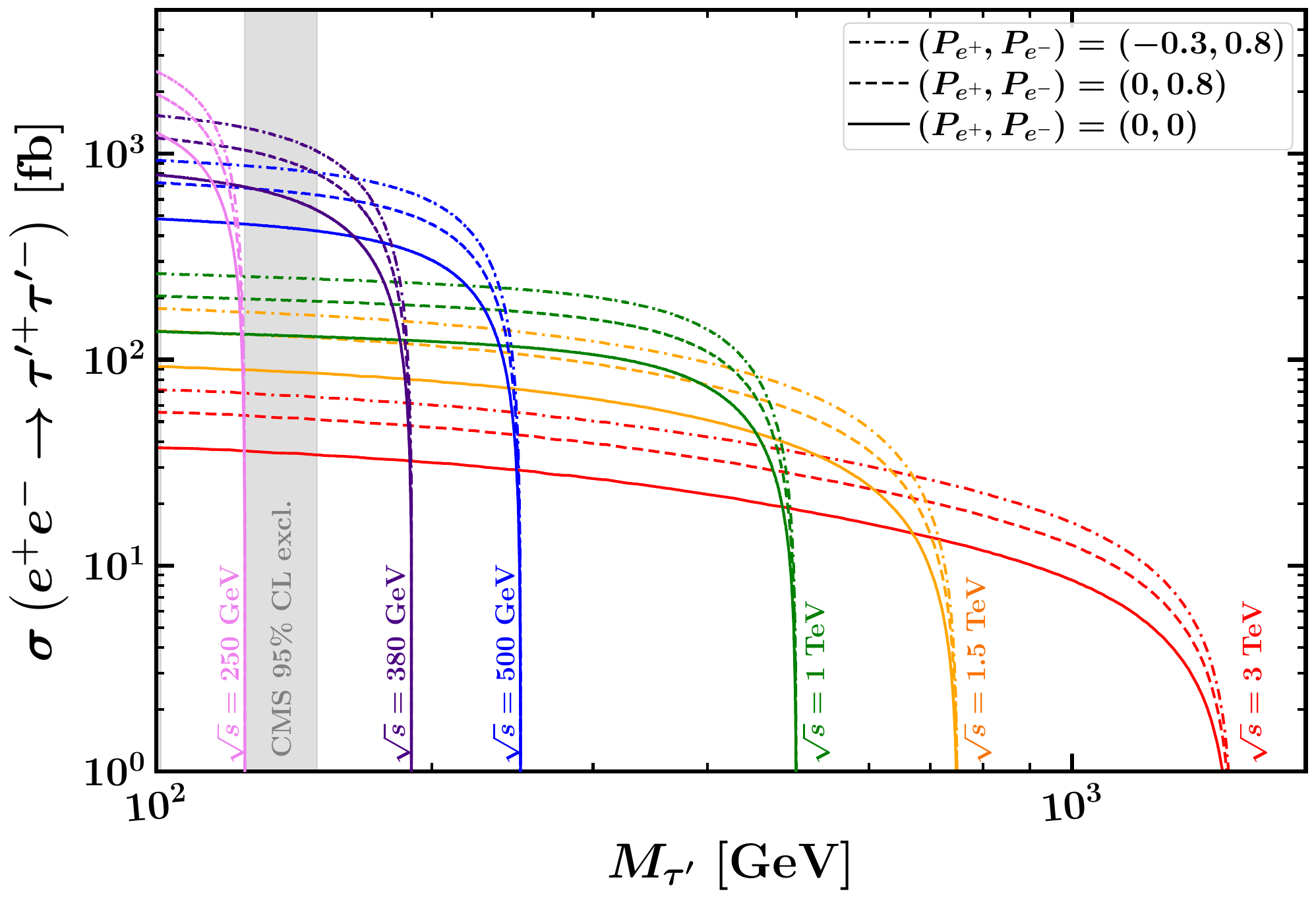}
 \end{center}
 \caption{Pair production cross-section of weak isosinglet vectorlike leptons $\tau^\prime$
 as a function of $M_{\tau^\prime}$ at $e^+ e^-$ colliders, accounting for initial state radiation and beamstrahlung. Colors denote various choices of $\sqrt{s}$. Dash-dotted, dashed, and solid lines correspond to beam polarization choices of $(P_{e^+}, P_{e^-}) = $
 $(-0.3, 0.8)$, $(0, 0.8)$, and $(0, 0)$, respectively. The vertical gray shaded region is the 95\% CL exclusion obtained by the CMS collaboration \cite{CMS:2022nty} on the $\tau^\prime$ considered here.
\label{fig:Sigma}}
\end{figure}

At an $e^+e^-$ collider, the production of $\tau^{\prime +} \tau^{\prime -}$ pairs proceeds through
an $s$-channel photon and $Z$-boson. If the positron and electron beams have polarizations $P_{e^+}$ and $P_{e^-}$ respectively, with  $P = 1$ and $-1$ corresponding to pure right-handed and left-handed polarizations, respectively, then the total partonic cross-section is
\beq
\hat \sigma(e^+e^- \rightarrow \tau^{\prime +} \tau^{\prime -}) \>=\> 
\frac{2 \pi \alpha^2}{3} (\hat s + 2 M_{\tau^{\prime}}^2)\, \sqrt{1 - 4 M_{\tau^{\prime}}^2/\hat s} \,\Bigl [
|a_L|^2 (1 - P_{e^-})(1 + P_{e^+}) 
\phantom{I,} &&
\nonumber \\
+ 
|a_R|^2 (1 + P_{e^-})(1 - P_{e^+})
\Bigr ], &&
\label{eq:sigmahat}
\eeq
where the left-handed and right-handed amplitude coefficients are 
\beq
a_L &=& \frac{1}{\hat s} + \frac{1}{c_W^2} (s_W^2 - 1/2) \frac{1}{\hat s - M_Z^2}
,
\label{eq:aL}
\\
a_R &=& \frac{1}{\hat s} + \frac{s_W^2}{c_W^2} \frac{1}{\hat s - M_Z^2}
.
\label{eq:aR}
\eeq
Here $\hat{s}$ represents the center of mass energy after accounting for the spread of beam energies; $\hat{s} < s$.
Since $|a_L| < |a_R|$ for $\sqrt{\hat s} > 93$ GeV, we see that the production cross-section is maximized when $P_{e^-}$ is positive (and, if available, when $P_{e^+}$ is negative). The production cross-section for $e^+ e^- \rightarrow \tau^{\prime +} \tau^{\prime -}$ including the effects of initial state radiation (ISR) and beamstrahlung is shown as a function of $M_{\tau^\prime}$ in Figure~\ref{fig:Sigma}, for various $\sqrt{s}$ and beam polarization choices $(P_{e^+}, P_{e^-}) = (-0.3, 0.8)$, $(0, 0.8)$, and $(0, 0)$.  The first two of these choices would maximize the cross-sections for ILC and CLIC designs, respectively. In what follows, we assume unpolarized beams, $(P_{e^+}, P_{e^-}) = (0, 0)$, and note that results can generally be re-scaled by accounting for the modified  cross sections.

\section{Event simulation and peak reconstruction\label{sec:peakreco}}
\setcounter{equation}{0}
\setcounter{figure}{0}
\setcounter{table}{0}
\setcounter{footnote}{1}

\subsection{Event simulation}
To simulate the production and the decay of $\tau^\prime$, we produced a {\it Universal FeynRules Output} (UFO) \cite{Degrande:2011ua} file containing the Feynman rules for vertices involving $\tau^\prime$ from an input
Lagrangian, using {\sc FeynRules} v2.3 \cite{Alloul:2013bka}\footnote{The UFO and {\sc FeynRules} files for isosinglet vectorlike leptons are available in a github repository \cite{VLL_UFOs}.}. The UFO file can be readily imported into Monte Carlo event generators.
We used {\sc Whizard} v3.1.2 \cite{Kilian:2007gr,Moretti:2001zz} to generate
parton-level events for various signal and background processes at leading order while accounting
for ISR (using {\sc Whizard}'s built-in implementation of the lepton ISR structure function) and
beamstrahlung (using the {\sc Circe2} subpackage~\cite{Ohl:1996fi,Ohl:circe2}).
We used beam spectra for future $e^+ e^-$ colliders from Ref.~\cite{Ohl:spectra}.
We then used {\sc Pythia} v8.306 \cite{Bierlich:2022pfr,Sjostrand:2014zea} for showering and
hadronization, and {\sc Delphes} v3.5.0 \cite{deFavereau:2013fsa} for detector simulation. For modeling detector response at $e^+ e^-$ colliders
with $\sqrt{s} =$ 0.25, 0.5, and 1 TeV (0.38, 1.5, and 3 TeV),
we used a generic
ILC \cite{ILDConceptGroup:2020sfq,DelphesILCgen}
(CLIC\footnote{When using the CLIC detector model for {\sc Delphes}, we chose the cone size parameter $R$ for the jet algorithm to be 1.5, 1.2, and 1.0 for $e^+ e^-$ collisions at 0.38, 1.5, and 3 TeV, respectively, and we checked that using slightly different $R$ gives similar results.
Furthermore, we also accounted for the effects of beam induced $\gamma \gamma \rightarrow$ hadrons background for $\sqrt{s} = $ 1.5 and 3 TeV colliders where its impact is more prominent.}
\cite{Leogrande:2019qbe,CLICdp:2018vnx,CLICdp:2017vju}) detector model
for {\sc Delphes}, operating at the ``medium" $b$-tagging point and
utilizing exclusive jet clustering employing the Durham \cite{Catani:1991hj} (Valencia \cite{Boronat:2016tgd,Boronat:2014hva} with $\beta = \gamma = 1$)
algorithm as implemented using {\sc FastJet} v3.4.0 \cite{Cacciari:2011ma}.

Pair-produced $\tau^\prime$ lead to final states with 2 tau leptons:
\beq
Z Z \tau^+ \tau^-, \qquad
h h \tau^+ \tau^-, \qquad
Z h \tau^+ \tau^- ,
\label{eq:finalstats_2tau}
\eeq
as well as 1 tau lepton and $\cancel{E}$ final states (with the $\cancel{E}$ arising from a $\nu_{\tau}$):
\beq
Z W^\pm \tau^\mp + \cancel{E}, \qquad
h W^\pm \tau^\mp + \cancel{E},
\label{eq:finalstats_1tau}
\eeq
and with 0$\tau$ + $\cancel{E}$:
\beq
W^+ W^- + \cancel{E}.
\label{eq:finalstats_0tau}
\eeq
For simulating the corresponding signal events, we forced
$Z$-boson decays to ($e^+e^-/\mu^+\mu^-$, $\tau^+\tau^-$, $b \overline{b}$, $j j$, $\nu \overline{\nu}$),
$h$ boson decays to ($\tau^+\tau^-$, $b \overline{b}$, $g g$, $W {W}^*/Z Z^*$), and
$W$ boson decays to ($e \nu_e/\mu \nu_\mu$, $\tau \nu_\tau$, $j j$)
in the above final states, and simulated at least $10^5$ events for each of the forced signal components. We then weighted each final state by the appropriate branching ratio.
Here, ``$j$" refers to a non-$b$ jet.

\begin{figure}[!t]
 \begin{center}
   \includegraphics[width=15cm]{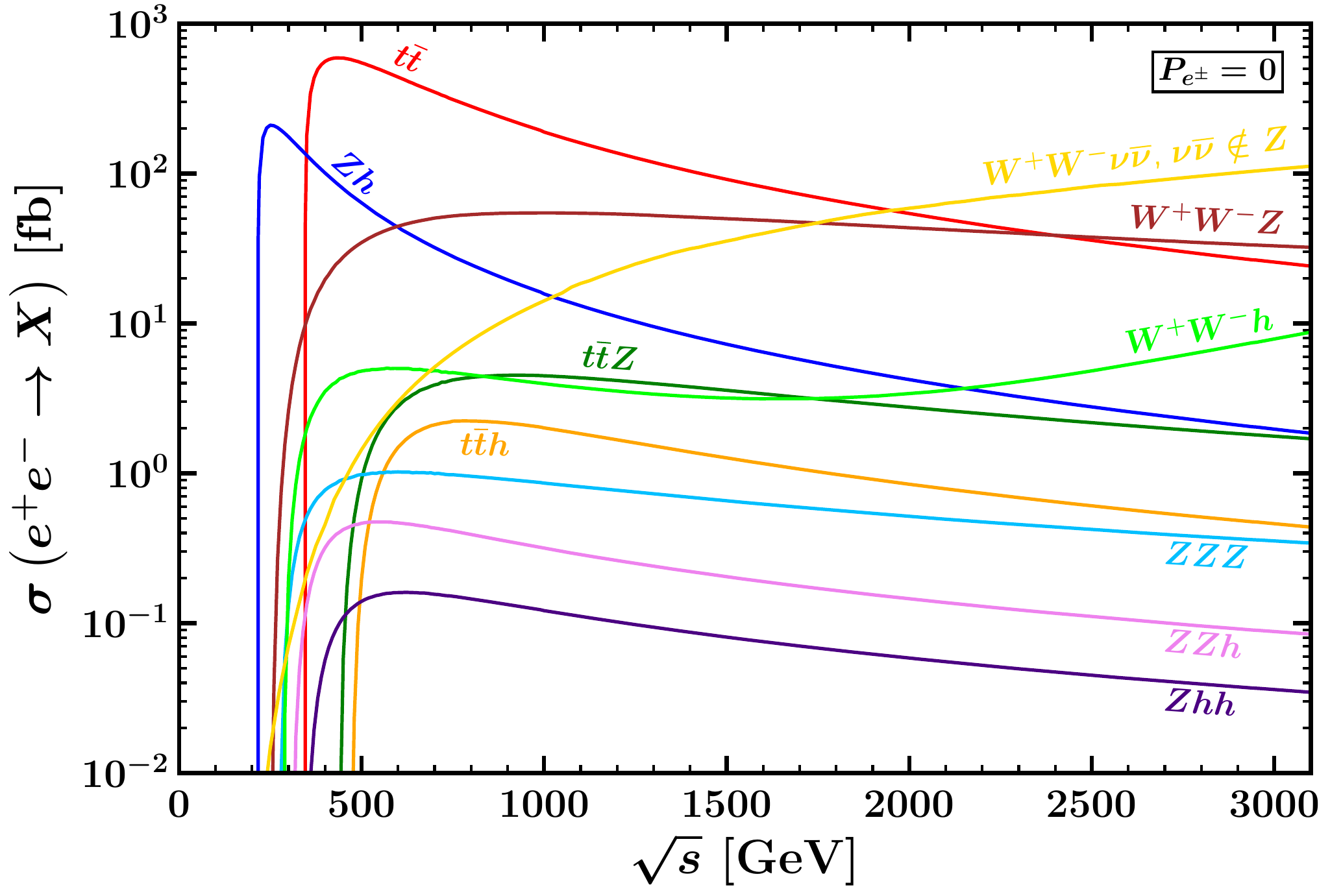}
 \end{center}
 \caption{Cross-sections for various SM background processes at leading order as a function of $\sqrt{s}$ at an $e^+ e^-$ collider with unpolarized beams while accounting for initial state radiation (but not beamstrahlung due to the availability of the beam spectra only at a few $\sqrt{s}$).
\label{fig:SigmaBG}}
\end{figure}

For backgrounds, we considered each of the following SM processes (as kinematically allowed
at a chosen $\sqrt{s}$): $t\overline t$, $t\overline t Z$, $t\overline t h$, $Zh$, $Zhh$, $ZZh$,
$ZZZ$, $W^+W^-h$, $W^+W^-Z$, and $W^+W^-\nu\overline{\nu}$ with $\nu \overline{\nu} \notin Z$
(to avoid double counting with $W^+ W^- Z, Z \rightarrow \nu \overline{\nu}$).
We simulated at least $10^6$ events for each background process.
Leading order cross-sections with ISR for these SM processes are shown in Figure~\ref{fig:SigmaBG} as a function of $\sqrt{s}$ for $e^+ e^-$ collisions with unpolarized beams.
Due to the availability of the beam spectra only at a few $\sqrt{s}$,
Figure~\ref{fig:SigmaBG} does not include the effects of the beamstrahlung,
but we note that the beam effects make a noticeable difference only at higher $\sqrt{s}$.
We do, however, account for beamstrahlung in our simulations at various collider options below.

\subsection{Reconstruction of the mass peak}

In this section, we outline our strategy for reconstruction of the $\tau^{\prime}$ mass peak.  We first give definitions of the particle objects used and define the signal regions where we attempt to reconstruct mass peaks.  We then discuss how we reconstruct candidate bosons ($Z$, $h$, $W$) that will ultimately be combined with a $\tau$ or $\nu_\tau$  to give the $\tau^{\prime}$ leptons.  Finally, we discuss our peak reconstruction algorithm: its task is to determine which boson gets paired with which $\tau$ and to effectively restore missing momenta from neutrinos.

\noindent
\emph{Physics Object Definition}:
In classifying the particles to determine our signal regions, 
we require all leptons and jets to satisfy a cut on $p_T$ and pseudo-rapidity $\eta$:  
\beq
p_T &>& 5 \text{ GeV},\label{eq:pTcut}\\
|\eta| &<& 3.\label{eq:etacut}
\eeq
We additionally require leptons 
to satisfy the following isolation cuts:
\beq
\Delta R_{L, L^\prime} &>& 0.1 \text{ (for each $L,L^\prime = e, \mu, \tau$)},\label{eq:ll_isolation}\\
\Delta R_{L, J} &>& 0.3 \text{ (for each $J = j, b$ and $L = e, \mu, \tau$)},\label{eq:lj_isolation}
\label{eq:leptonisolation}
\eeq
where $\Delta R = \sqrt{(\Delta \eta)^2 + (\Delta \phi)^2}$.
Here and from now on, ($\tau$, $j$, $b$) refers to a tagged hadronically decaying tau, non $b$-tagged jet, $b$-tagged jet), respectively.

\noindent
\emph{Signal Regions:} We consider various signal regions (SRs) that have:
\beq
N_{\ell} + N_j + N_b &=& 4,\label{eqn:howmanyleptonjets} \\ 
N_\tau &=& 1\>\,\text{or}\>\,2.\label{eqn:howmanytaus}
\eeq
The fraction of events with one or more photons ranges from a few percent (for lower energies) to $\sim$45\% (for $\sqrt{s}=3$ TeV). We have also included these events in our mass peak reconstructions.
Denoting $\ell \equiv e, \mu$, we consider the following SRs with exactly 2$\tau$ leptons of opposite sign, targeting the final states in Eq.~(\ref{eq:finalstats_2tau}):  
\beq
4 \ell + 2 \tau,\label{eq:4lta}\\
2 \ell + 2 j + 2 \tau,\label{eq:2l2j2ta}\\
2 \ell + 1 j + 1 b + 2 \tau,\label{eq:2l1j1b2ta}\\
2 \ell + 2 b + 2 \tau,\label{eq:2l2b2ta}\\
4 j + 2 \tau,\label{eq:4j2ta}\\
3 j + 1 b + 2 \tau,\label{eq:3j1b2ta}\\
2 j + 2 b + 2 \tau,\label{eq:2j2b2ta}\\
1 j + 3 b + 2 \tau,\label{eq:1j3b2ta}\\
4 b + 2 \tau.\label{eq:4b2ta}
\eeq
Targeting the final states in Eq.~(\ref{eq:finalstats_1tau}) we consider the following SRs with exactly 1$\tau$ lepton:
\beq
2 \ell + 2 j + 1 \tau,\\
2 \ell + 1 j + 1 b + 1 \tau,\label{eq:2l1j1b1ta}\\
4 j + 1 \tau,\label{eq:4j1ta}\\
3 j + 1 b + 1 \tau,\label{eq:3j1b1ta}\\
2 j + 2 b + 1 \tau,\\
1 j + 3 b + 1 \tau.\label{eq:1j3b1ta}
\eeq
Note that the signal regions in Eqs.~(\ref{eq:2l1j1b1ta}), (\ref{eq:3j1b1ta}), and (\ref{eq:1j3b1ta})
capture the possibility where one of the jets from the $W$-boson decay fakes a $b$ jet.  Recall, in our reconstruction, we do not allow the possibility that both jets from a $W$-boson decay fake $b$ jets, and in any case the probability of this would be small.  We do not discuss the $W^+ W^- + \cancel{E}$ final state of Eq.~(\ref{eq:finalstats_0tau}) where multiple neutrinos complicate the reconstruction of a mass peak. In each of the signal regions in Eqs.~(\ref{eq:4lta}) - (\ref{eq:1j3b1ta}), the jet multiplicity for exclusive jet clustering with the Durham/Valencia algorithm is required to be
$N_j + N_b + N_\tau$.

\noindent
\emph{Boson Reconstruction:}
Candidate $Z$ bosons are reconstructed from $e^+e^-/\mu^+\mu^-$,
from $jj$, or $jb$ (in SRs
with an odd number of tagged $b$ jets) in mass windows.  The invariant masses are constrained to lie in the range
\begin{equation}
M_{\ell \ell} = \left.M_Z\right\vert^{+5 \text{ GeV}}_{-5 \text{ GeV}}, \qquad
M_{jj/jb} = \left.M_Z\right\vert^{+7.5 \text{ GeV}}_{-15 \text{ GeV}}.
\label{eq:Zreco}
\end{equation}
Candidate $h$ bosons are reconstructed from a pair of $b$ jets (or in SRs
with an odd number of tagged $b$ jets, from $jb$) with an invariant mass
\begin{equation}
M_{bb/jb} = \left.M_h\right\vert^{+10 \text{ GeV}}_{-25 \text{ GeV}}.
\label{eq:hreco}
\end{equation}
This approach to Higgs reconstruction will also catch some $h \rightarrow gg$ events where the gluon fakes a $b$ jet, as well as some $h \rightarrow V V^{\ast}$ decays where the vector bosons cluster into fat jets.\footnote{The invariant mass regions chosen for the hadronic decays of the $Z$ and $h$ do not overlap.  Different choices of the invariant mass window might admit more events (and improve the statistics of the mass peaks), but at some cost in the purity of the different signal regions with respect to the final state that contributes to a given region (See related discussion in Sec.~\ref{subsec:Results_500GeV},  particularly Table~\ref{tab:SignalEfficiencies}.)}
Additionally, when reconstructing mass peaks for 100 GeV $< M_{\tau^\prime} <$ 125 GeV where $\tau^\prime$ only decays to $Z\tau$ and $W\nu_\tau$, we reconstruct candidate $Z$ bosons (instead of candidate $h$ bosons)
from a $b$-jet pair with an invariant mass
\beq
M_{bb} = \left.M_Z\right\vert^{+7.5 \text{ GeV}}_{-15 \text{ GeV}}.
\label{eq:Zbbreco}
\eeq

In the SRs with exactly $1\tau$, $W$ bosons are also a target. There, we reconstruct candidate $W$ bosons from jet pairs in the window\footnote{In the SRs with exactly 1$\tau$, the invariant mass regions chosen for the hadronic decays of the $Z$ and $W$ overlap in the region $M_Z - 15\text{ GeV} < M_{jj/jb} < M_W + 5\text{ GeV}$.
In this region, we consider the reconstructed boson both as a candidate $Z$ and a candidate $W$, and proceed with our algorithm below.}
\begin{equation}
M_{jj/jb} = \left.M_W\right\vert^{+5 \text{ GeV}}_{-25 \text{ GeV}}.
\label{eq:Wreco}
\end{equation}
While we nominally do not expect a $b$ jet from
the decays of $W$ boson,
in SRs with an odd number of tagged $b$ jets,
we will see below that there is a non-trivial contribution from reconstructing a $W$ boson in the cases where a
$c$ jet (or $u/d/s$-quark initiated jet in rare cases) fakes a $b$ jet.

Note that when reconstructing candidate $Z$, $h$, and $W$ bosons from jet pairs, we allow a broader invariant mass range below their respective masses, as can be seen in Eqs.~(\ref{eq:Zreco}) - (\ref{eq:Wreco}). This is because the invariant mass reconstructions of these bosons have more support in the low tail.

\noindent
\emph{Peak reconstruction algorithm:} 
When reconstructing the mass peaks in events with $\tau$ leptons in the final state, one must deal with the fact that some fraction of each $\tau$ is carried away by neutrinos. To help restore this momenta, we make use of the collinear approximation, which says that in the detector frame, the three-momentum of each neutrino is in nearly the same direction as the parent $\tau$. This approximation should be excellent for $\tau^{\prime}$ masses much larger than the bosons of the SM. Let us denote the hadronic $\tau$ 3-momentum visible in the detector as $\xvec{p}_{\tau}^{\rm vis}$.   
We would like to estimate the 3-momentum of the $\tau$ lepton before it decayed, which we denote $\xvec{p}^{\rm est}$. In the collinear approximation, they are related by
\beq
\xvec{p}_{\tau}^{\rm est} &=& r \xvec{p}_{\tau}^{\rm vis},
\eeq
which defines $r$ such that the momentum fraction carried by the visible decay products, $1/r$, is between 0 and 1.

We then employ the following algorithm to reconstruct the mass of $\tau^\prime$ in each event:
\begin{itemize}
    \item Compute the total visible four-momentum in an event $p^\mu_{\rm vis}$ only from
    $\ell$, $j$, $b$, and $\tau$
  as specified in Eqs.~(\ref{eqn:howmanyleptonjets}) and (\ref{eqn:howmanytaus}), ignoring the visible momenta of photons and/or objects that did not pass the relevant object cuts in Eqs.~(\ref{eq:pTcut})-(\ref{eq:lj_isolation}).
    
    \item Then, compute the total missing four-momentum using $\cancel{p}^\mu = p^\mu_{\rm tot} - p^\mu_{\rm vis}$, where $p^\mu_{\rm tot}$ is the
    total four-momentum in the event.
    Here, we simply take $p^\mu_{\rm tot} \simeq (\sqrt{s}, \xvec{0})$,
    ignoring the impact of the ISR due to the lack of a
    reliable way of estimating the effective center-of-mass energy
    on an event-by-event basis.  
    (This approximation, as can be seen for example in Figure~\ref{fig:Mtaup_sqrts500}, will lead to reconstructed mass distributions with tails above the mass peak $\sim M_{\tau^\prime}$.)
    \item Reconstruct --as described above-- all candidate $Z/h$ bosons, $B_\alpha$, in any of the SRs,
    in addition to reconstructing candidate $W$ bosons, $W_\beta$, in SRs with $1\tau$.
    The criteria in Eqs.~(\ref{eq:Zreco})-(\ref{eq:Wreco}) to reconstruct the candidate bosons can lead to the exclusion of a notable fraction of events.
    \item Find all the possible (tau, boson) pairings that reconstruct a $\tau^\prime$ pair:
    \beq
    \tau^\prime_1 \supset (\tau_1, \nu_1, B_\alpha)
    \text{ and }
    \tau^\prime_2 \supset
    \begin{cases}
        (\tau_2, \nu_2, B_\beta) \text{ in SRs with exactly } 2\tau\\
        (\nu_2, W_\beta) \text{ in SRs with exactly } 1\tau
    \end{cases}
    \label{eq:pairing}
    \eeq
    such that the bosons
    in $\tau_1^\prime$ and $\tau_2^\prime$
    are distinct. In SRs with $2\tau$, $\tau_1$ is usually taken to be the $\tau$ with highest energy.  However, in the case that there is exactly one $Z$ reconstructed from leptons, we instead  relabel as $\tau_1$ the $\tau$ in each pairing that is being paired with the leptonically decaying $Z$.
    We label  the neutrino from $\tau_1$ decay as $\nu_1$.  In SR with two $\tau$ leptons,   $\nu_2$ is the neutrino
    from $\tau_2$ decay, while in the $1 \tau$ SR, $\nu_{2}$ is the neutrino that is produced in association with the $W$ in the decay of the $\tau^{\prime}$. 
    \item If any of the candidate $Z$, $h$, or $W$ bosons in a pairing are reconstructed from (possibly $b$-tagged) jets, rescale the four-momentum of the two jets that formed the candidate boson by a common factor such that their invariant mass is exactly $M_Z$, $M_h$, or $M_W$, respectively. This modifies $p^\mu_{\rm vis}$; therefore $\cancel{p}^\mu$ is
    adjusted such that 
    $p^\mu_{\rm vis} + \cancel{p}^\mu = p^\mu_{\rm tot}$.
    \item Use the collinear approximation for the neutrino $\nu_1$ in the decay of $\tau_1$:
    \beq
    E_{\nu_1} = |\xvec{p}_{\nu_1}|, \quad \xvec{p}_{\nu_1} = (r - 1) \xvec{p}_{\tau_1},
    \label{eq:pmu_nu1}
    \eeq
    and obtain the four-momentum of the other neutrino using:
    \beq
    E_{\nu_2} = \cancel{E} - E_{\nu_1}, \qquad \xvec{p}_{\nu_2} = \frac{E_{\nu_2}}{|\xvec{\cancel{p}} - \xvec{p}_{\nu_1}|} \ \left(\xvec{\cancel{p}} - \xvec{p}_{\nu_1}\right)
    ,
    \label{eq:pmu_nu2}
    \eeq
    such that both $\nu_1$ and $\nu_2$ are on-shell.
    \item For each pairing, as in Eq.~(\ref{eq:pairing}), solve for $r$ by imposing that the reconstructed $\tau^\prime$ masses are equal:
    \beq
    p^2_{\tau^\prime_1} &=&
    p^2_{\tau^\prime_2},
    \eeq
    which, after using four-momentum conservation
    $p^\mu_{\tau^\prime_2} = p^\mu_{\rm tot} - p^\mu_{\tau^\prime_1}$,
    simplifies to
    \beq
    p^2_{\rm tot} &=& 2 p_{\rm tot} \cdot p_{\tau^\prime_1}.
    \eeq
    With $p^\mu_{\tau^\prime_1} = p^\mu_{B_\alpha} + p^\mu_{\tau_1} + p^\mu_{\nu_1}$ and
    $p^\mu_{\rm tot} \simeq (\sqrt{s}, \xvec{0})$, we obtain
    \beq
    r &=& 1 + \frac{1}{\left | \xvec{p}_{\tau_1} \right |} \left ( \frac{\sqrt{s}}{2} - E_{B_\alpha} - E_{\tau_1}\right).
    \label{eq:r}
    \eeq

    We impose $E_{\nu_1} \geq 0$ and $E_{\nu_2} \geq 0$ by requiring:
    \beq
    1 \leq r \leq 1 + \frac{\cancel{E}}{\left| \xvec{p}_{\tau_1} \right|}.
    \eeq
    A pairing is rejected if the corresponding $r$ does not satisfy this requirement.
    This therefore excludes some events  if no pairing survives.  We find this especially occurs for smaller $M_{\tau^{\prime}}$ where the collinear approximation may not be as good.
     
    \item Assuming that multiple candidate pairings for the event remain, the ambiguity is resolved by choosing the pairing (with its attendant calculated $r$)  that minimizes the magnitude of the total three-momentum in the event:
    \beq
    \xvec{p}_\text{total} = \xvec{p}_\text{vis} + \xvec{p}_{\nu_1} + \xvec{p}_{\nu_2}.
    \eeq
    This pairing is then used to  compute the reconstructed $\tau^{\prime}$ mass
    \beq
    M^{\rm reco}_{\tau^\prime} &=& \sqrt{p^2_{\tau^\prime_1}}.
    \eeq 
\end{itemize}
Note that if there is exactly one leptonic $Z$ boson in the event, it is always used for the $M^{\rm reco}_{\tau'}$ returned by the algorithm. This takes advantage of the improved mass resolution in the leptonic channel. Otherwise, the reconstructed $\tau'$ mass used is the one for which the $\tau$ has the higher energy in $2\tau$ events, or the one associated with the $\tau$ in $1\tau$ events.

\section{Results at $e^+ e^-$ colliders\label{sec:Results}}
\setcounter{equation}{0}
\setcounter{figure}{0}
\setcounter{table}{0}
\setcounter{footnote}{1}
In this section, we present results for planned future $e^+ e^-$ colliders.
We begin with results for a $\sqrt{s} =$ 500 GeV collider, and discuss in detail the mass reconstruction
prospects for a wide range of $\tau^\prime$ masses in various signal regions. We then present results for two potential colliders with $\sqrt{s} =$ 250 and 380 GeV, and we show that despite their relatively low center of mass energies, such machines can improve on the present LHC exclusion. We then move to higher energies, presenting results for $\sqrt{s} =$ 1, 1.5, and 3 TeV.
In all cases, our results show that the discovery of the $\tau^{\prime}$ up to very near the kinematic limit should usually be straightforward, although for a 3 TeV machine statistics can be a limiting factor for masses near threshold. 

\subsection{$\sqrt{s} =$ 500 GeV\label{subsec:Results_500GeV}}

\begin{figure}[!h]
  \begin{center}
    \begin{minipage}[]{0.495\linewidth}
     \includegraphics[width=8cm]{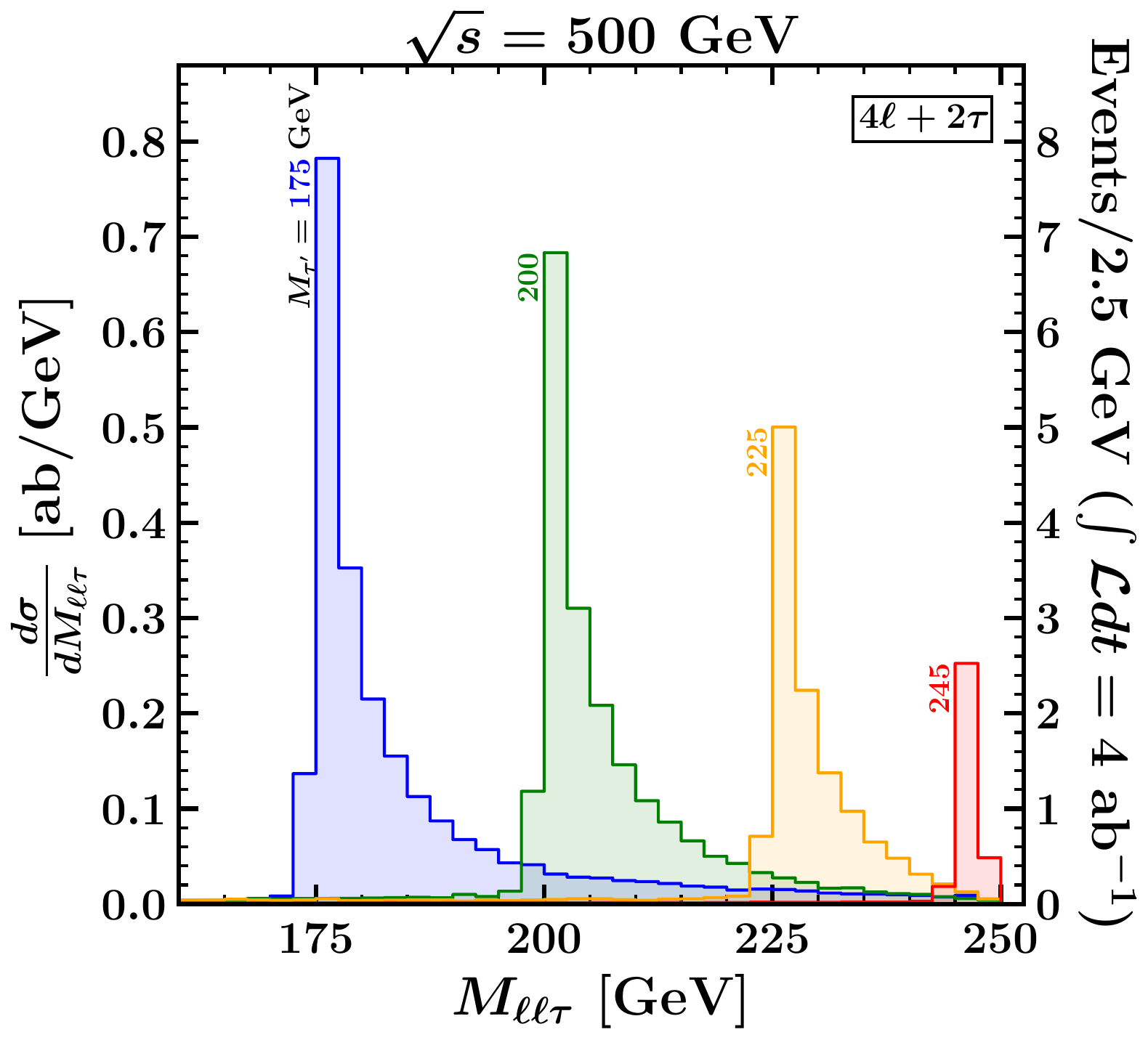}
   \end{minipage}
  \end{center}
  \begin{minipage}[]{0.495\linewidth}
    \includegraphics[width=8cm]{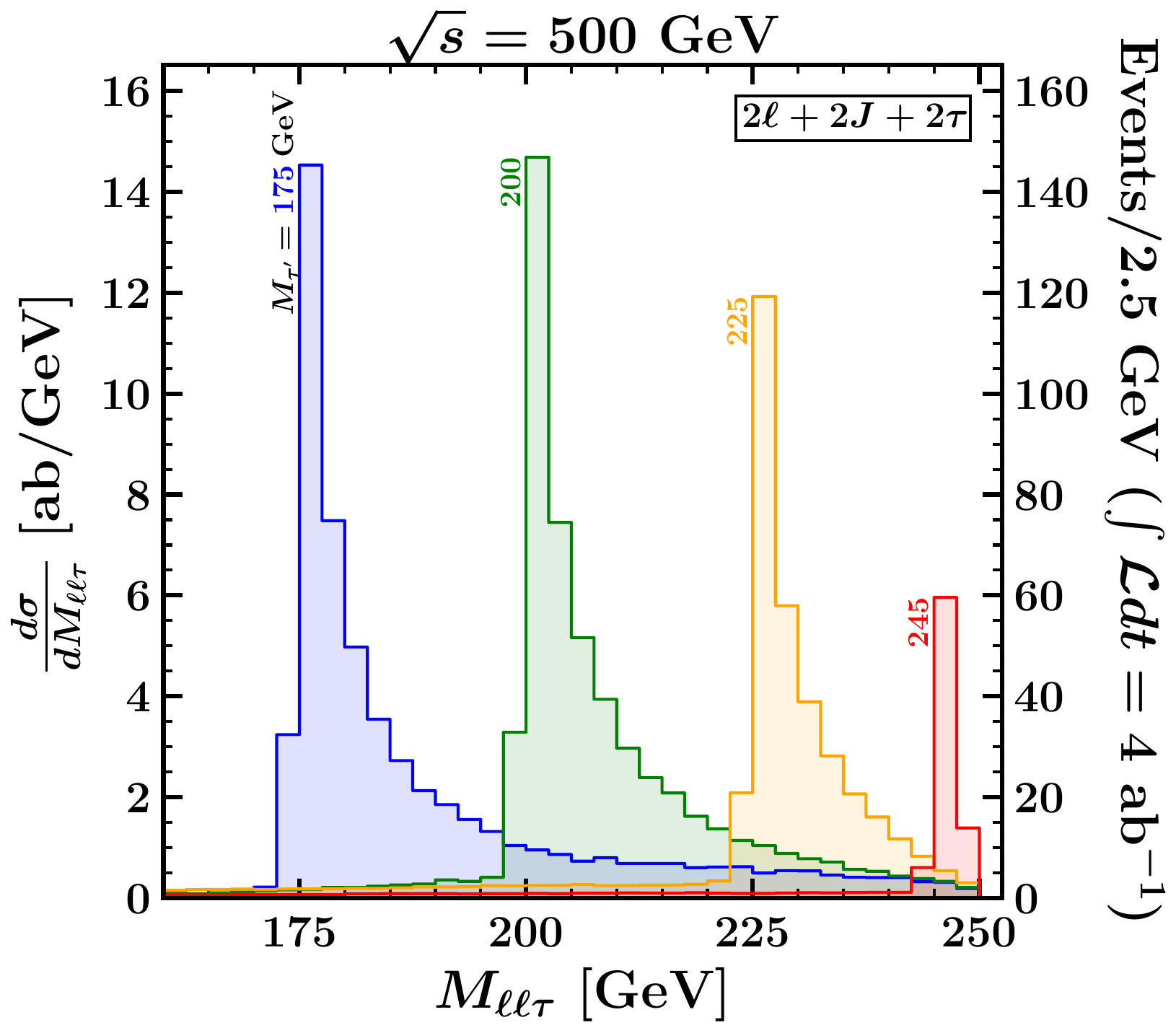}
  \end{minipage}
  \begin{minipage}[]{0.495\linewidth}
    \includegraphics[width=8cm]{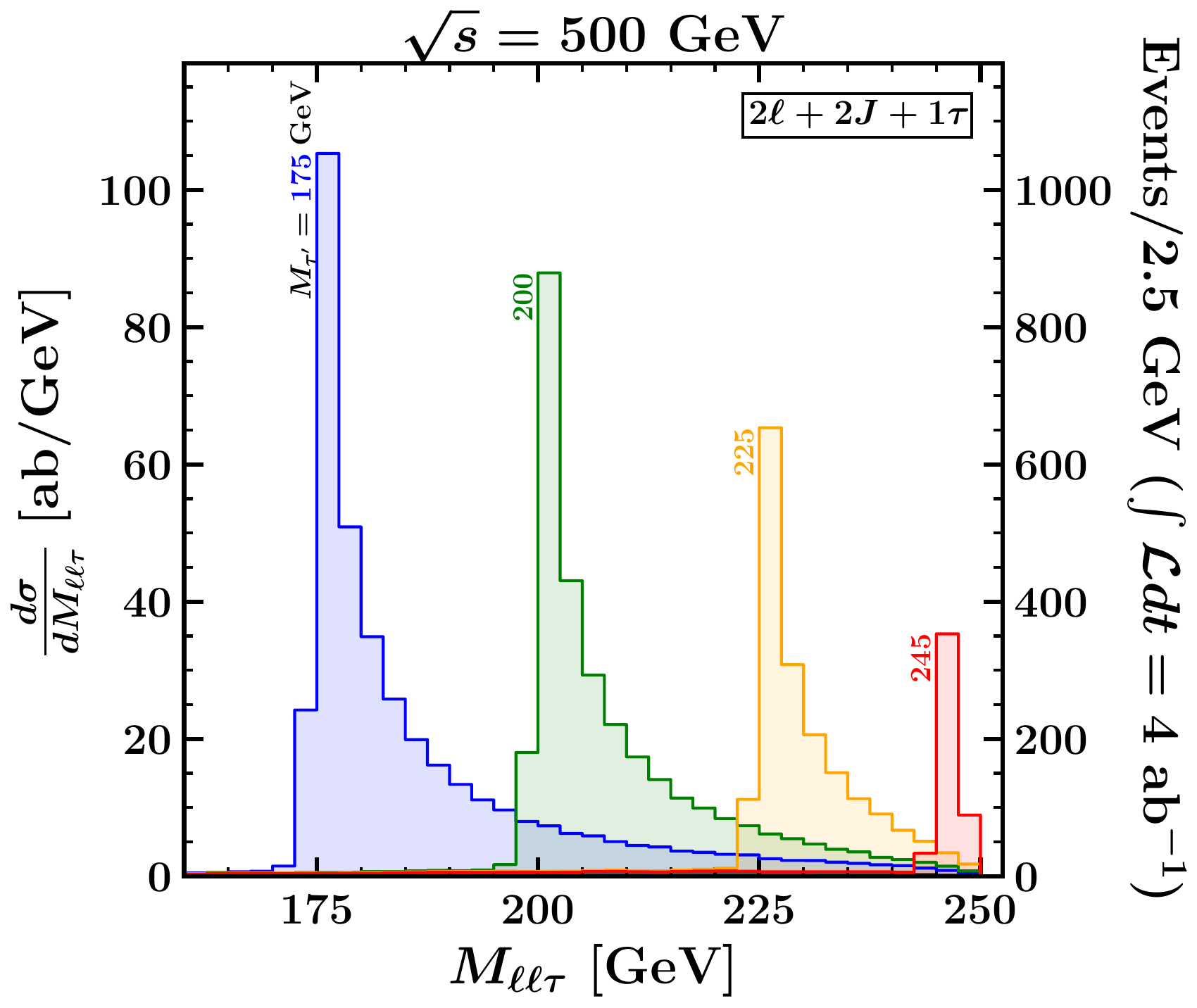}
  \end{minipage}
  \vspace{0.2cm}
  \begin{minipage}[]{0.495\linewidth}
    \includegraphics[width=8cm]{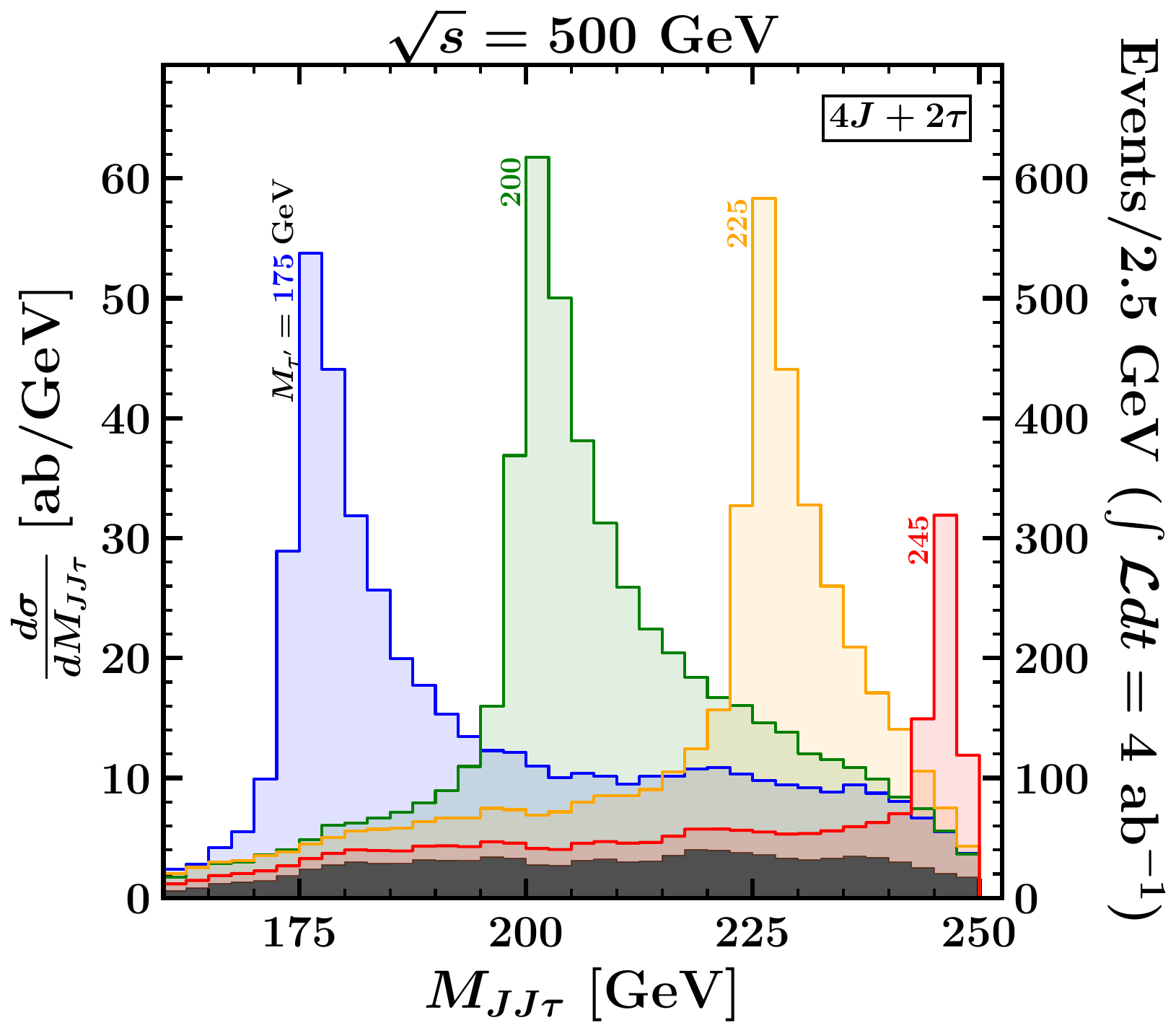}
  \end{minipage}
    \begin{minipage}[]{0.495\linewidth}
    \includegraphics[width=8cm]{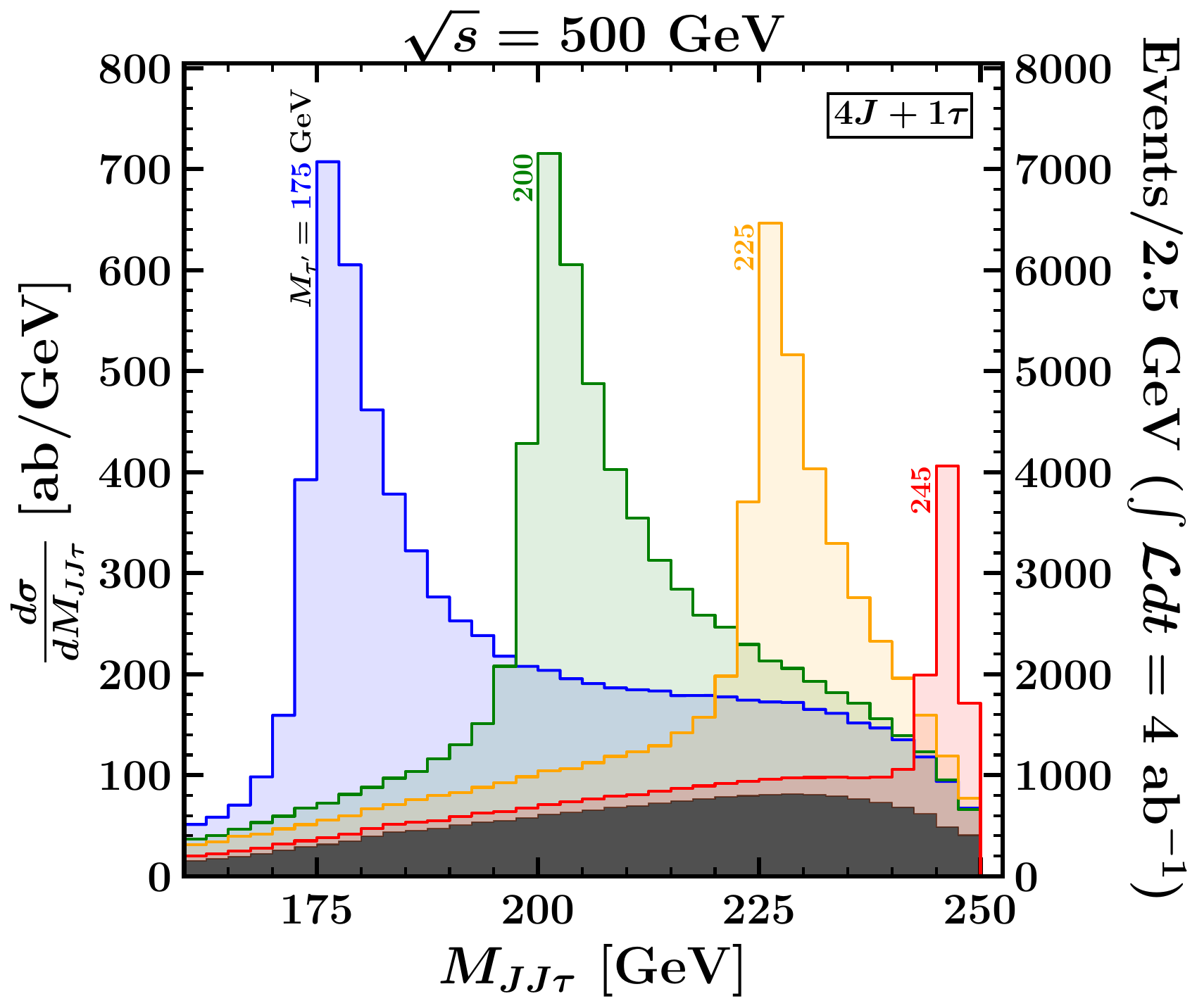}
  \end{minipage}
 \caption{Reconstructed $\tau^\prime$ mass peaks in five different signal region
 combinations, for four different $M_{\tau^\prime} =$ 175, 200, 225, and 245 GeV as labeled, for $e^+ e^-$ collisions at $\sqrt{s} = 500$ GeV. Backgrounds are shown as gray-shaded histograms stacked together with the signal histograms.
\label{fig:Mtaup_sqrts500}}
\end{figure}

In Fig.~\ref{fig:Mtaup_sqrts500}
we show the $\tau^\prime$ mass peaks, reconstructed using the algorithm described in the previous section, at a $\sqrt{s} = 500$ GeV $e^+e^-$ collider. The different colored peaks, stacked together with the backgrounds (gray-shaded histograms), correspond to four different choices $M_{\tau^{\prime}} = 175$, 200, 225, and 245 GeV, as labeled. On each right vertical axis is shown the events per (2.5 GeV) bin, under the assumption of an integrated luminosity of 4 ab$^{-1}$. Clear mass peaks are visible, even for $M_{\tau^\prime}=245$ GeV at a 500 GeV collider. The first panel shows results in which two leptonically decaying $Z$ bosons were reconstructed, resulting in narrow $\tau'$ mass peaks and negligible backgrounds but poor signal statistics with the assumed integrated luminosity. The two panels in the middle row show signal regions that feature exactly one leptonically decaying $Z$ boson, which is used to reconstruct the $\tau'$ mass peak. Again the backgrounds are essentially negligible, and the signal mass peaks are narrow, with good statistics. 
The last two panels of Figure \ref{fig:Mtaup_sqrts500} show signal regions without leptonically decaying $Z$ bosons occurring in the reconstruction. This gives the best signal statistics, but significantly wider mass peak resolutions, and backgrounds that are non-negligible but still clearly under control.

Note that while we pay attention to the $b$ tagging during our reconstruction algorithm---to decide, for example, whether to try to reconstruct a Higgs boson--- in the last four panels of Figure \ref{fig:Mtaup_sqrts500} we have combined non-overlapping signal regions with both $b$-tagged and non-$b$-tagged jets, as indicated by the notation $J$, which stands for $j$ or $b$.  For example, in the bottom-right panel the signal regions of Eqs.~(\ref{eq:4j1ta})-(\ref{eq:1j3b1ta}) are summed. 
We find that the largest contribution to the background for SRs with $b$ jets is from $t\overline{t}$, while backgrounds in SRs without $b$ jets are dominated by $W^+ W^- Z$ production. These backgrounds are smoothly varying as a function of the reconstructed mass, so an effective subtraction should be possible.

\begin{figure}[!h]
  \begin{minipage}[]{0.495\linewidth}
    \includegraphics[width=8cm]{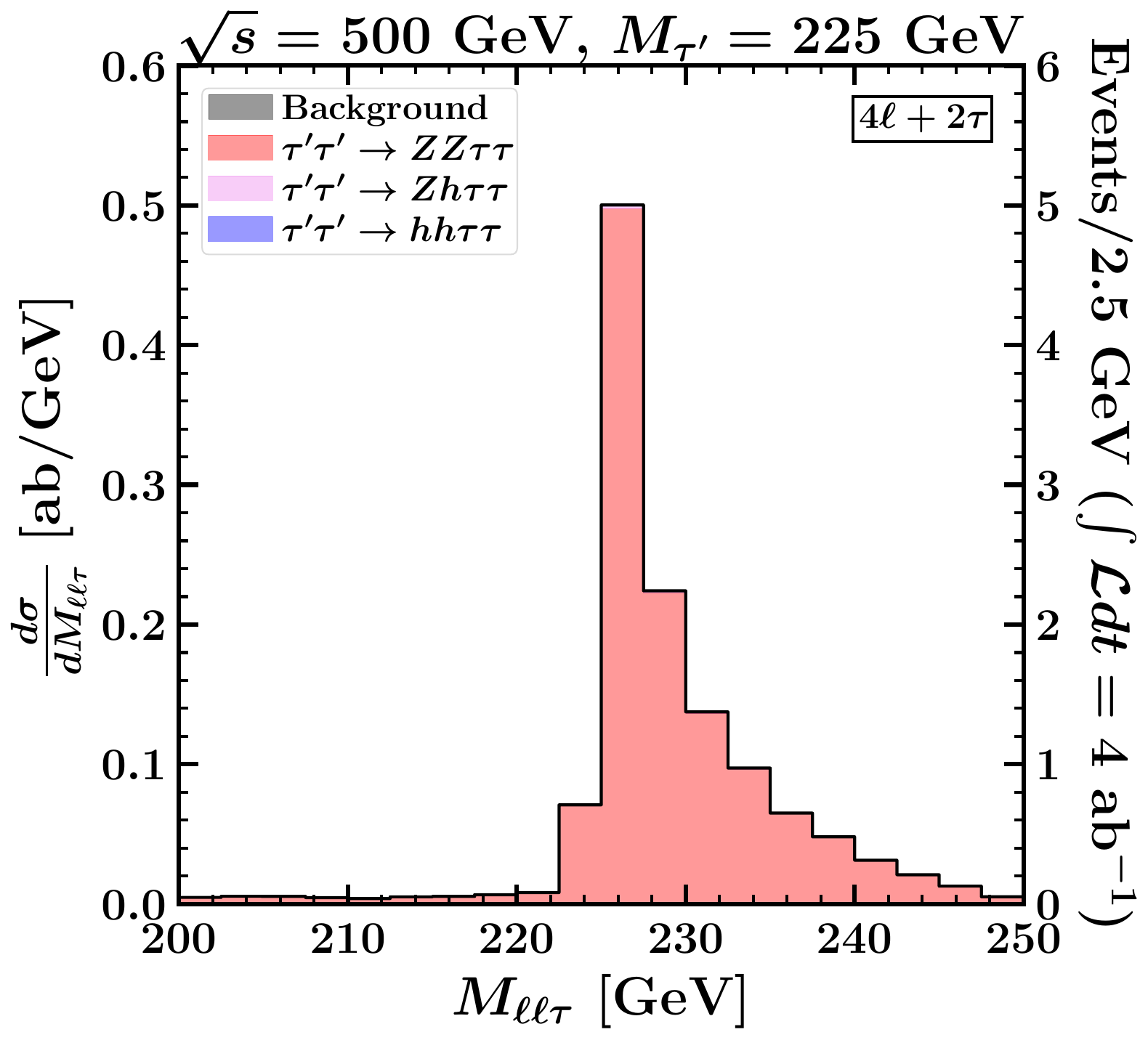}
  \end{minipage}
  \begin{minipage}[]{0.495\linewidth}
    \includegraphics[width=8cm]{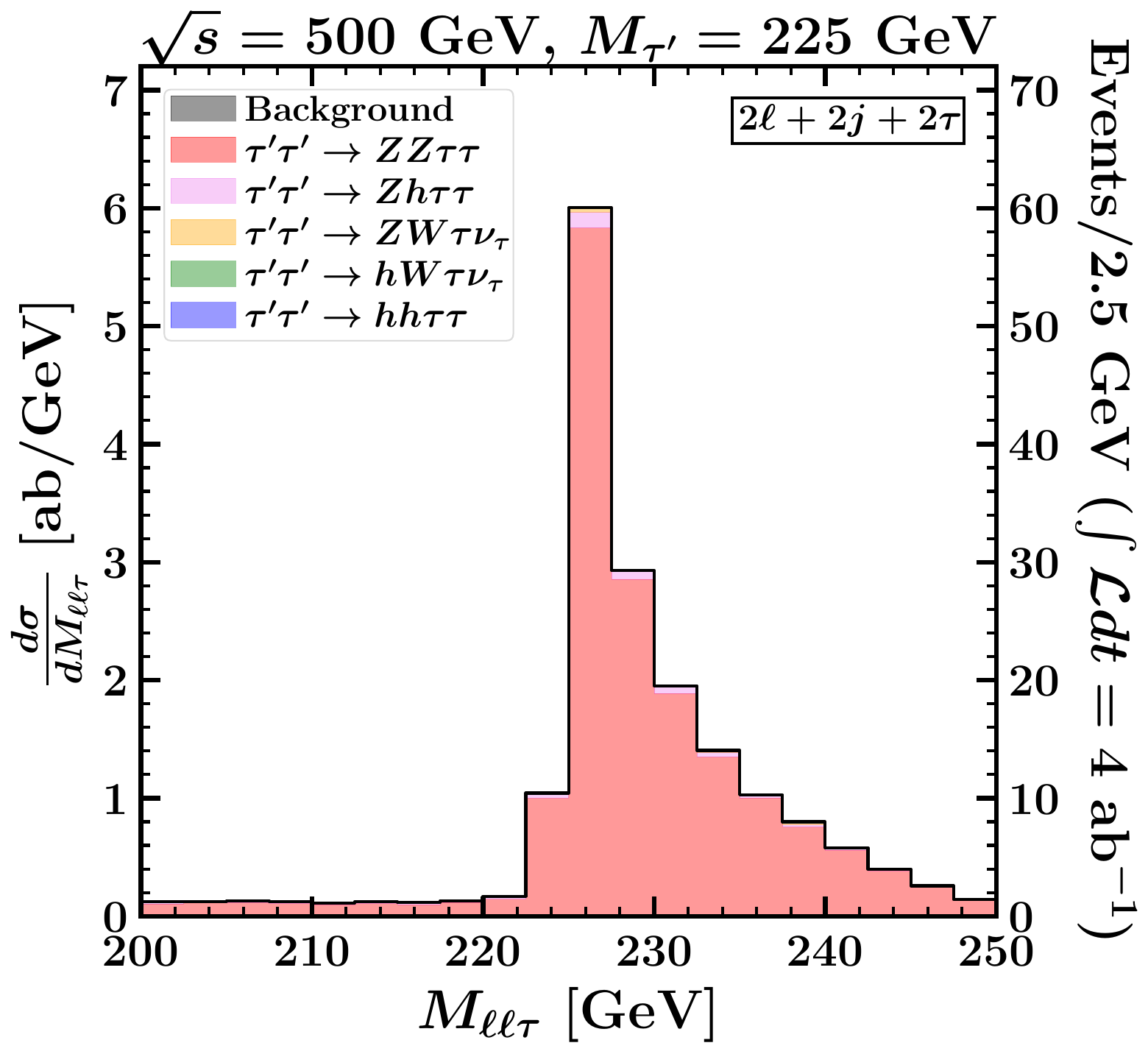}
  \end{minipage}
  \vspace{0.2cm}
  \begin{minipage}[]{0.495\linewidth}
    \includegraphics[width=8cm]{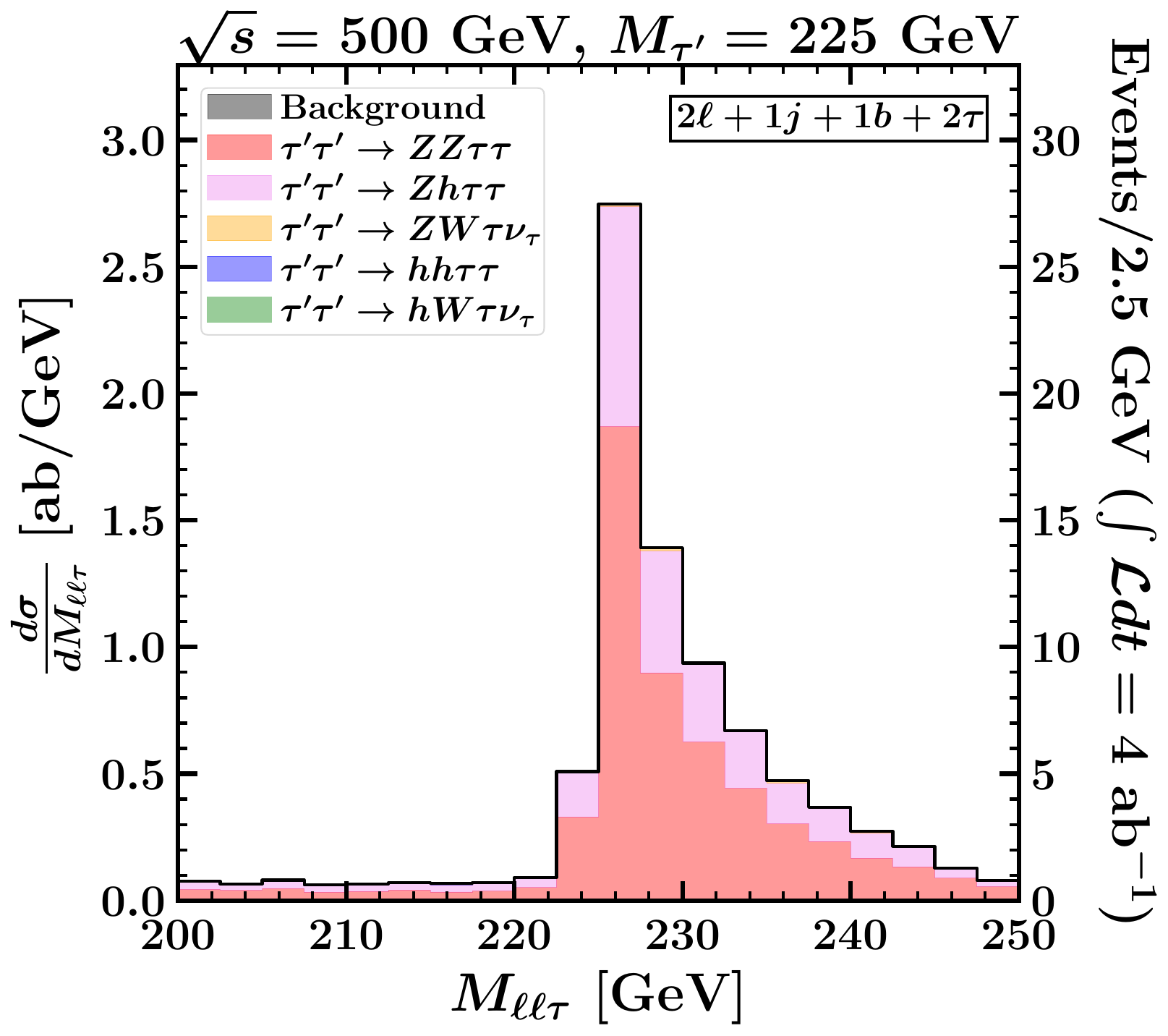}
  \end{minipage}
  \begin{minipage}[]{0.495\linewidth}
    \includegraphics[width=8cm]{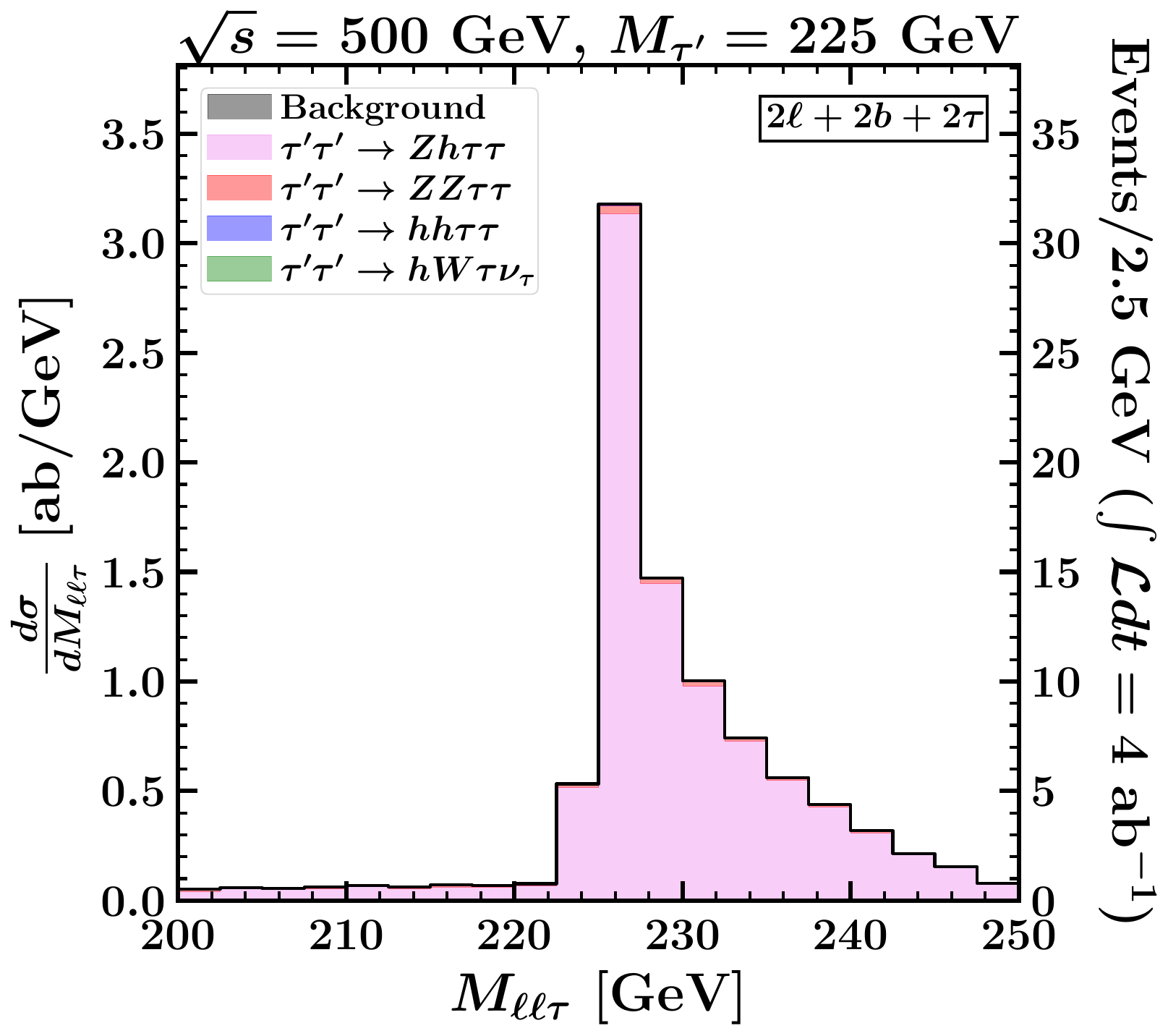}
  \end{minipage}
 \caption{Signal final states contributing to the mass peak
 for $M_{\tau^\prime} = 225$ GeV in four different signal regions with $2\tau$ where at least one of the two $\tau^\prime$ particles is reconstructed from a leptonically decaying $Z$.
 Backgrounds are too small to be visible.
\label{fig:Mtaup_sqrts500_Breakdown_SR2tau_leptons}}
\end{figure}

\begin{figure}[!h]
  \begin{center}
   \begin{minipage}[]{0.495\linewidth}
     \includegraphics[width=8cm]{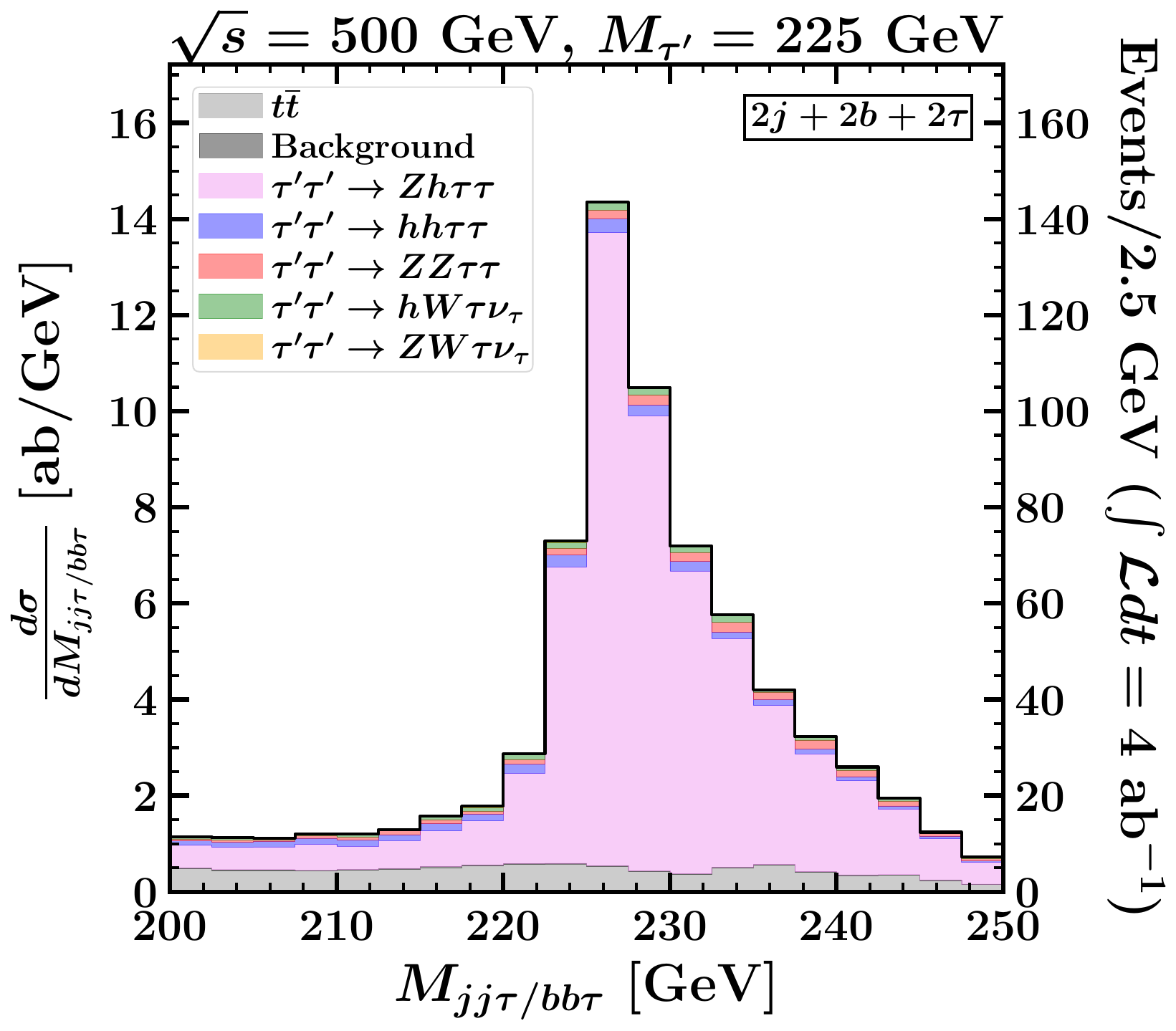}
   \end{minipage}
  \end{center}
  \vspace{0.2cm}
  \begin{minipage}[]{0.495\linewidth}
    \includegraphics[width=8cm]{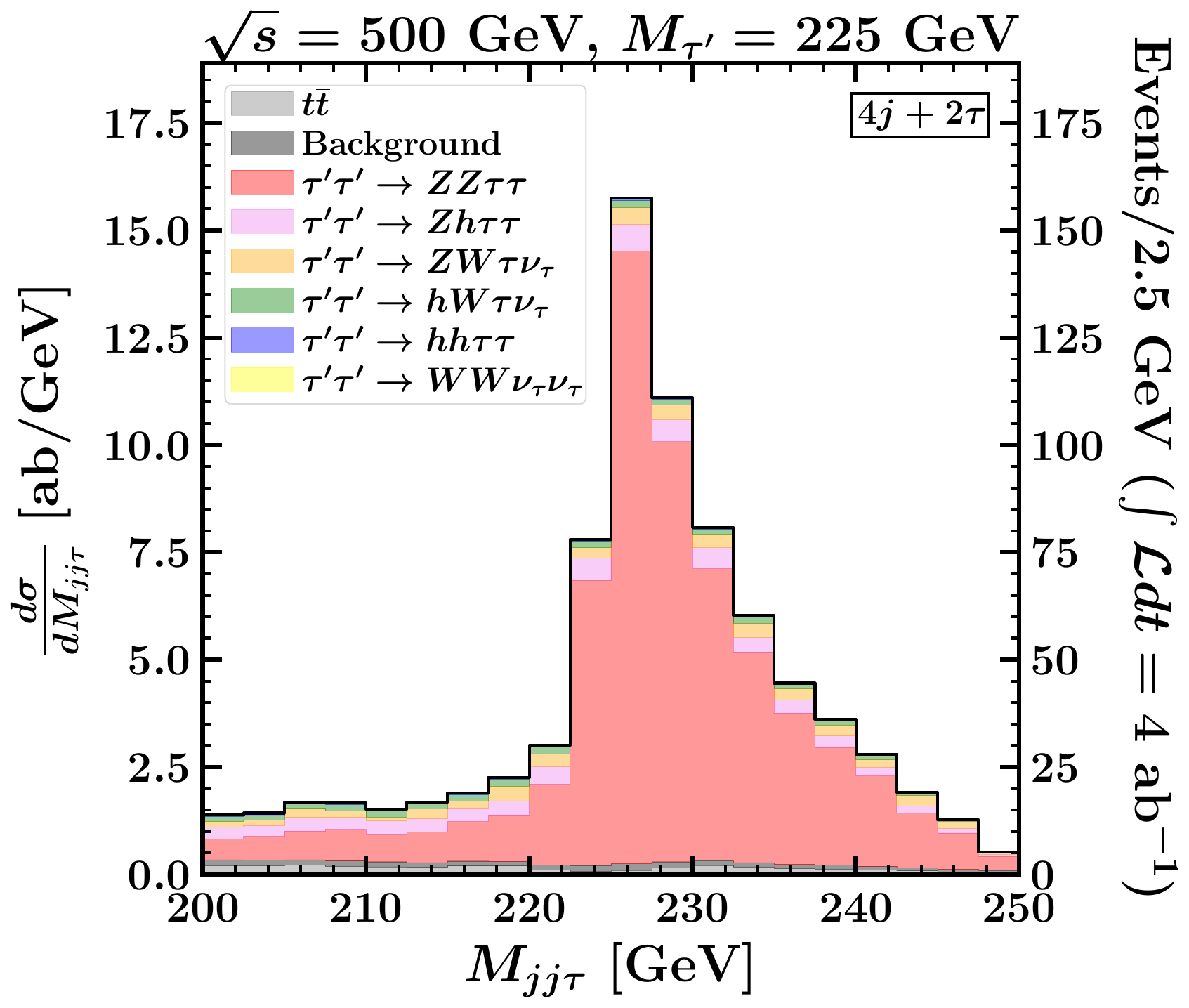}
  \end{minipage}
  \begin{minipage}[]{0.495\linewidth}
    \includegraphics[width=8cm]{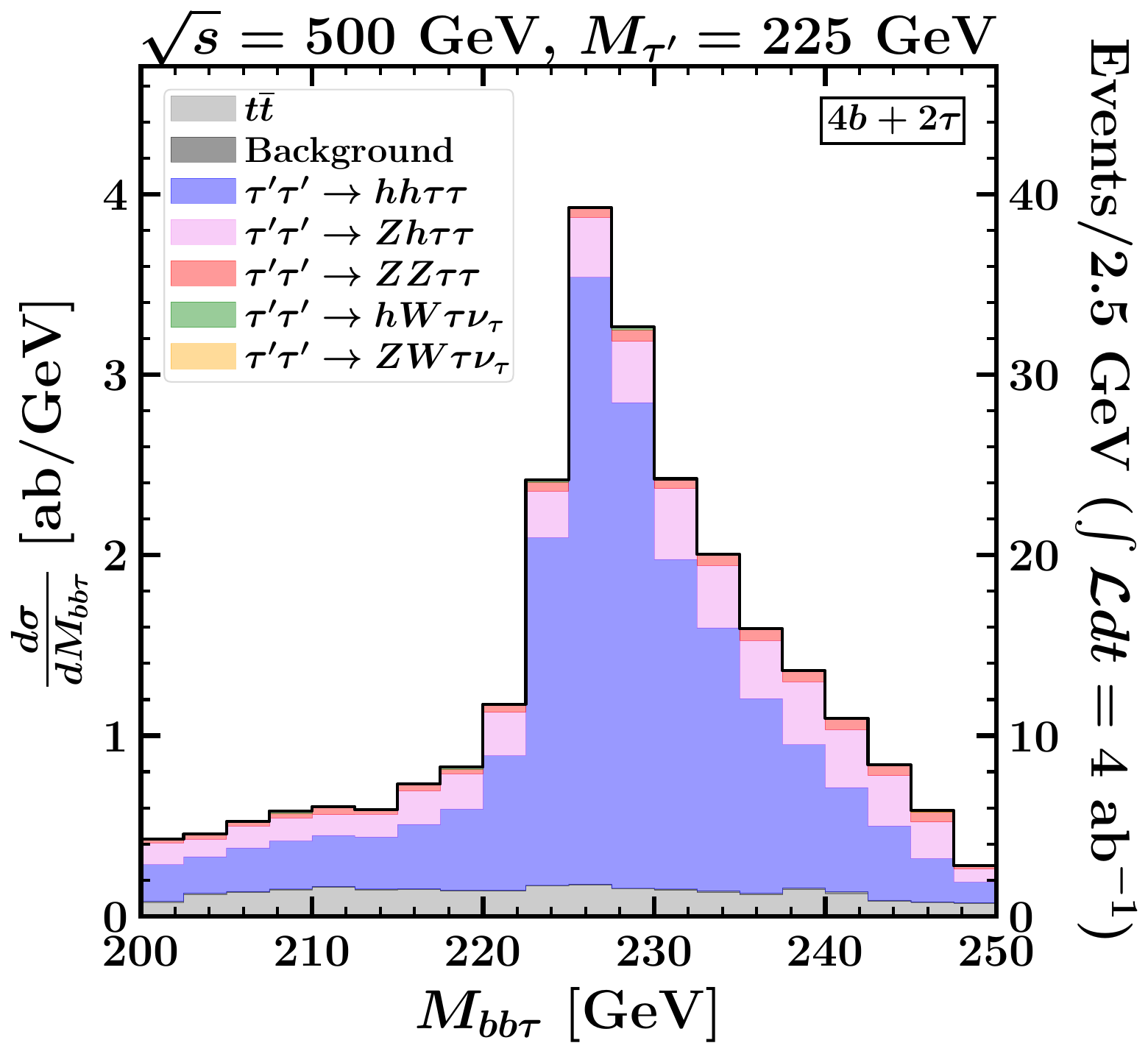}
  \end{minipage}
  \vspace{0.2cm}
  \begin{minipage}[]{0.495\linewidth}
    \includegraphics[width=8cm]{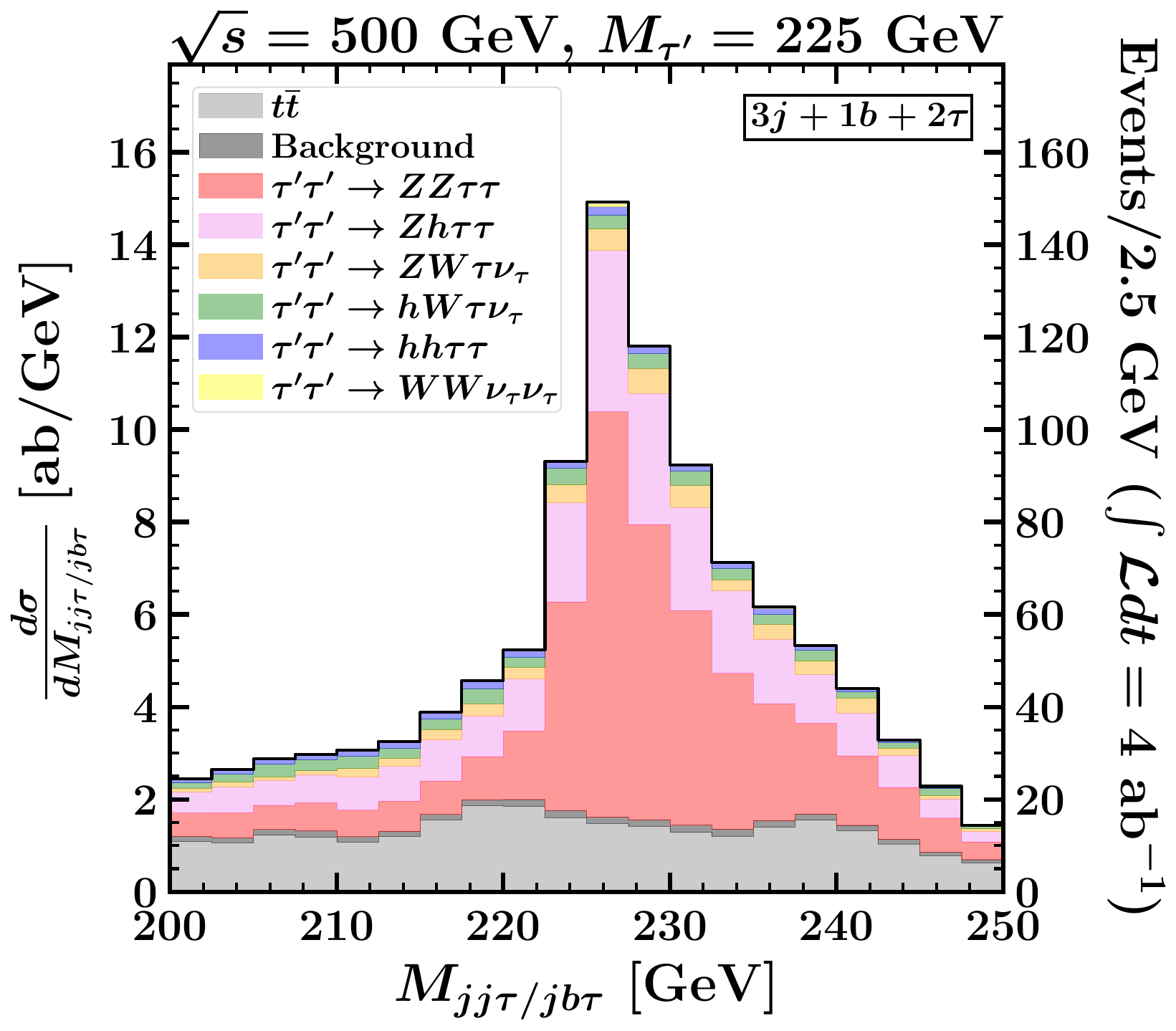}
  \end{minipage}
  \begin{minipage}[]{0.495\linewidth}
    \includegraphics[width=8cm]{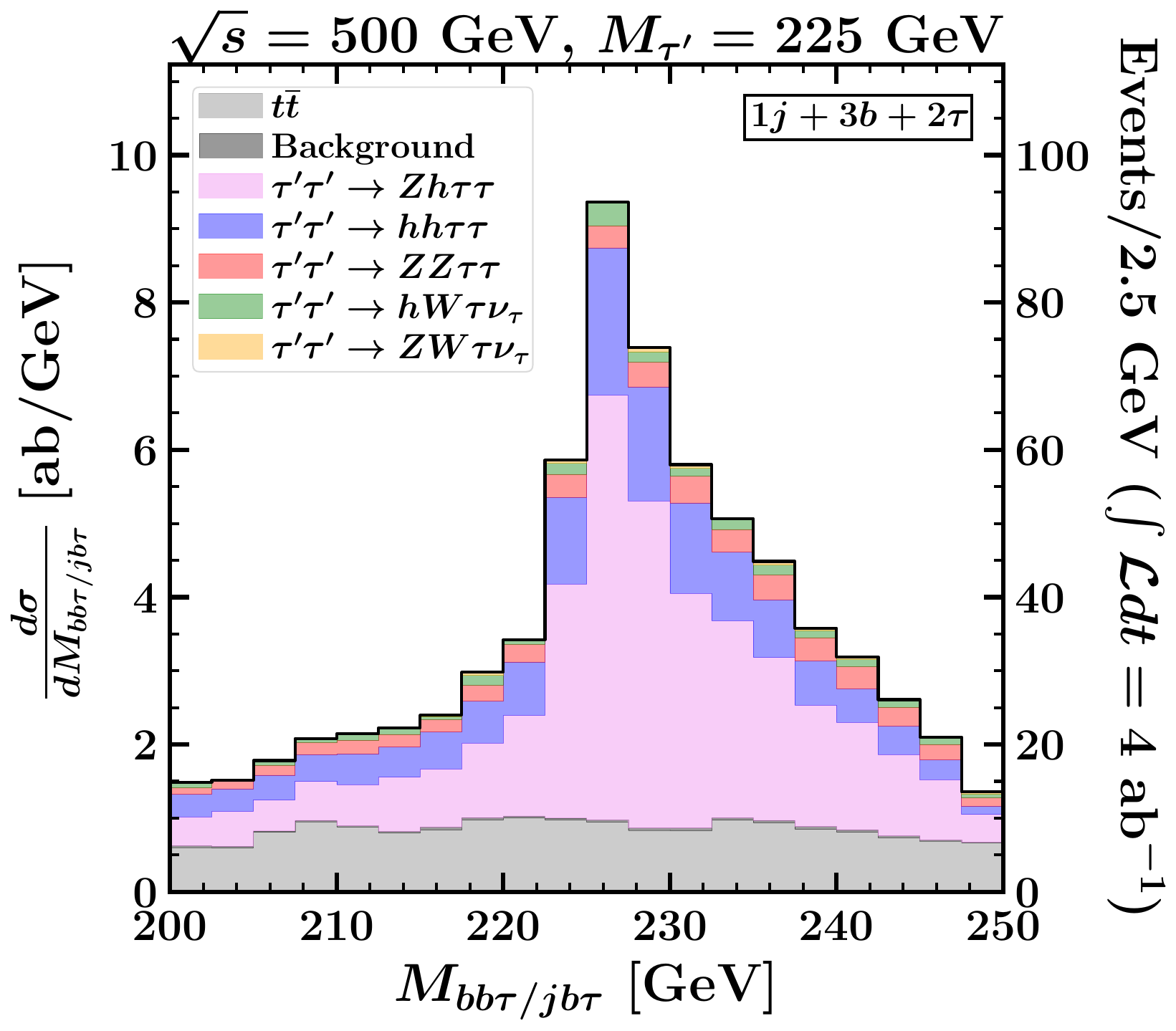}
  \end{minipage}
 \caption{Signal final states contributing to the mass peak
 for $M_{\tau^\prime} = 225$ GeV in five different signal regions with $2\tau$ where both $\tau^\prime$ leptons are reconstructed from hadronically decaying $Z/h$.
 $t\overline{t}$ and the other backgrounds are shown as gray and black shaded histograms,
 respectively, stacked together with the signal contributions.
\label{fig:Mtaup_sqrts500_Breakdown_SR2tau_jets}}
\end{figure}

\begin{figure}[!h]
  \begin{minipage}[]{0.495\linewidth}
    \includegraphics[width=8cm]{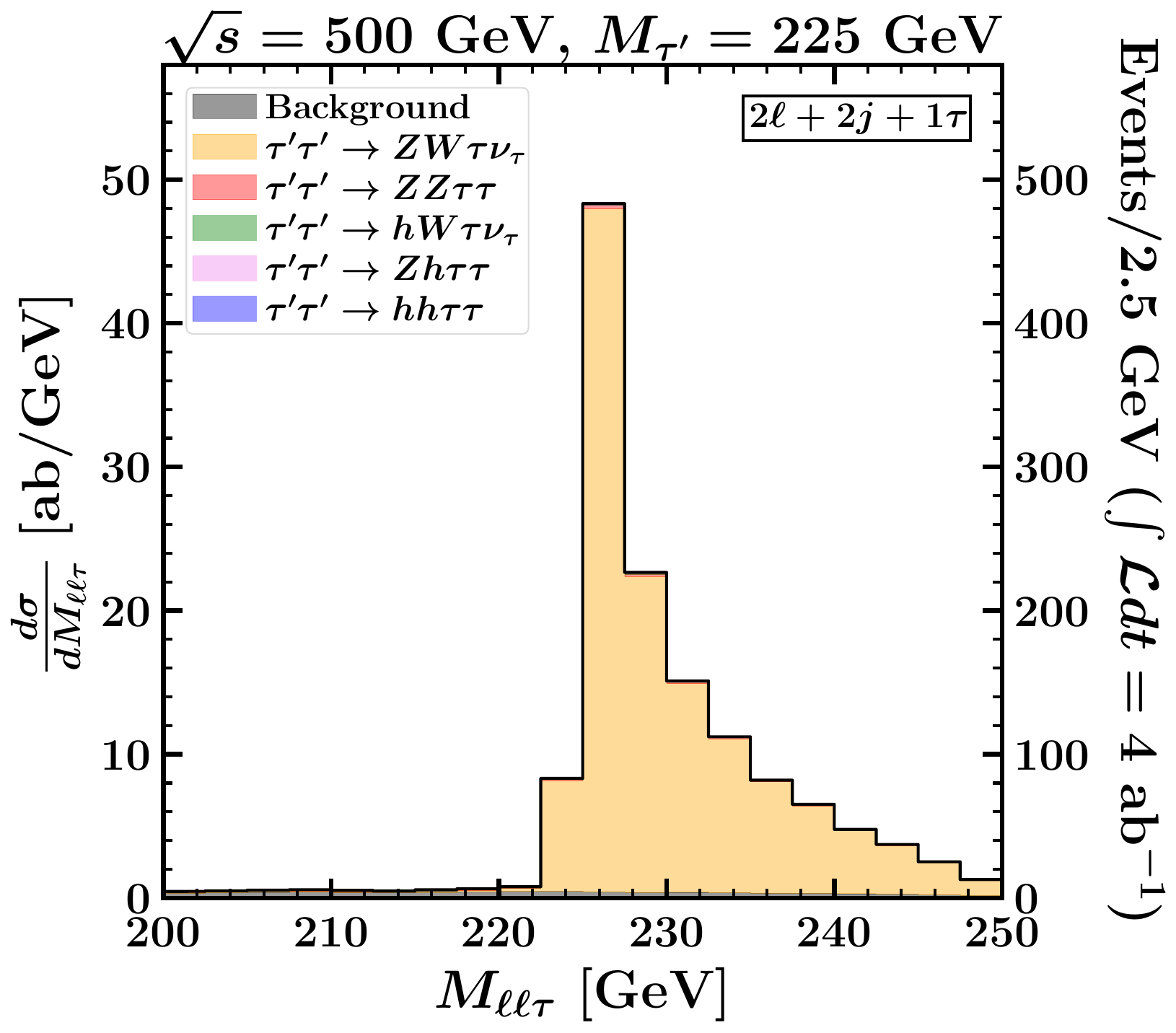}
  \end{minipage}
  \begin{minipage}[]{0.495\linewidth}
    \includegraphics[width=8cm]{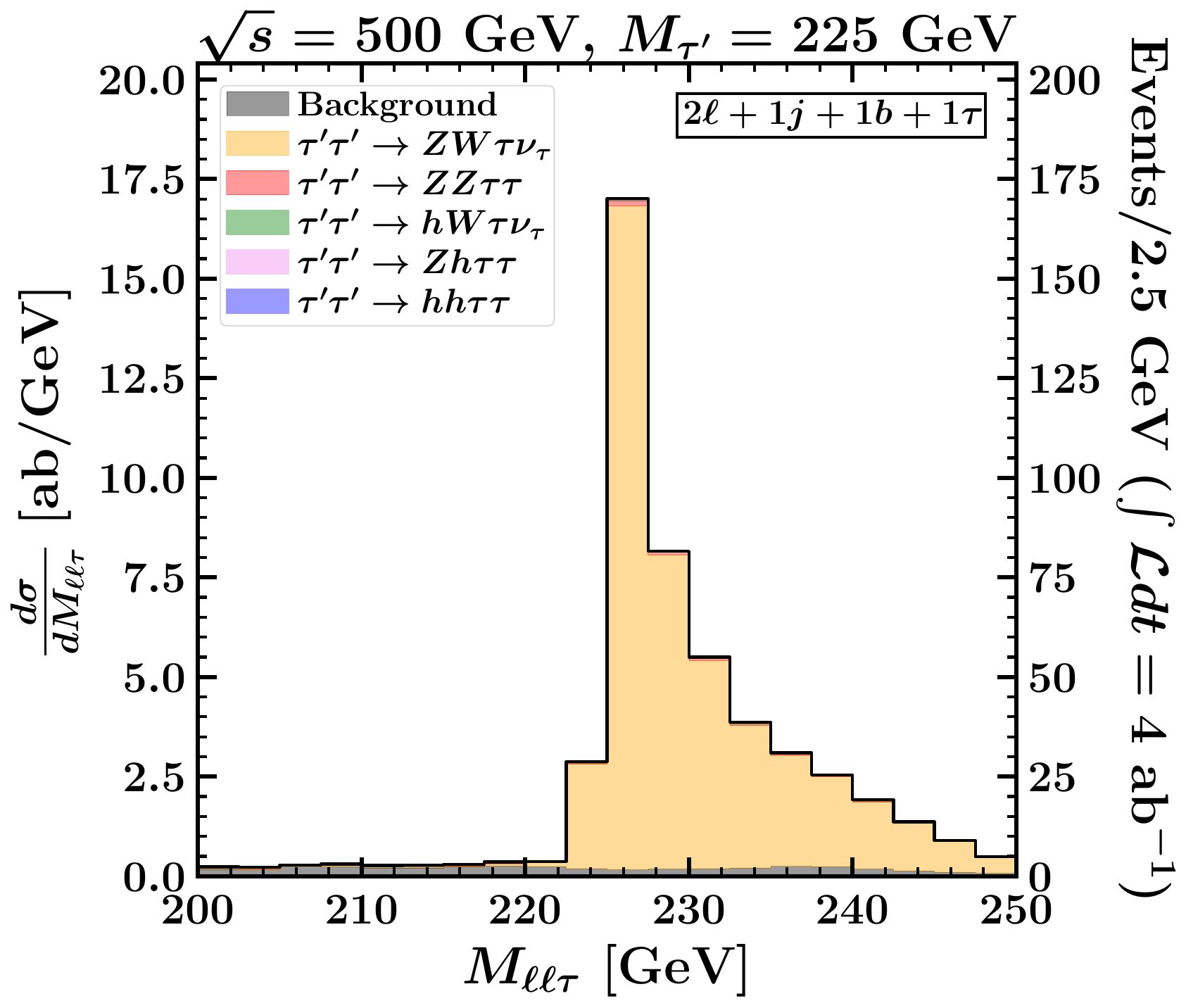}
  \end{minipage}
  \vspace{0.2cm}
  \begin{minipage}[]{0.495\linewidth}
    \includegraphics[width=8cm]{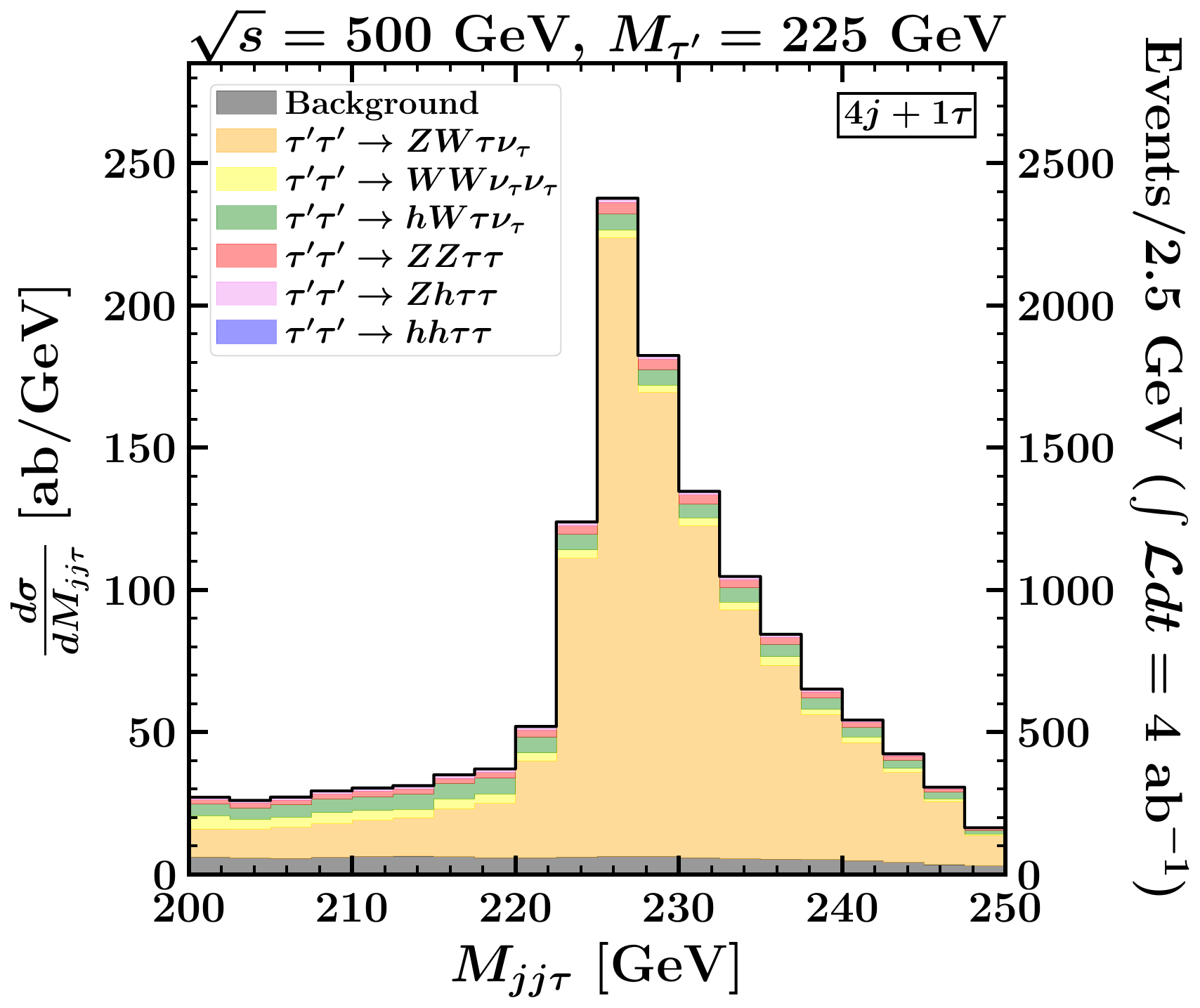}
  \end{minipage}
  \begin{minipage}[]{0.495\linewidth}
    \includegraphics[width=8cm]{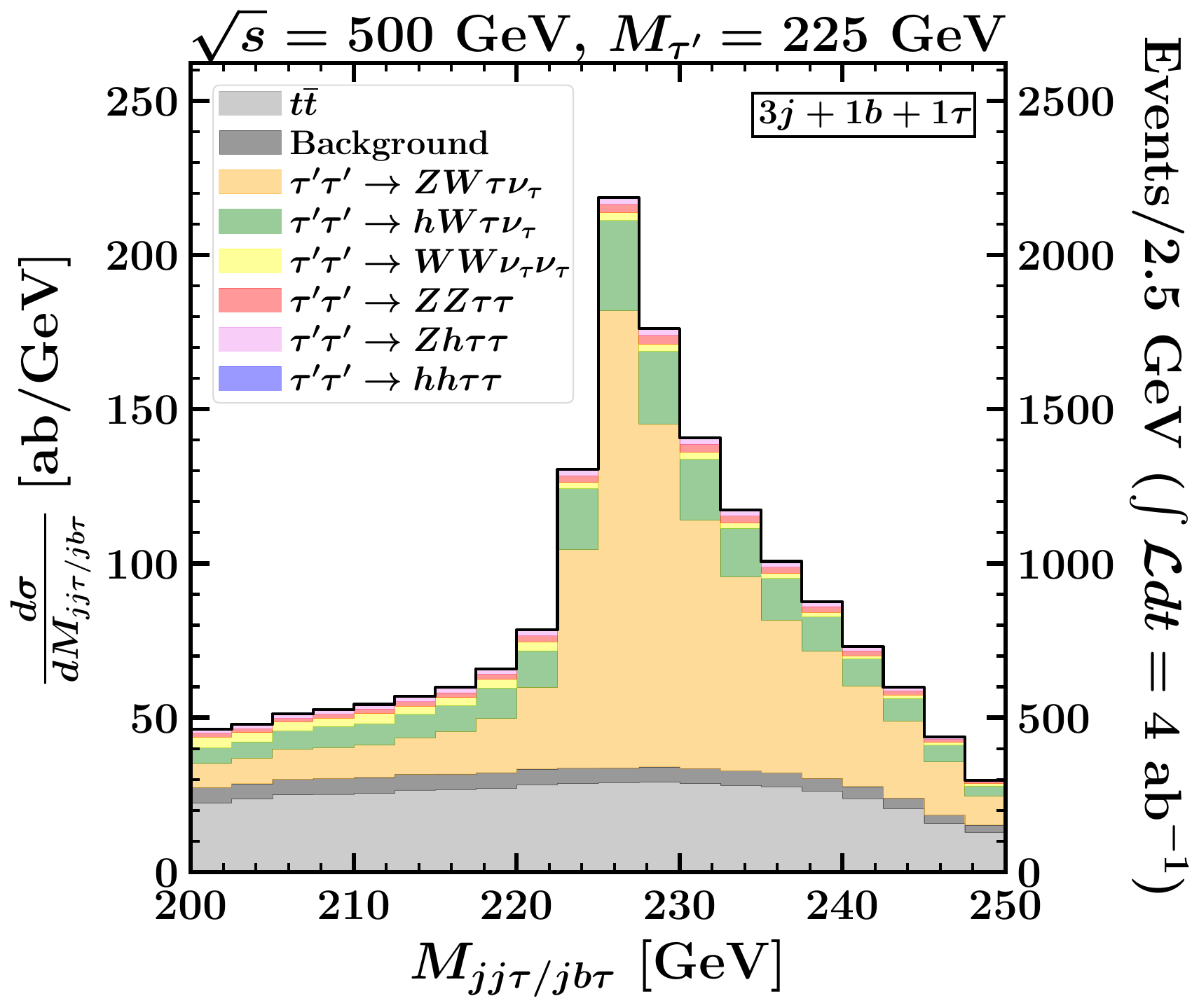}
  \end{minipage}
  \vspace{0.2cm}
  \begin{minipage}[]{0.495\linewidth}
    \includegraphics[width=8cm]{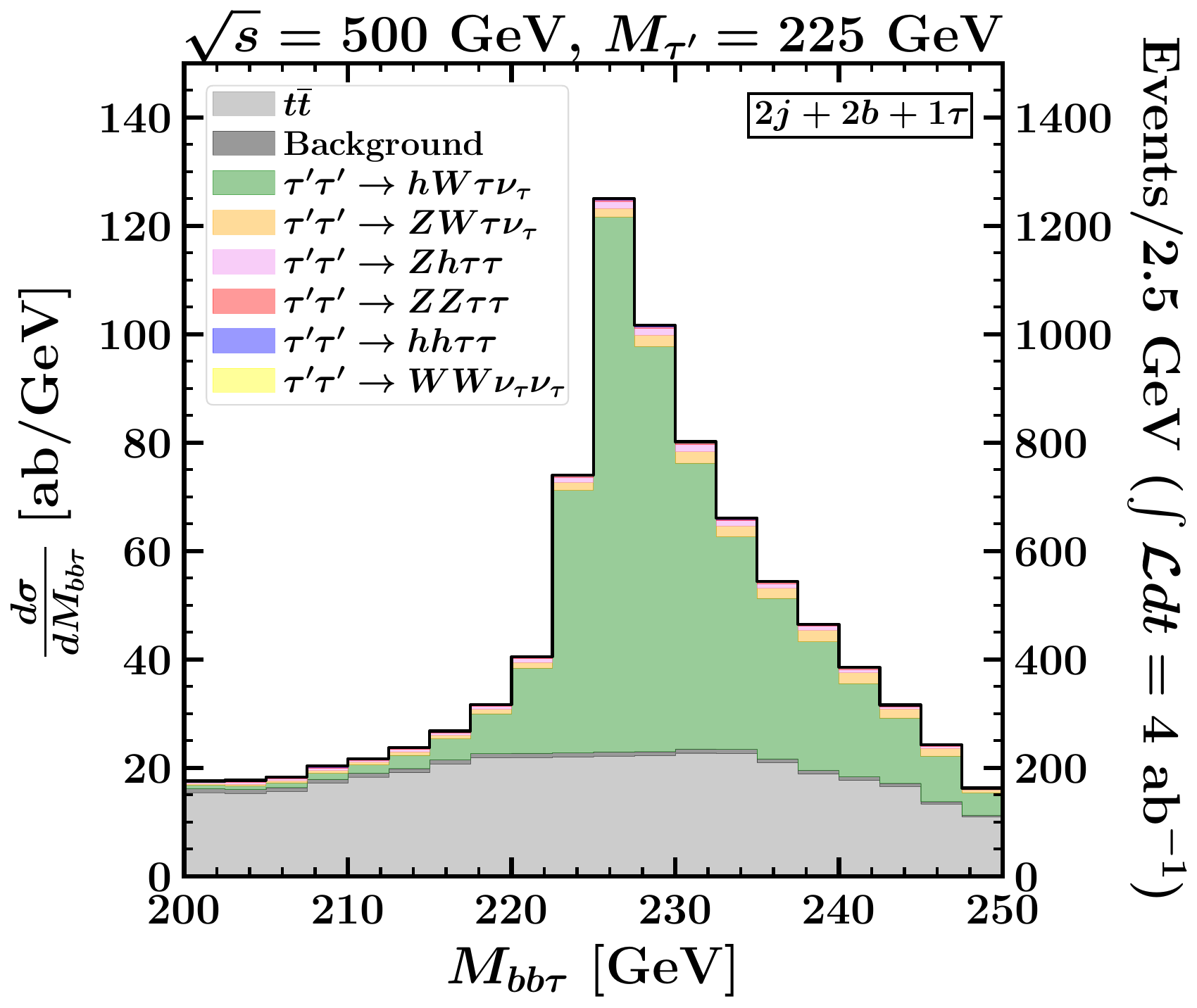}
  \end{minipage}
  \begin{minipage}[]{0.495\linewidth}
    \includegraphics[width=8cm]{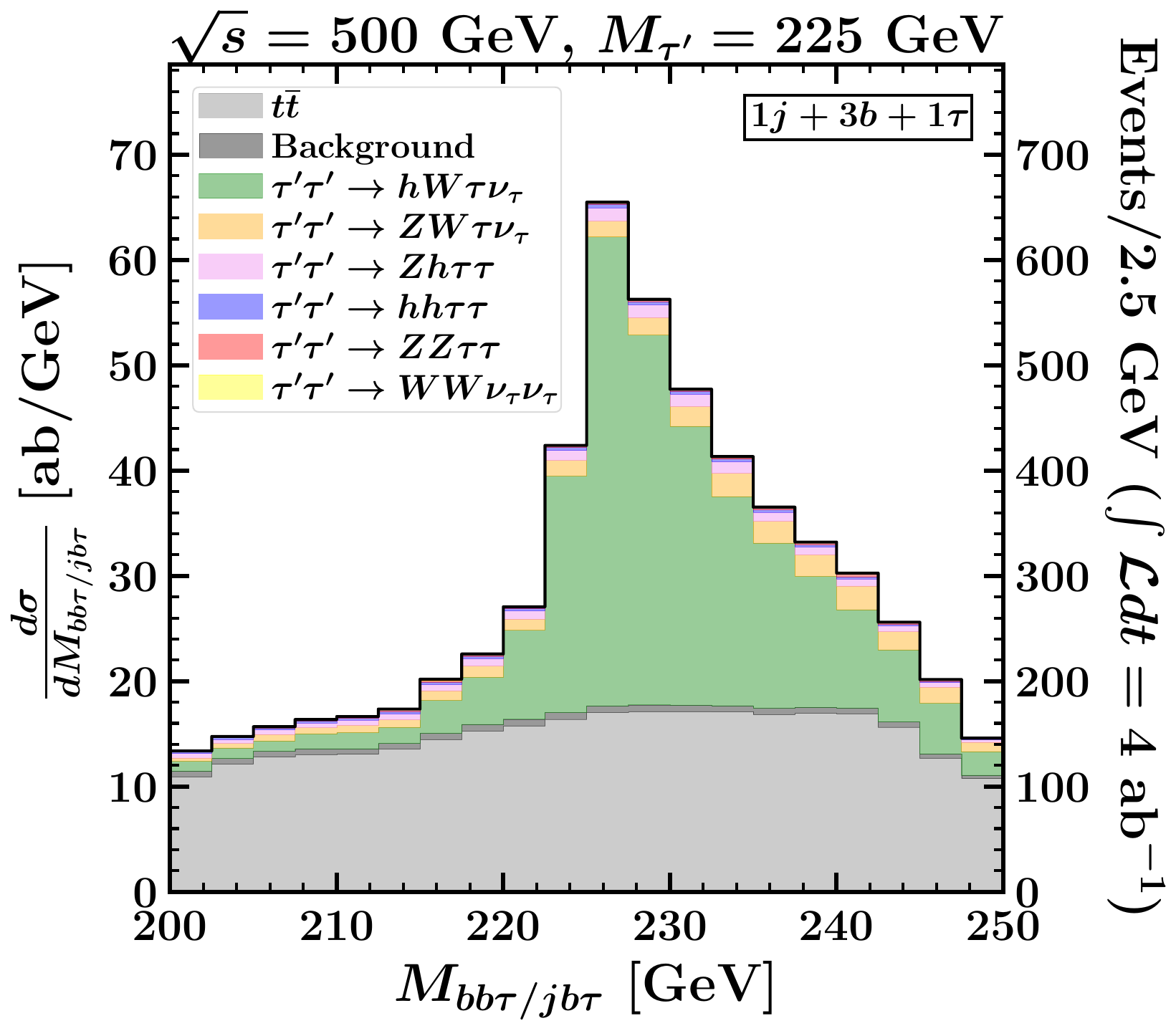}
  \end{minipage}
 \caption{Signal final states contributing to the mass peak
 for $M_{\tau^\prime} = 225$ GeV in five different signal regions with $1\tau$.
 $t\overline{t}$ and the other backgrounds are shown as gray and black shaded histograms,
 respectively, stacked together with the signal contributions.
\label{fig:Mtaup_sqrts500_Breakdown_SR1tau}}
\end{figure}

Next we study the contributions of different final states to the mass peaks in different signal regions.  This information could be utilized by an experiment to fit for the branching ratios of the $\tau^{\prime}$ and verify the relations of Eqs.~(\ref{eq:BR1})--(\ref{eq:BR3}). As an illustrative case, we specialize to $M_{\tau^{\prime}} = 225$ GeV. 
In Fig.~\ref{fig:Mtaup_sqrts500_Breakdown_SR2tau_leptons} we examine signal regions where at least one of the two $\tau^{\prime}$ leptons is reconstructed from a leptonically decaying $Z$ boson.  In the upper two panels, we observe that the $4\ell+2\tau$ and $2\ell+2j+2\tau$ signal regions provide pure samples of the $ZZ \tau \tau$ final state. Similarly, the $2\ell+2b+2\tau$ final state (lower right panel)  effectively provides a pure sample of the $Zh \tau \tau$ final state. Not surprisingly, the peak from $2\ell+1j+1b+2\tau$ (lower left panel) contains a mixture of $ZZ\tau\tau$ and $Zh\tau\tau$.

We now turn to $2 \tau$ final states where both $\tau^{\prime}$ leptons are reconstructed with hadronically decaying $Z$ or Higgs bosons.  The results are shown in Fig.~\ref{fig:Mtaup_sqrts500_Breakdown_SR2tau_jets}.  In the top panel, we see that the $2j+2b+2\tau$ signal region effectively selects for the $Zh \tau \tau$ final state.  The dominant background (shown in gray) arises from $t \bar{t}$ production, but is small. The $4j+2 \tau$ topology (middle row, left panel) provides a relatively pure sample from the $ZZ \tau \tau$ final state, while the $4b +2 \tau$ topology (middle row, right) provides a fairly pure sample of $hh \tau \tau$ events.  Topologies with an odd number of $b$ jets (lower two panels) show clear mass peaks, but with non--trivial contributions from multiple final states.  We also note that choosing $b$ tagging at different operating points will cause events to migrate between the signal regions shown in these panels.  In practice, using these signal regions to help fit for the branching ratios of the $\tau^{\prime}$ would be possible with a thorough understanding of $b$-tagging efficiency.

In Fig.~\ref{fig:Mtaup_sqrts500_Breakdown_SR1tau} we turn to the breakdown of final state sources for the $1 \tau$ signal regions. In the first two panels, we see that the $1\tau$ topologies where the candidate $Z$ decays leptonically provides a pure sample of the $ZW \tau \nu_{\tau}$ final state with essentially no background and good statistics.  This is similarly true for the $4j + 1 \tau$ 
signal region (middle row, left panel), where the $Z$ boson is reconstructed hadronically from non-$b$ tagged jets, albeit with a broader mass peak.  The signal region with 4 jets including a single $b$ tag (middle row, right panel) also predominantly contains events from $ZW \tau \nu_{\tau}$ but with some contribution from  $hW \tau \nu_{\tau}$ and a non-trivial (but smooth) background from $t\bar{t}$ (light gray).  Signal regions with 1$\tau$ and at least two $b$ jets (lower two panels) give nearly pure samples of $hW \tau \nu_{\tau}$, again with background from $t \bar{t}$ that is non-trivial but likely easily subtracted.

We summarize the results of these figures in Table \ref{tab:SignalEfficiencies}. There we show for each signal region the percentage of signal events that derive from a given $\tau^{\prime} \tau^{\prime}$ decay topology
(e.g. $ZZ\tau\tau$, $Zh\tau\tau$ etc.). 
As noted above, for many signal regions, the number of background events is very small, and even in cases where it is larger, it is
smoothly varying and should be able to be subtracted  effectively. The first section of the table shows the purest signal regions, where roughly 90\% of the events (or more) result from one particular final state. The existence of such signal regions with large contributions from a single final state should make it particularly straightforward to extract the branching ratios of the $\tau^{\prime}$ to different final states. As noted above, different choices of $b$-tagging operation point can cause events to migrate between signal regions.  However, independent of operating point, the signal regions with high purity ($\gtrsim 90\%$) retain this characteristic.   

\FloatBarrier

\begin{table}
\caption{
The percentages of events in each signal region that come from a given $\tau^{\prime+}\tau^{\prime-}$ final state, for $M_{\tau^\prime} = 225$ GeV at $\sqrt{s} = 500$ GeV with medium $b$-tagging. In each row, the results for the most important final state, defined as the one that contributes most to the given signal region, are highlighted in bold.
These numbers can therefore be interpreted as the ``purity" of each signal region with respect to its most important contributing final state. Generally, the amount of SM background under the mass peak will be small and/or easily subtracted.
}
\label{tab:SignalEfficiencies}
\begin{center}
\begin{tabular}{
        | c |
        | c | c | c | c | c |
    }
\hline
    ~Signal region~ &
    ~$ZZ\tau\tau$~ & ~$Zh\tau\tau$~ & ~$hh\tau\tau$~ & ~$ZW\tau\nu_\tau$~ & ~$hW\tau\nu_\tau$~
    \\[1pt]
\hline
\hline
    ~$4 \ell + 2 \tau$~ &
    ~{\bf 98.7}~ & ~1.3~ & ~0.0~ & ~0.0~ & ~0.0~
    \\[1pt]
    ~$2 \ell + 2 j + 2 \tau$~ &
    ~{\bf 95.4}~ & ~3.9~ & ~0.0~ & ~0.5~ & ~0.1~
    \\[1pt]
    ~$2 \ell + 2 b + 2 \tau$~ &
    ~2.2~ & ~{\bf 97.2}~ & ~0.6~ & ~0.0~ & ~0.0~
    \\[1pt]
    ~$2 j + 2 b + 2 \tau$~ &
    ~3.8~ & ~{\bf 88.5}~ & ~4.9~ & ~0.3~ & ~2.6~
    \\[1pt]
    ~$2 \ell + 2 j + 1 \tau$~ &
    ~1.3~ & ~0.6~ & ~0.0~ & ~{\bf 97.5}~ & ~0.7~
    \\[1pt]
    ~$2 \ell + 1 j + 1 b + 1 \tau$~ &
    ~1.6~ & ~0.8~ & ~0.0~ & ~{\bf 96.7}~ & ~0.9~
    \\[1pt]
    ~$2 j + 2 b + 1 \tau$~ &
    ~0.8~ & ~3.1~ & ~0.7~ & ~5.2~ & ~{\bf 89.9}~
    \\[1pt]
\hline
\hline
    ~$4 j + 2 \tau$~ &
    ~{\bf 77.6}~ & ~9.9~ & ~1.0~ & ~7.0~ & ~4.5~
    \\[1pt]
    ~$4 b + 2 \tau$~ &
    ~3.7~ & ~20.1~ & ~{\bf 75.9}~ & ~0.1~ & ~0.3~
    \\[1pt]
    ~$4 j + 1 \tau$~ &
    ~4.1~ & ~1.9~ & ~0.2~ & ~{\bf 76.6}~ & ~8.5~
    \\[1pt]
    ~$1 j + 3 b + 1 \tau$~ &
    ~1.4~ & ~6.1~ & ~1.9~ & ~9.7~ & ~{\bf 80.7}~
    \\[1pt]
\hline
\hline
    ~$2 \ell + 1 j + 1 b + 2 \tau$~ &
    ~{\bf 65.4}~ & ~33.6~ & ~0.3~ & ~0.4~ & ~0.1~
    \\[1pt]
    ~$3 j + 1 b + 2 \tau$~ &
    ~{\bf 52.2}~ & ~30.9~ & ~3.3~ & ~6.7~ & ~6.7~
    \\[1pt]
    ~$1 j + 3 b + 2 \tau$~ &
    ~8.8~ & ~{\bf 61.0}~ & ~25.4~ & ~0.9~ & ~4.0~
    \\[1pt]
    ~$3 j + 1 b + 1 \tau$~ &
    ~3.5~ & ~2.8~ & ~0.4~ & ~{\bf 65.0}~ & ~20.6~
    \\[1pt]
\hline
\end{tabular}
\end{center}
\end{table}

\FloatBarrier

\subsection{$\sqrt{s} =$ 250 and 380 GeV \label{subsec:Results_250GeV}}

We now turn to results for lower energy colliders which might be available on a shorter time scale.  We consider $\sqrt{s}=250$ GeV, potentially relevant for a Higgs factory, and $\sqrt{s}=380$ GeV, a top factory.   

At a $\sqrt{s}=250$ GeV machine,  $\tau^{\prime}$ pair production is only possible for $M_{\tau^\prime} < 125$ GeV.  However, there is a gap between the LEP limits and the CMS bound (see Sec.~\ref{sec:singletvll}).  It is of interest to see whether a $\sqrt{s}=250$
GeV machine could access a $\tau^{\prime}$ in  this window, and whether it could make measurements of its branching fraction.  To answer these questions, we fix $M_{\tau^{\prime}} = 115$ GeV and repeat the analysis of the previous section.  For this mass, the $\tau^{\prime} \rightarrow h \tau$ final state is not accessible, and there is also no $t\overline t$ background. Note that here we also reconstruct candidate $Z$ bosons from $b$ jets. Moreover, the branching ratio is dominantly (77\%) to $W \nu_{\tau}$. (Recall that this is, in part, what makes discovery of such a $\tau^{\prime}$ so challenging at the LHC.)  It follows that the signal regions with a single $\tau$ have much better statistics than those with two $\tau$ leptons. 

\begin{figure}[!h]
  \begin{center}
  \begin{minipage}[]{0.495\linewidth}
    \includegraphics[width=8cm]{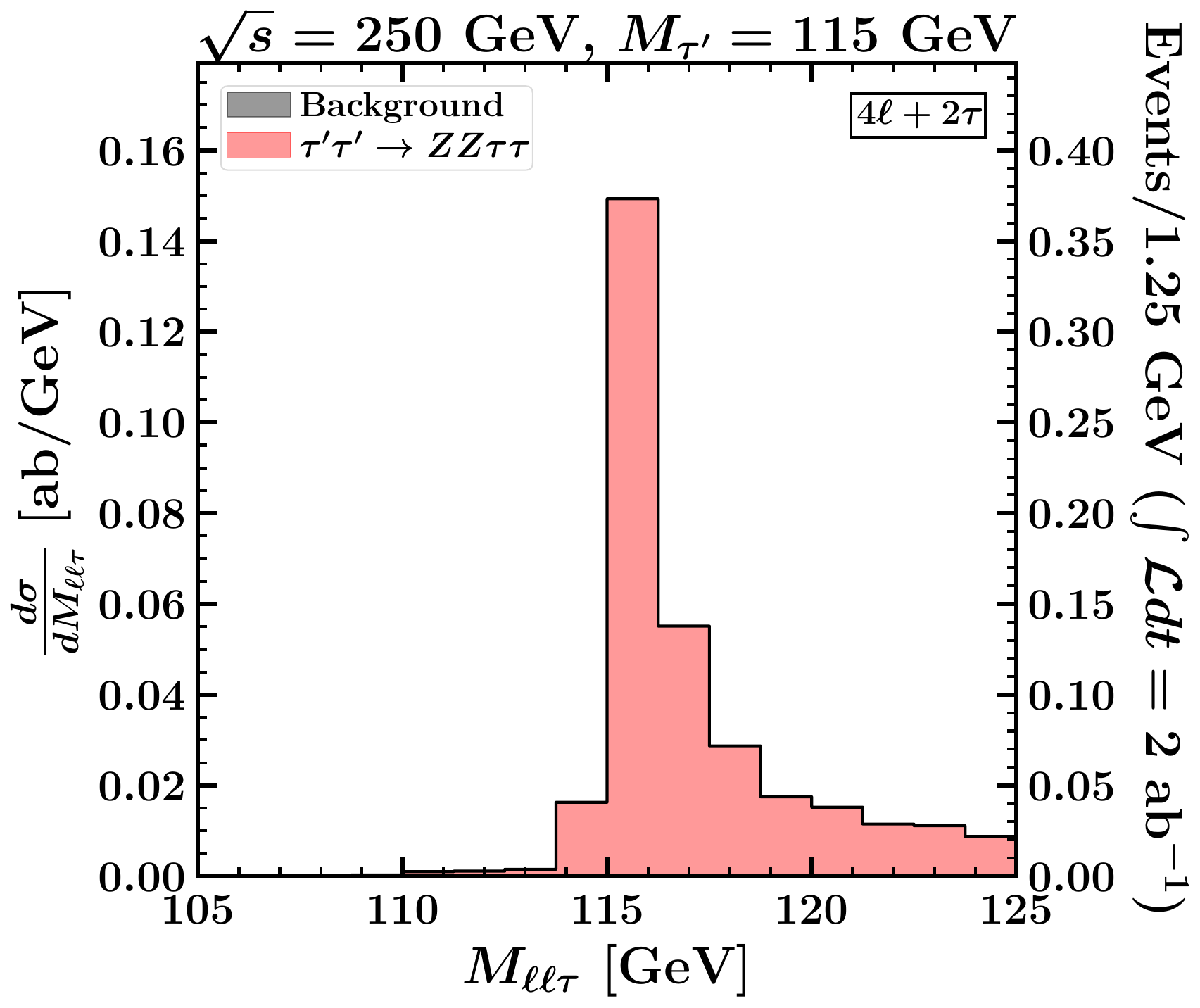}
  \end{minipage}   
  \end{center}
  \vspace{0.2cm}
  \begin{minipage}[]{0.495\linewidth}
    \includegraphics[width=8cm]{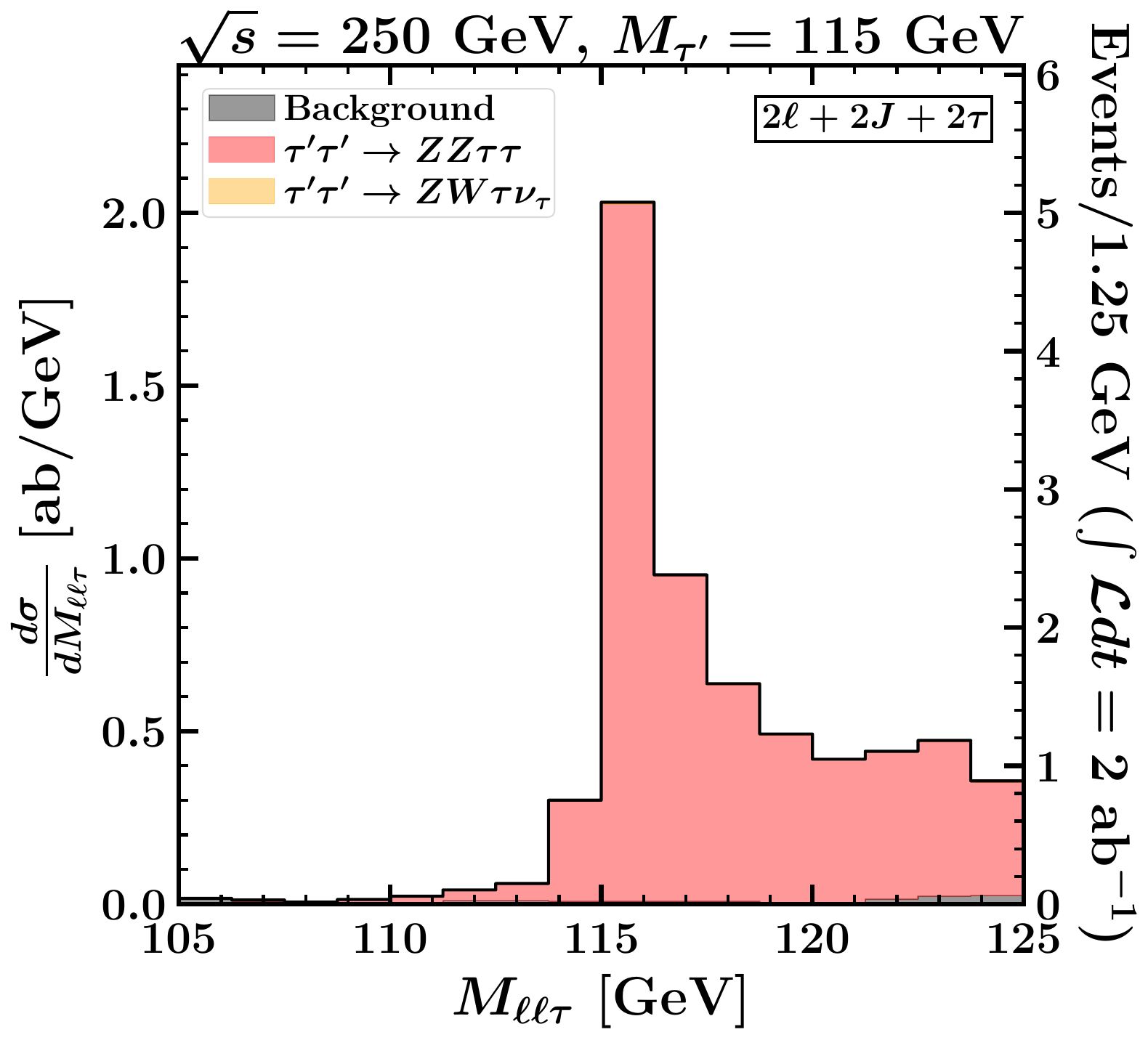}
  \end{minipage}
  \begin{minipage}[]{0.495\linewidth}
    \includegraphics[width=8cm]{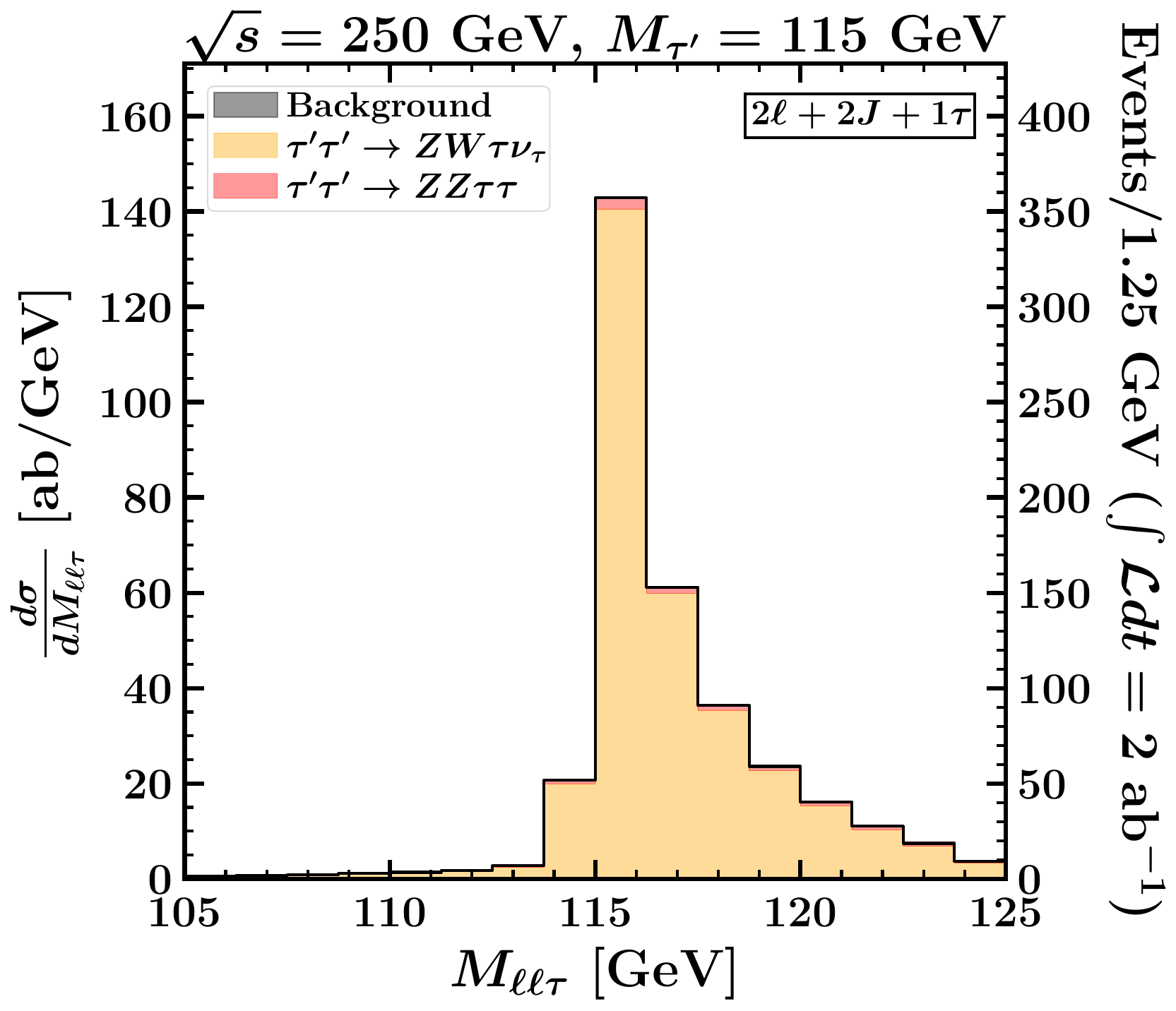}
  \end{minipage}
  \vspace{0.2cm}
  \begin{minipage}[]{0.495\linewidth}
    \includegraphics[width=8cm]{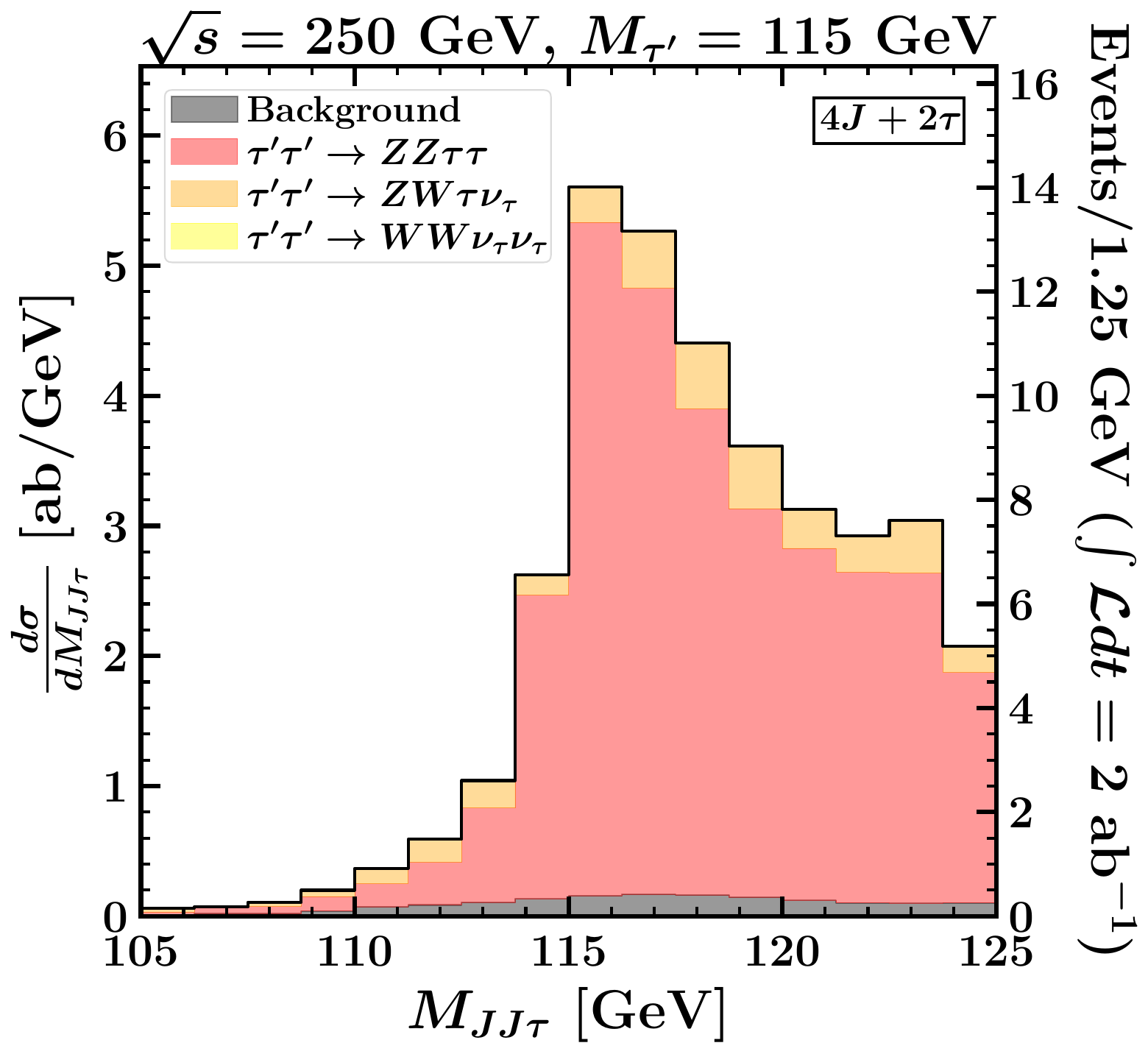}
  \end{minipage}
  \begin{minipage}[]{0.495\linewidth}
     \includegraphics[width=8cm]{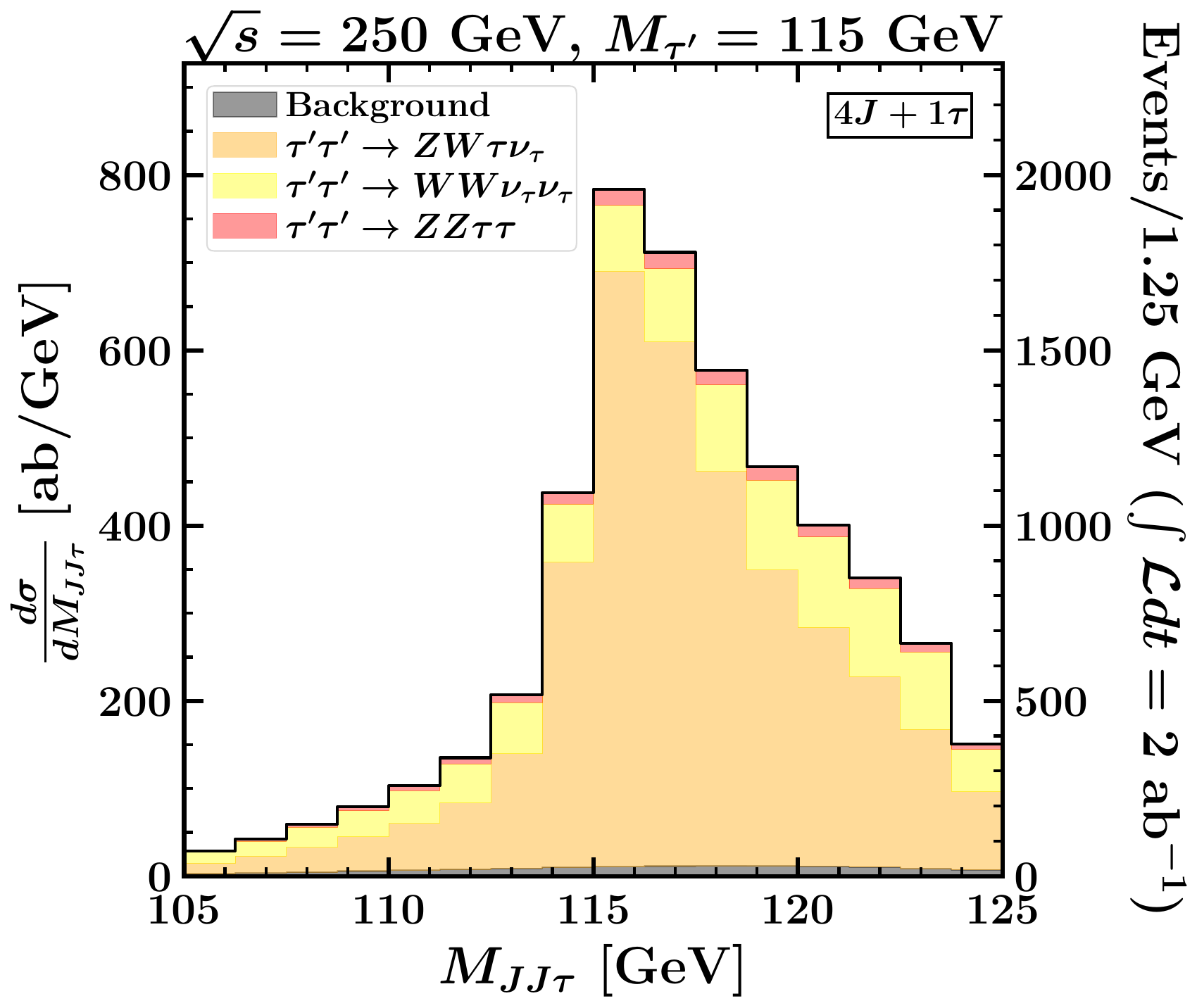}
\end{minipage}
\caption{Signal final states contributing to the mass peak
 for $M_{\tau^\prime} = 115$ GeV for $e^+ e^-$ collisions at $\sqrt{s} = 250$ GeV. Backgrounds are barely visible as gray-shaded histograms stacked together with the signal contributions.
\label{fig:Mtaup_sqrts250}}
\end{figure}

The results are shown in Fig.~\ref{fig:Mtaup_sqrts250}, in which we have separated the contributions in each panel by final-state source.  The signal region with $4\ell+ 2 \tau$ in the first panel provides a sharp mass peak but is not viable with 2 ab$^{-1}$ due to very limited statistics.  Signal regions where at least one of the two $Z$ bosons decays hadronically (middle row) are a more promising way to try to access the $ZZ \tau \tau$ final state, and thus make a determination of the branching ratios. A particularly sharp and significant mass peak is seen in the $2\ell+ 2J+1\tau$ signal region, providing access to a pure sample of the $ZW \tau \nu_{\tau}$ final state (middle row, right panel).  The last panel shows the $4J+1 \tau$ signal region, which gives a much broader mass peak than the case with a leptonically decaying $Z$, but has the best statistics. Note that, perhaps surprisingly, this signal region also receives a non-trivial contribution from the $WW \nu_{\tau} \nu_{\tau}$ final state. This is because our jet clustering algorithm exclusively clusters candidate event into 5 jets, and sometimes one of these jets is misidentified as a $\tau$.  Because of the relatively large branching ratio to the $WW \nu_{\tau} \nu_{\tau}$ final state, even a relatively small tau-misidentification rate allows this final state to have an impact. 
Taken together, these five signal regions should again allow a determination of $M_{\tau'}$ as well as the $\tau^{\prime}$ branching ratios.

\begin{figure}[!h]
  \begin{center}
   \begin{minipage}[]{0.495\linewidth}
    \includegraphics[width=8cm]{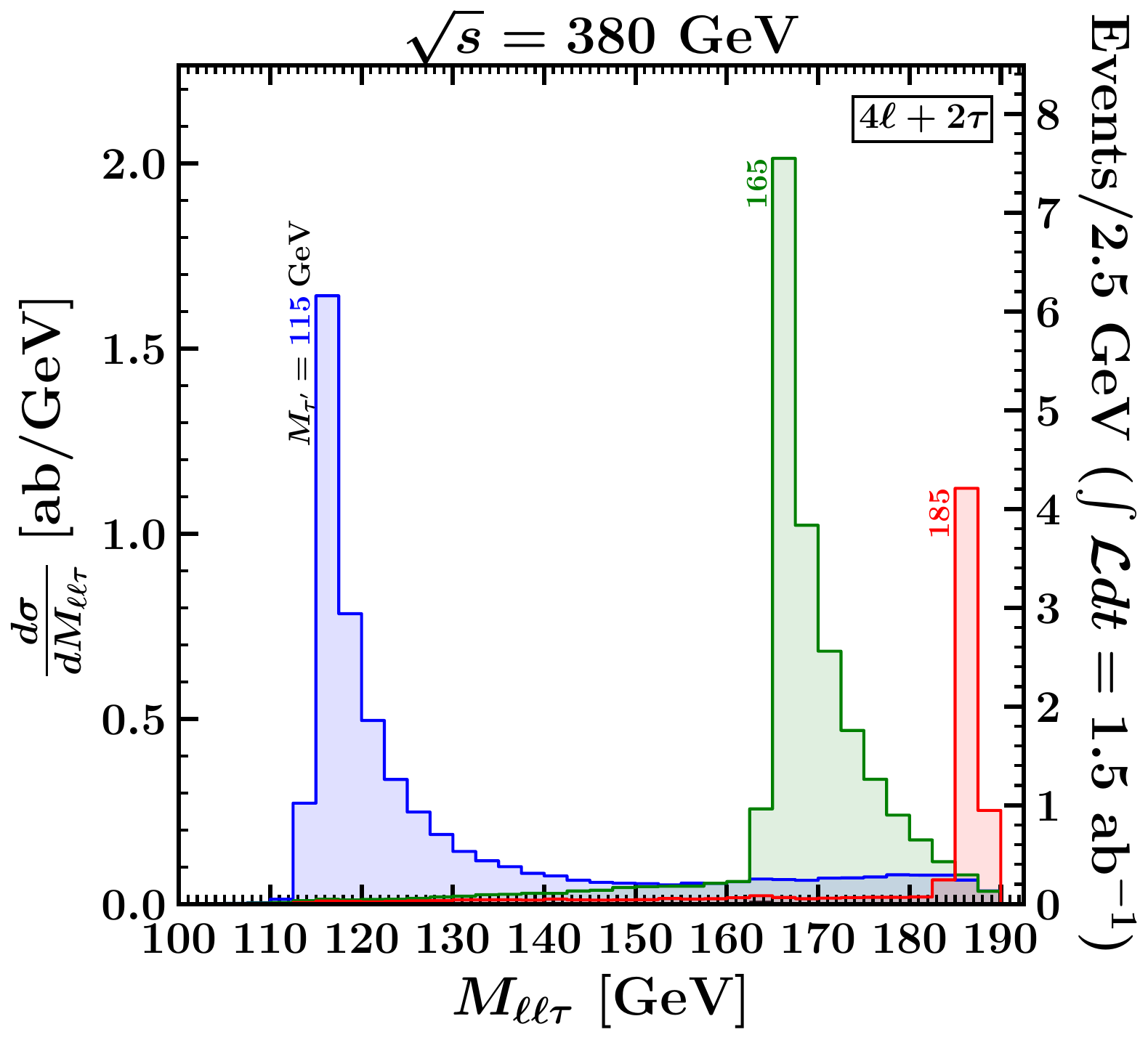}
  \end{minipage}
  \end{center}
  \vspace{0.2cm}
  \begin{minipage}[]{0.495\linewidth}
    \includegraphics[width=8cm]{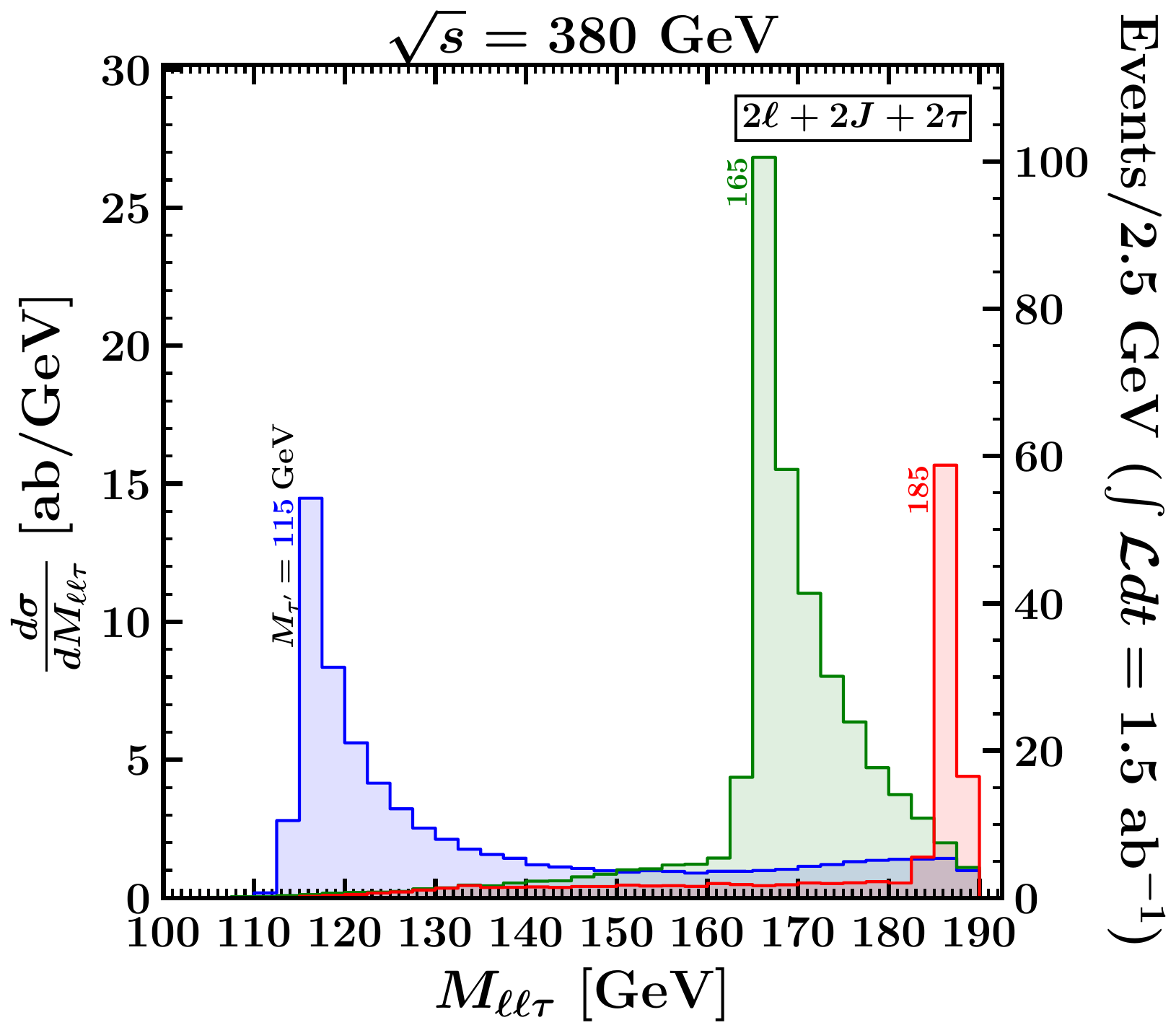}
  \end{minipage}
  \begin{minipage}[]{0.495\linewidth}
    \includegraphics[width=8cm]{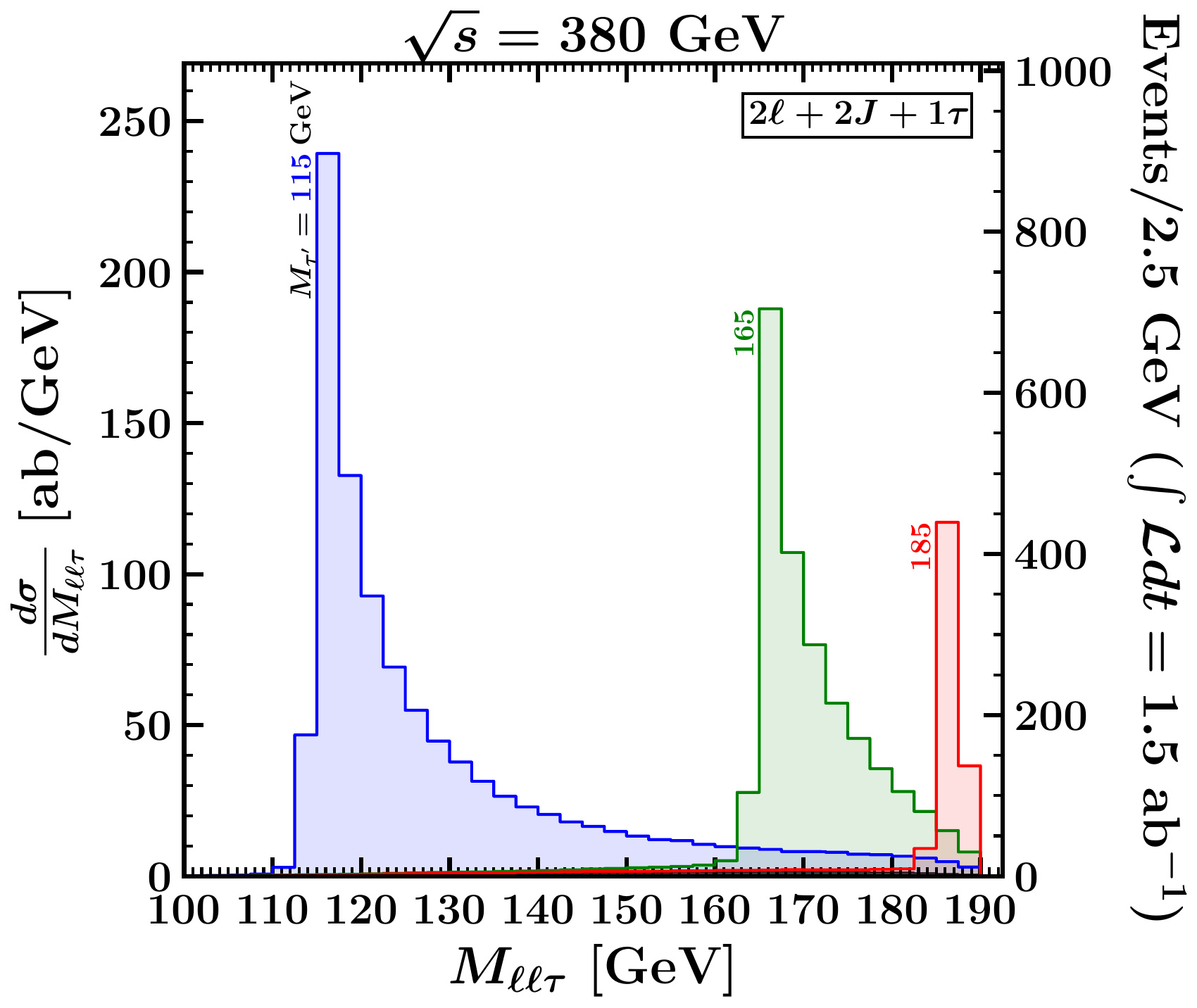}
  \end{minipage}
  \vspace{0.2cm}
  \begin{minipage}[]{0.495\linewidth}
    \includegraphics[width=8cm]{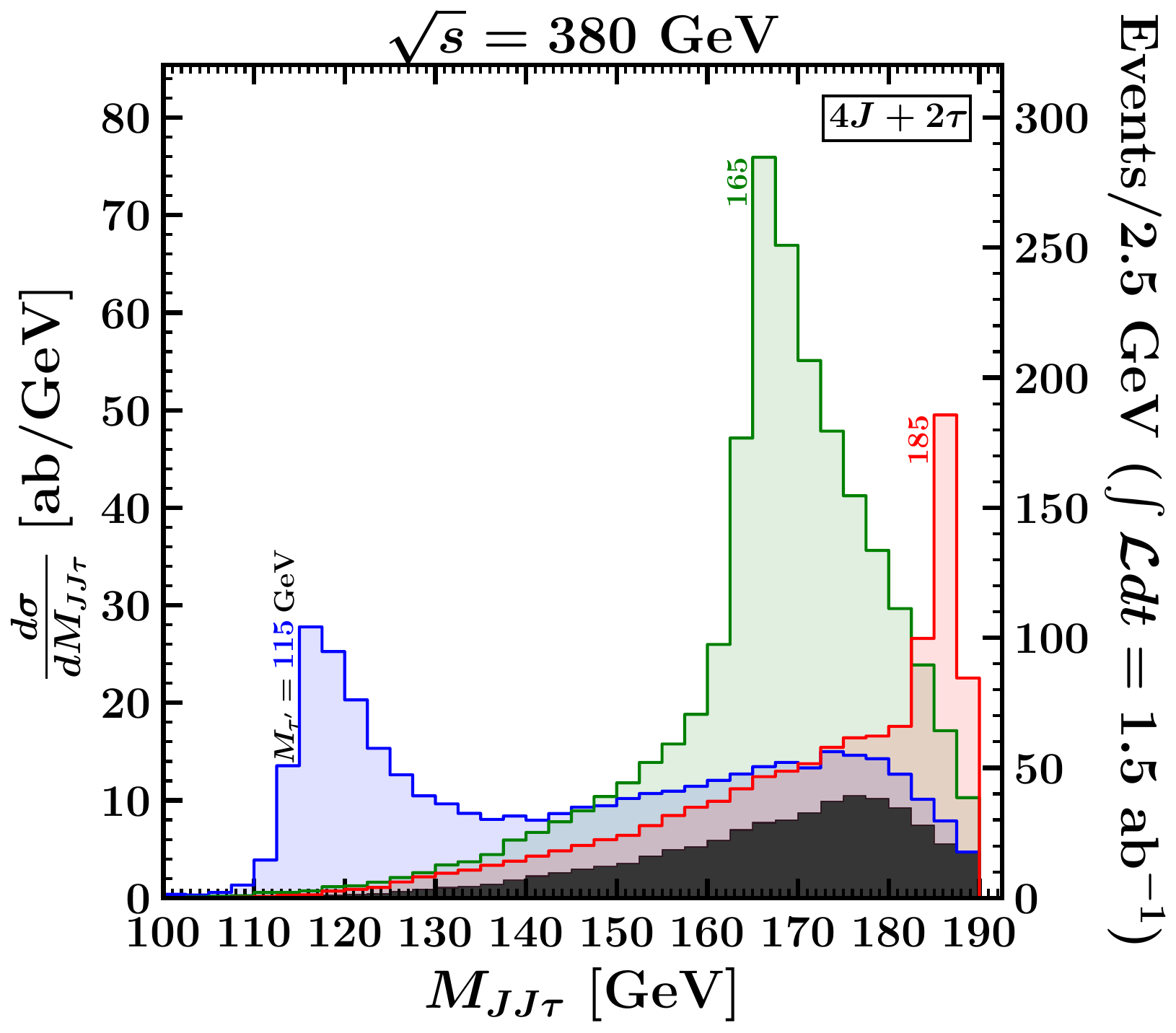}
  \end{minipage}
  \begin{minipage}[]{0.495\linewidth}
     \includegraphics[width=8cm]{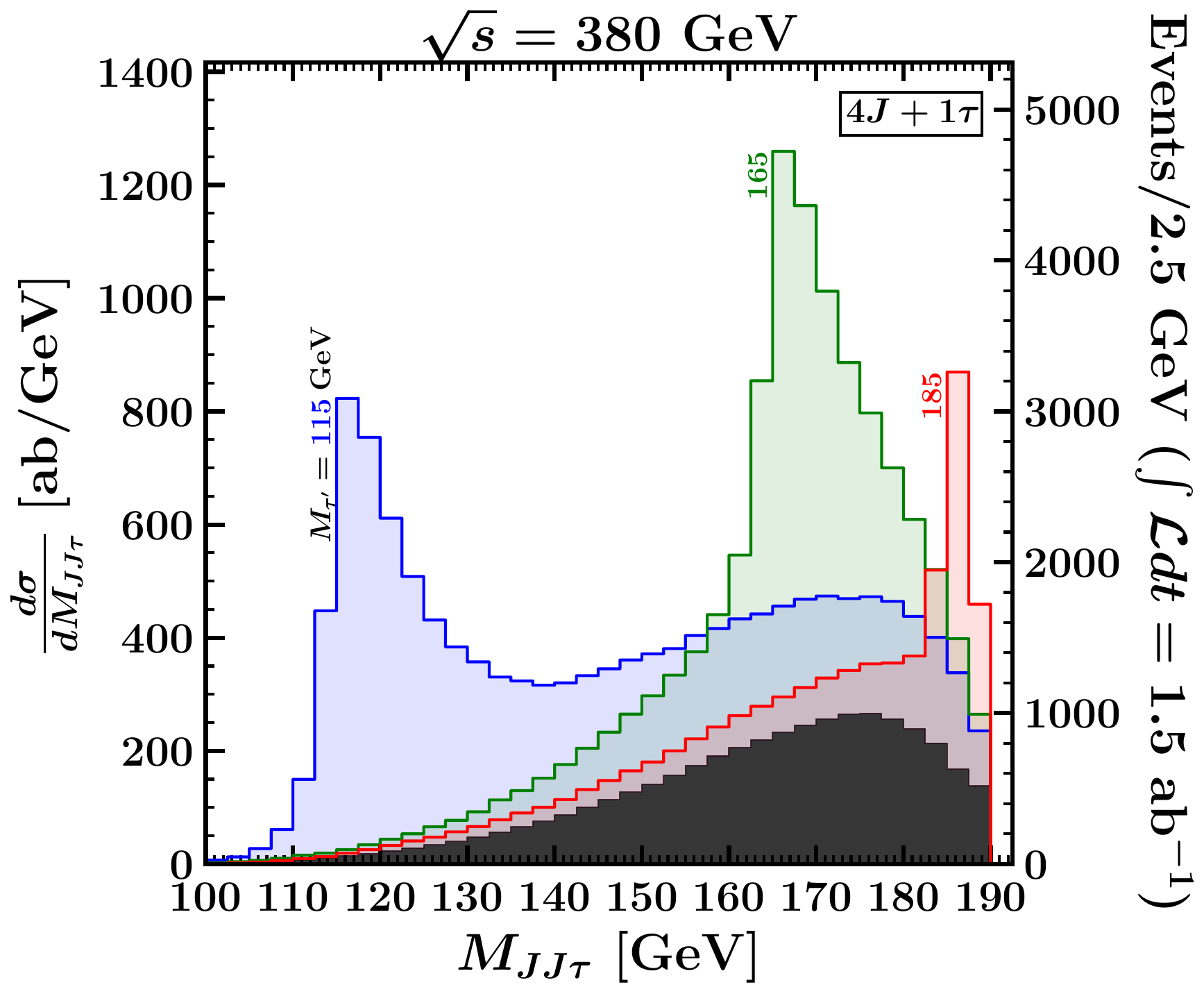}
   \end{minipage}
 \caption{Reconstructed $\tau^\prime$ mass peaks in five different signal regions for $M_{\tau^\prime} =$ 115, 165, and 185 GeV, as labeled, for $e^+ e^-$ collisions at $\sqrt{s} = 380$ GeV.
The background shown as the gray-shaded histogram is the one relevant for signal reconstruction with $M_{\tau^\prime} > M_h + M_\tau$. The background relevant for
$M_{\tau^\prime} < M_h + M_\tau$ is slightly smaller and is not shown. Each signal histogram contains the appropriate background.
\label{fig:Mtaup_sqrts380}}
\end{figure}

We next turn to a discussion of the $\tau^\prime$ at a top factory machine with a $\sqrt{s} = 380$ GeV.  Here, our focus is on demonstration of the reconstruction of the mass peaks in the various signal regions.  The results are shown in Fig.~\ref{fig:Mtaup_sqrts380}.  We have shown mass peaks for three choices of the $\tau^{\prime}$ mass, $M_{\tau^{\prime}} = 115$ GeV, 165 GeV, and 185 GeV.  Reconstruction of the mass peak is clearly possible in multiple signal regions with good statistics, except in the $4l+2\tau$ signal region shown in the first panel, where only a few events are expected with 1.5 ab$^{-1}$.  
As can be seen from Fig.~\ref{fig:BRs}, the branching ratios quickly reach their asymptotic values as $M_{\tau'}$ increases, so that the final state breakdown results for the $M_{\tau'} = 185$ GeV case should be similar to the examples discussed above in the $\sqrt{s}= 500$ GeV section. 

\FloatBarrier

\subsection{$\sqrt{s} =$ 1, 1.5, and 3 TeV\label{subsec:Results_1TeV}}

We now turn to higher energy machines.  Again, our emphasis is on the demonstration of the reconstruction of the mass peaks. The $\tau^{\prime} $ with masses shown in these plots should again have branching ratios close to their asymptotic values. As such, the breakdown by final state will be similar to that discussed in Sec.~\ref{subsec:Results_500GeV}.

\begin{figure}[!h]
  \begin{center}
   \begin{minipage}[]{0.495\linewidth}
    \includegraphics[width=8cm]{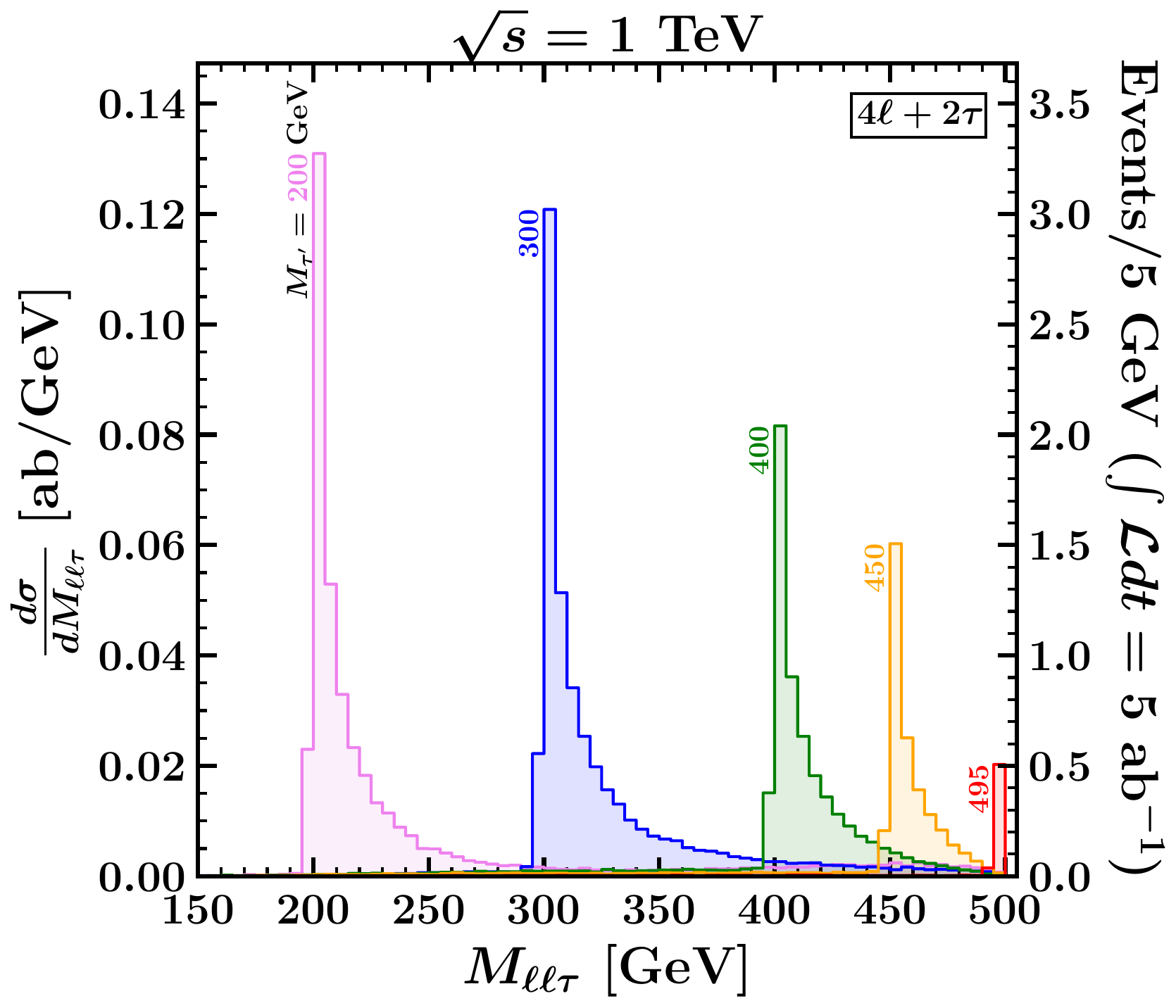}
  \end{minipage}
  \end{center}
  \vspace{0.2cm}
  \begin{minipage}[]{0.495\linewidth}
    \includegraphics[width=8cm]{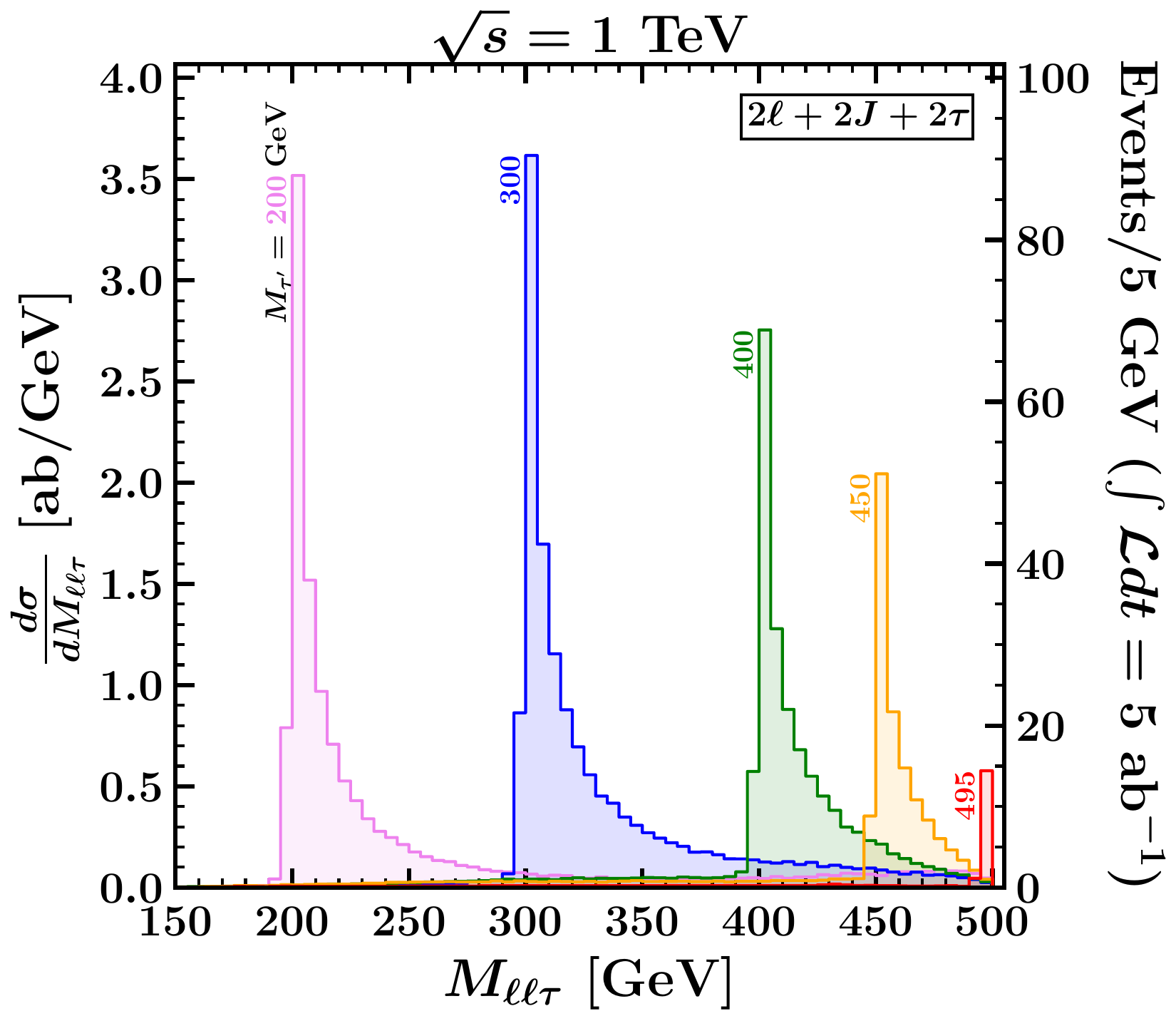}
  \end{minipage}
  \begin{minipage}[]{0.495\linewidth}
    \includegraphics[width=8cm]{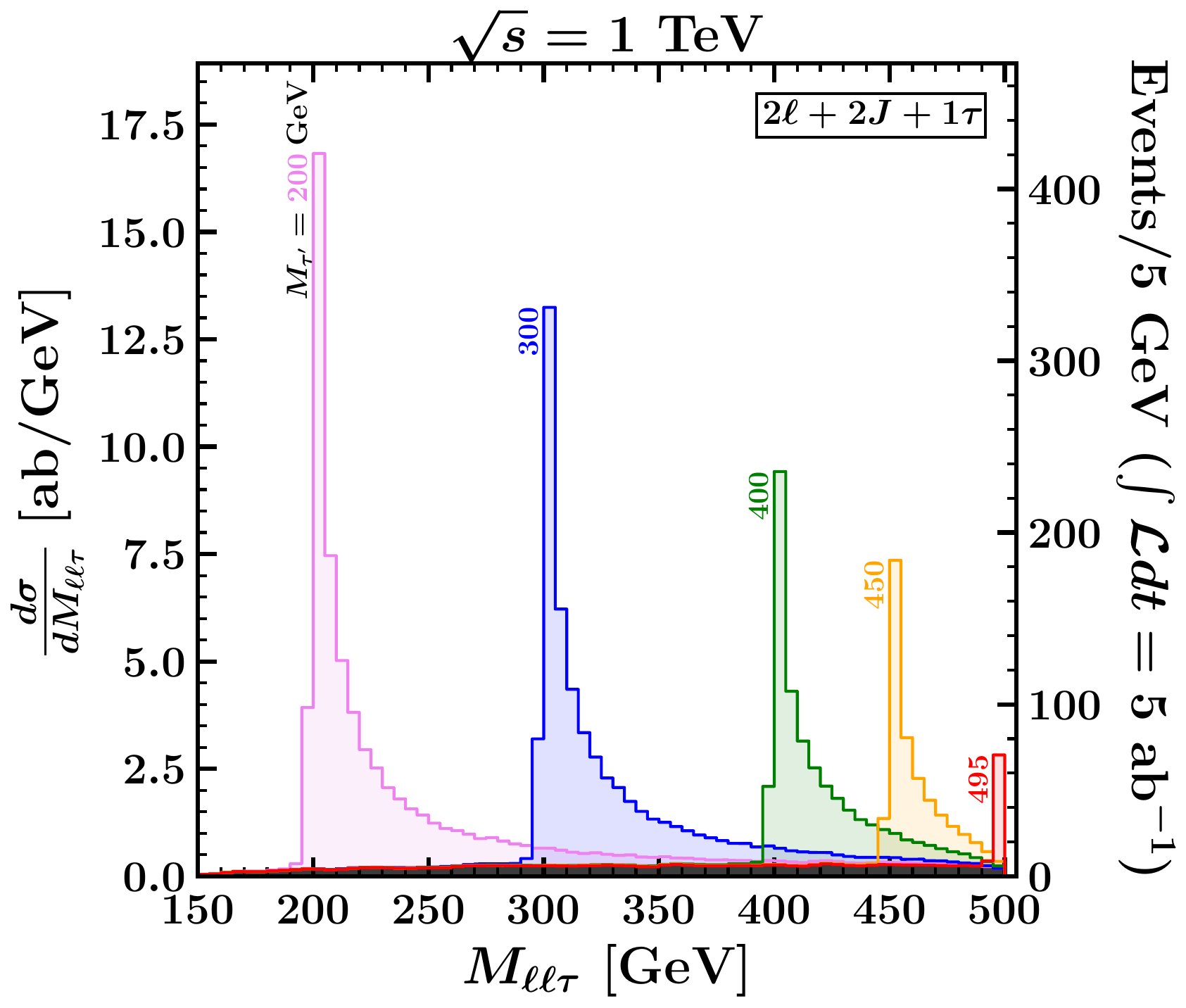}
  \end{minipage}
  \vspace{0.2cm}
  \begin{minipage}[]{0.495\linewidth}
    \includegraphics[width=8cm]{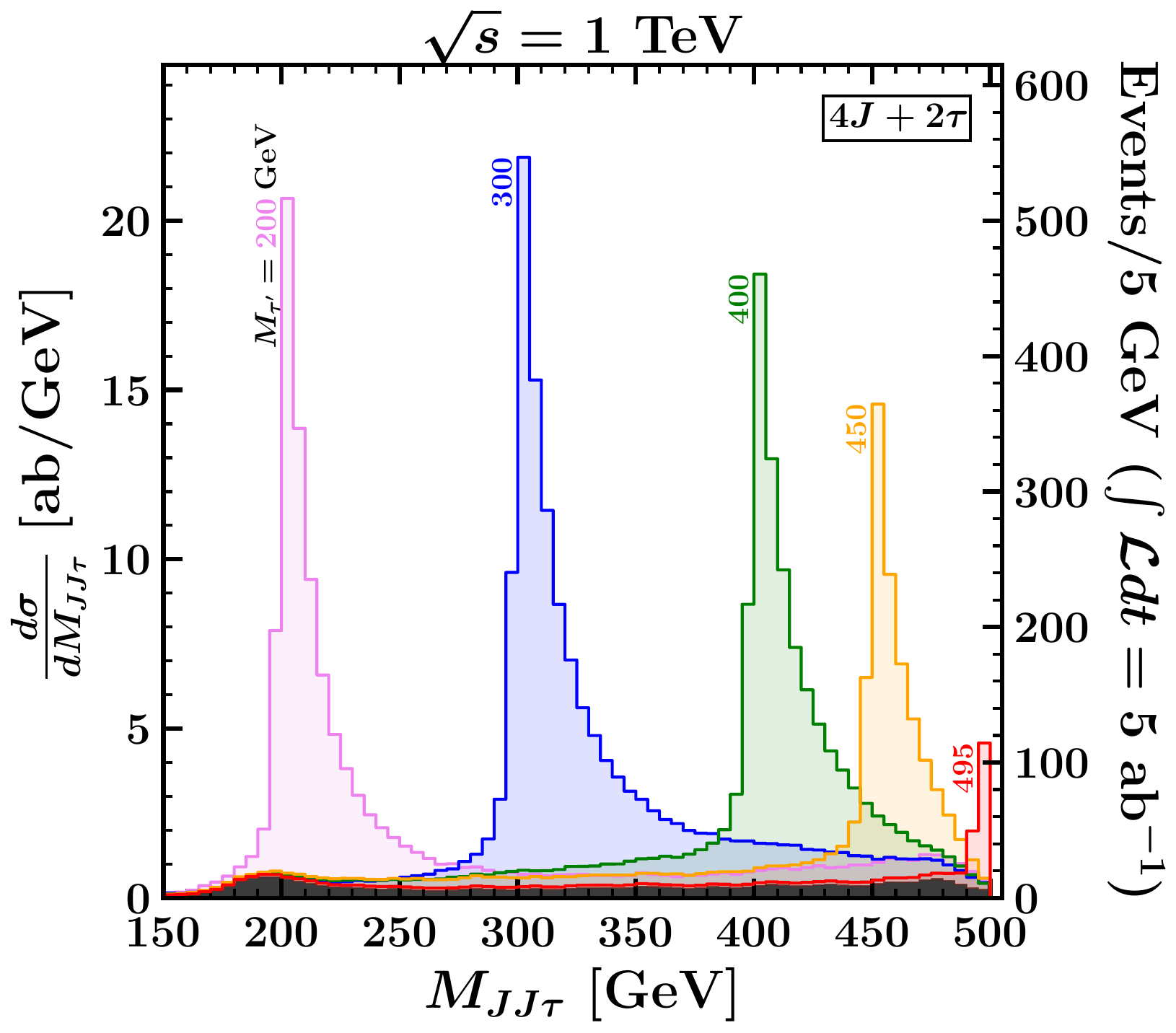}
  \end{minipage}
  \begin{minipage}[]{0.495\linewidth}
     \includegraphics[width=8cm]{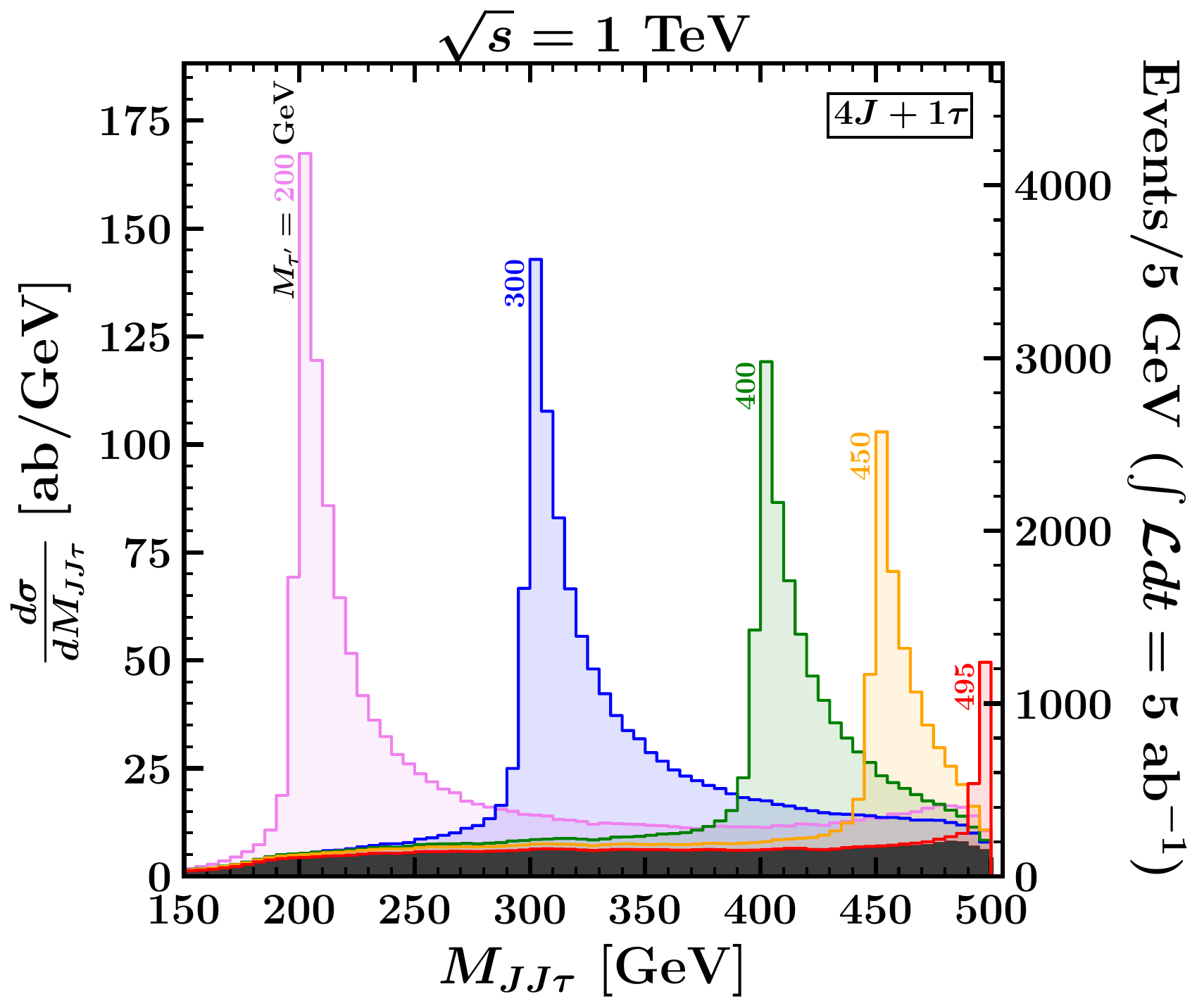}
\end{minipage}
 \caption{Reconstructed $\tau^\prime$ mass peaks in five different signal 
 regions, for $M_{\tau^\prime} =$ 200, 300, 400, 450, and 495 GeV as labeled, for $e^+ e^-$ collisions at $\sqrt{s} = 1$ TeV. Backgrounds are shown as gray-shaded histograms stacked together with the signal histograms.
\label{fig:Mtaup_sqrts1000}}
\end{figure}

\begin{figure}[!h]
  \begin{center}
   \begin{minipage}[]{0.495\linewidth}
    \includegraphics[width=8cm]{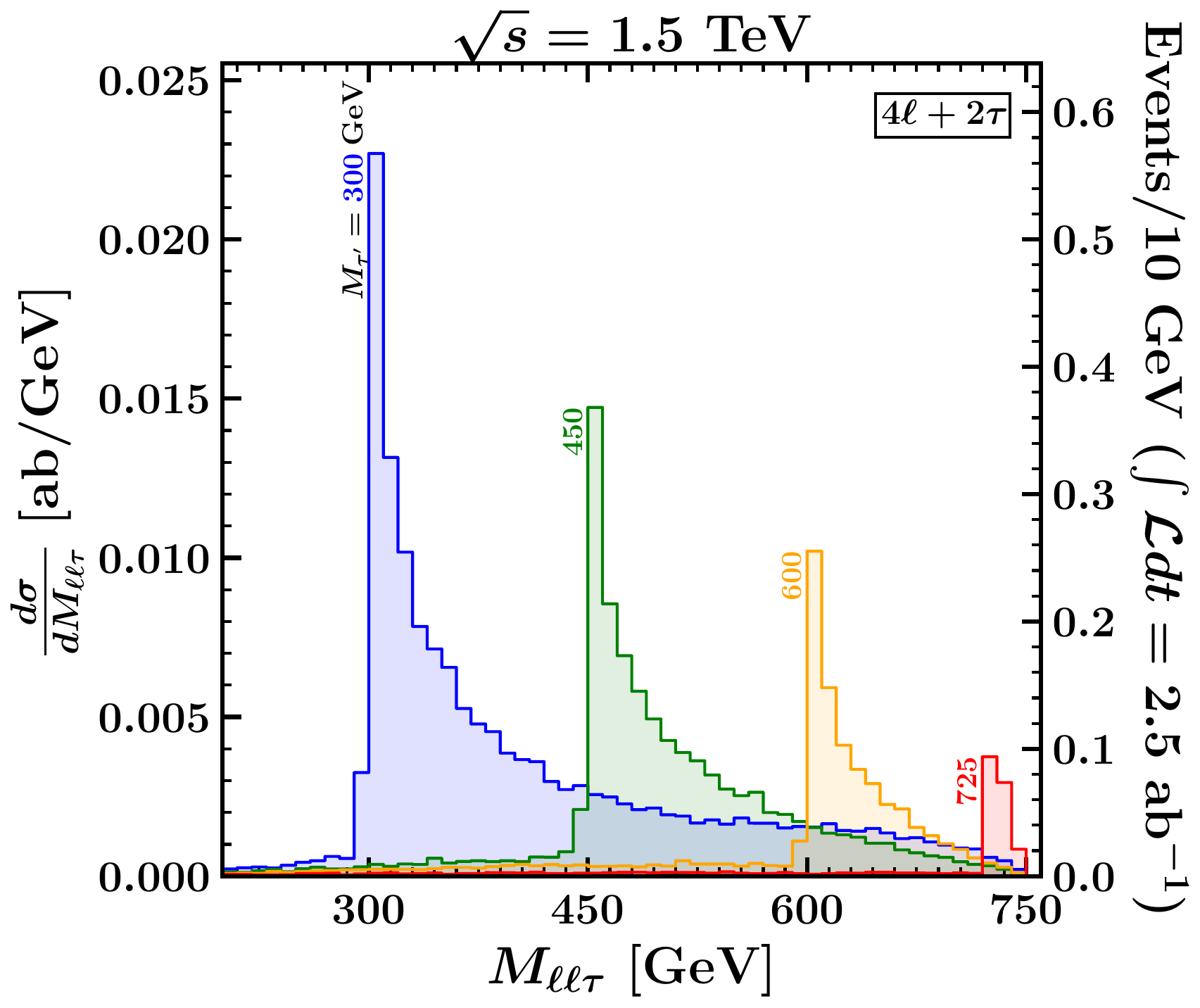}
  \end{minipage}
  \end{center}
  \vspace{0.2cm}
  \begin{minipage}[]{0.495\linewidth}
    \includegraphics[width=8cm]{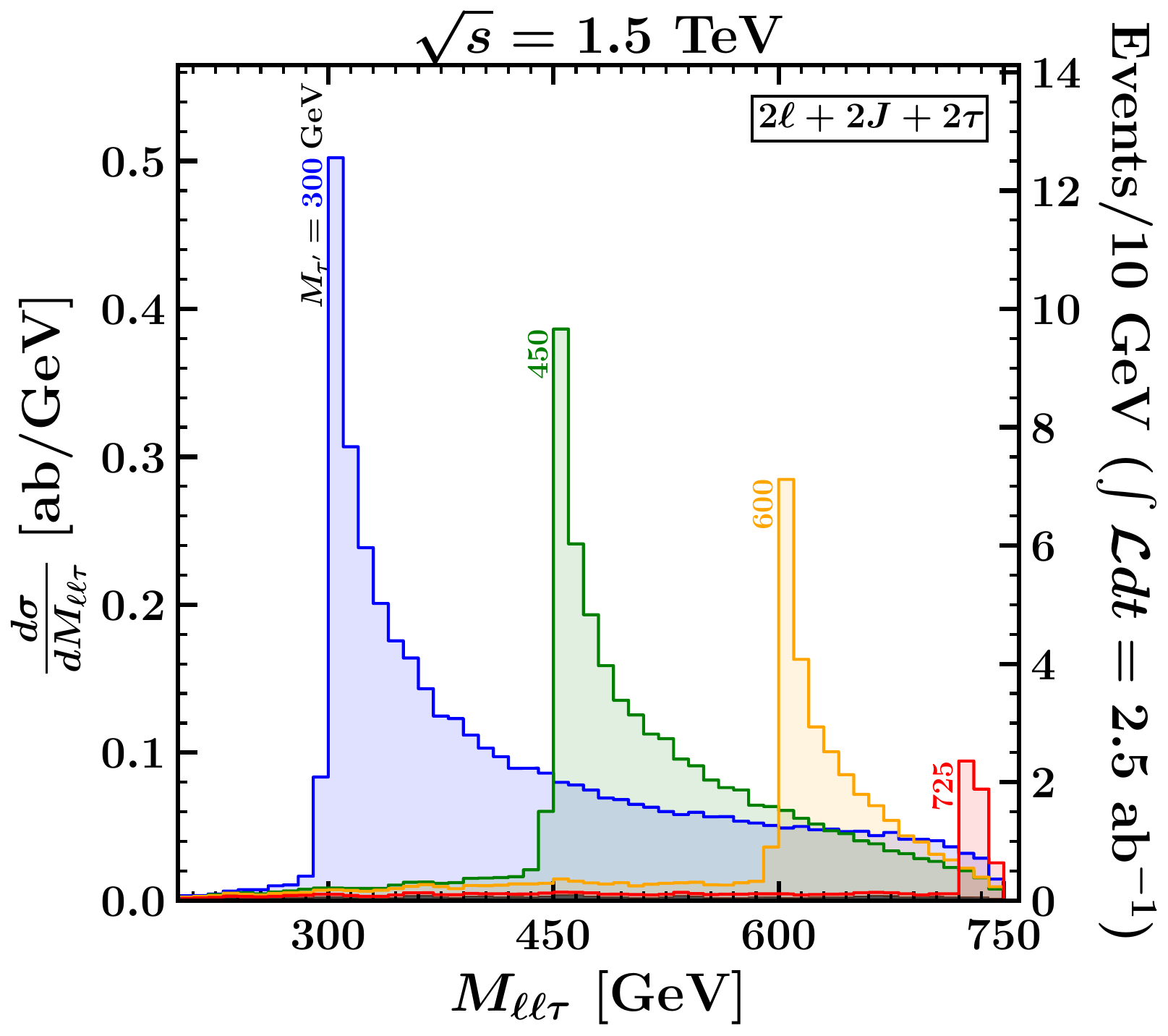}
  \end{minipage}
  \begin{minipage}[]{0.495\linewidth}
    \includegraphics[width=8cm]{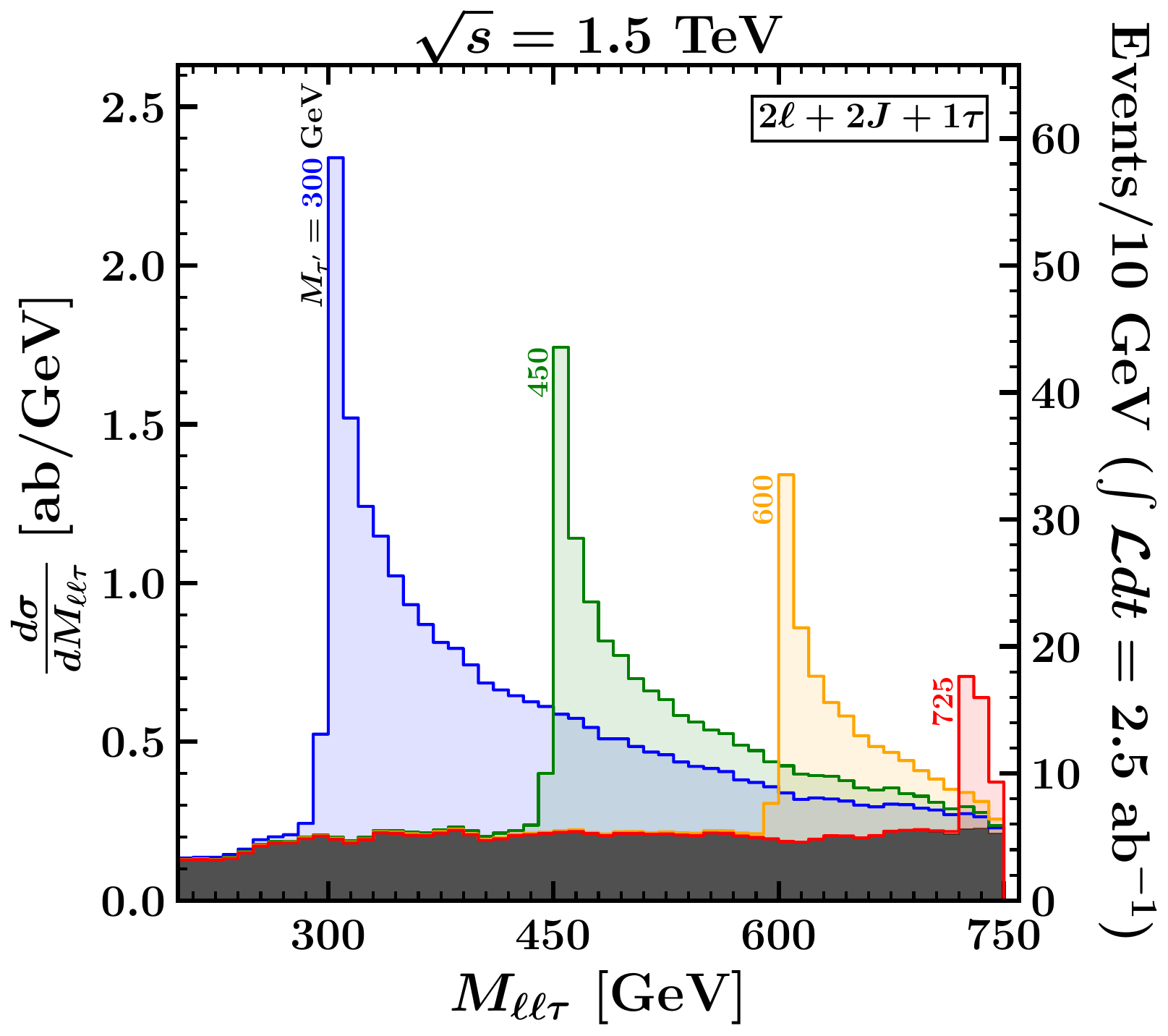}
  \end{minipage}
  \vspace{0.2cm}
  \begin{minipage}[]{0.495\linewidth}
    \includegraphics[width=8cm]{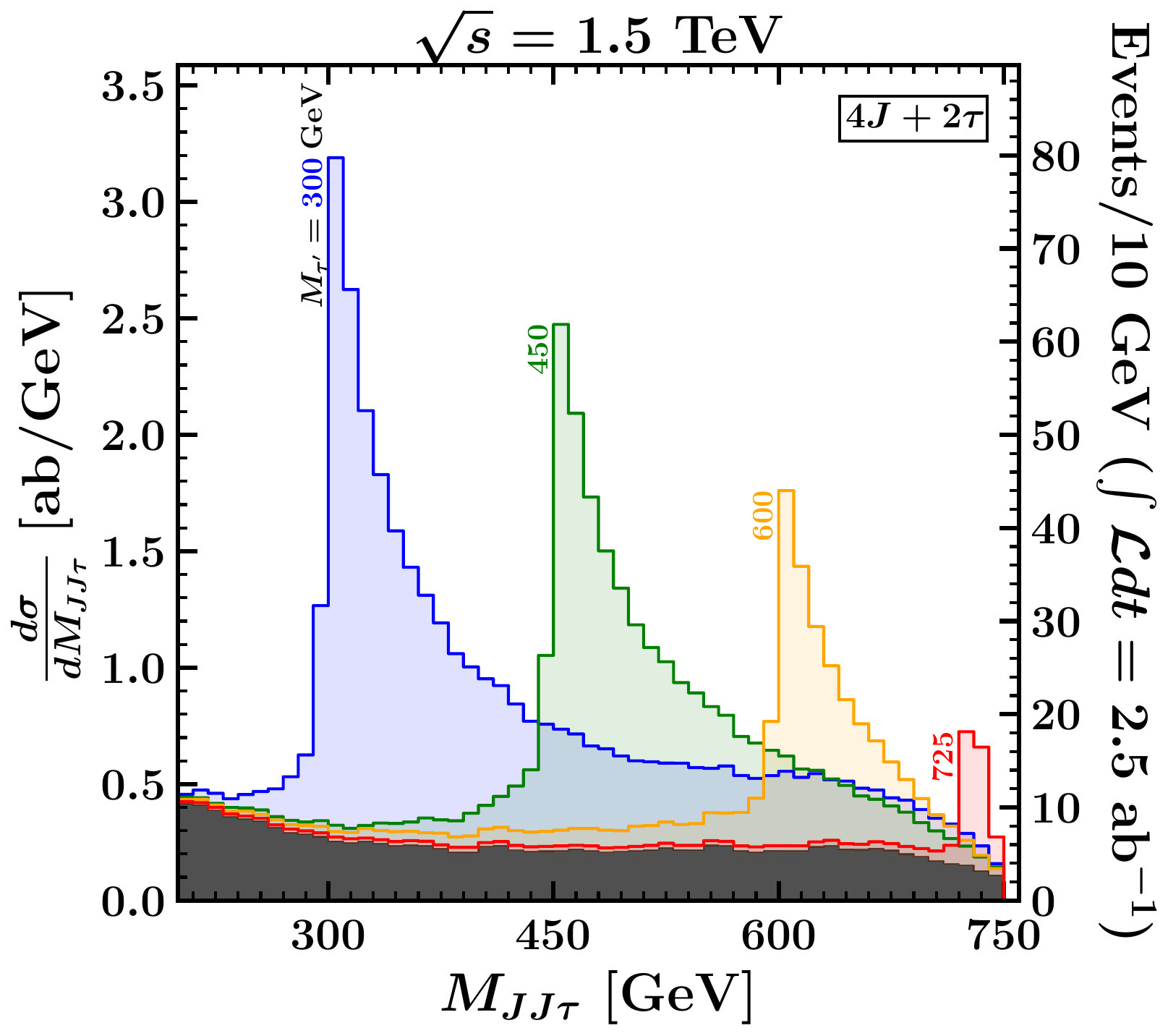}
  \end{minipage}
  \begin{minipage}[]{0.495\linewidth}
     \includegraphics[width=8cm]{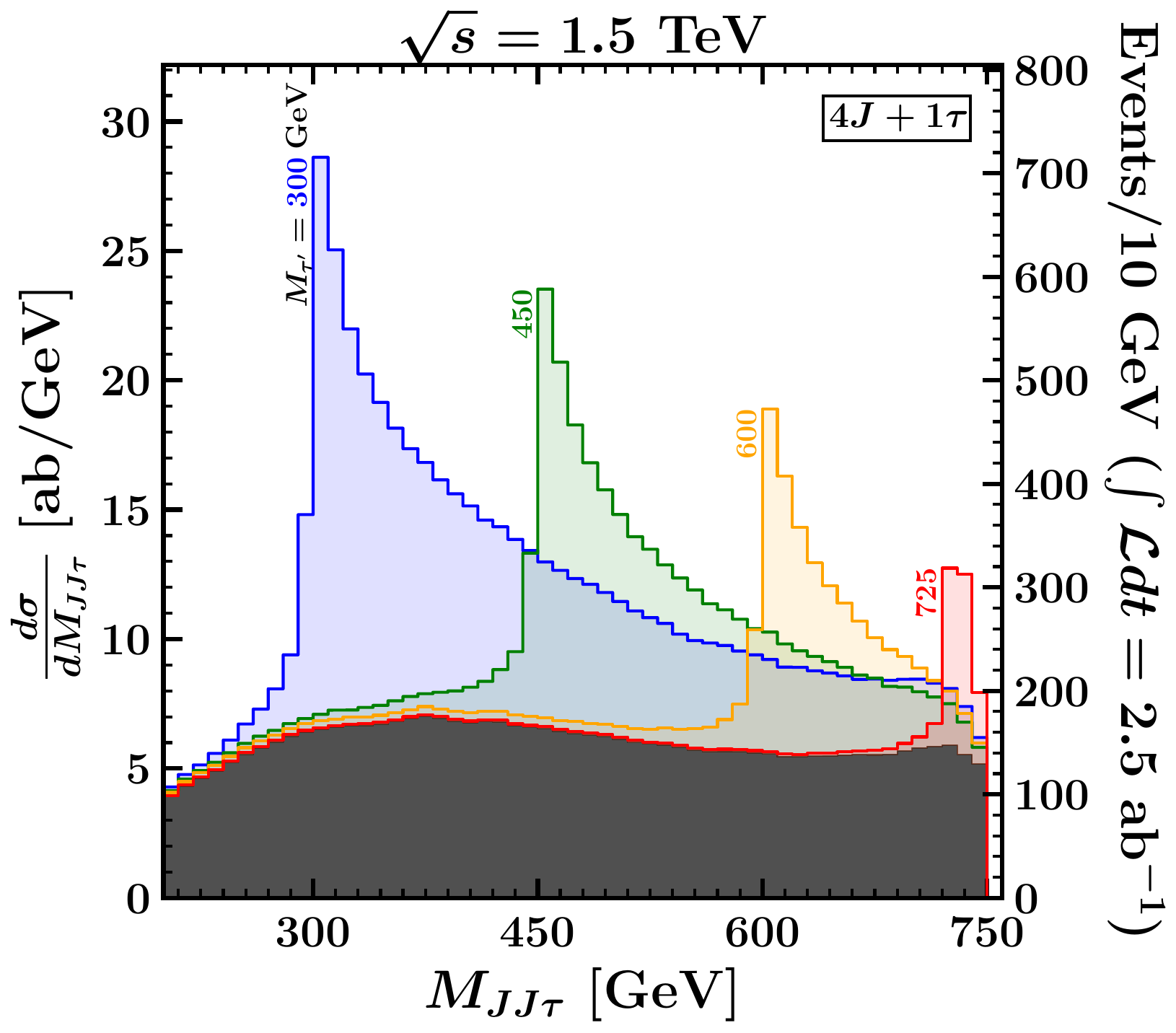}
   \end{minipage}
 \caption{Reconstructed $\tau^\prime$ mass peaks in five different signal 
 regions, for $M_{\tau^\prime} =$ 300, 450, 600, and 725 GeV as labeled, for $e^+ e^-$ collisions at $\sqrt{s} = 1.5$ TeV. Backgrounds are shown as gray-shaded histograms stacked together with the signal histograms.
\label{fig:Mtaup_sqrts1500}}
\end{figure}

\begin{figure}[!h]
  \begin{center}
   \begin{minipage}[]{0.495\linewidth}
    \includegraphics[width=8cm]{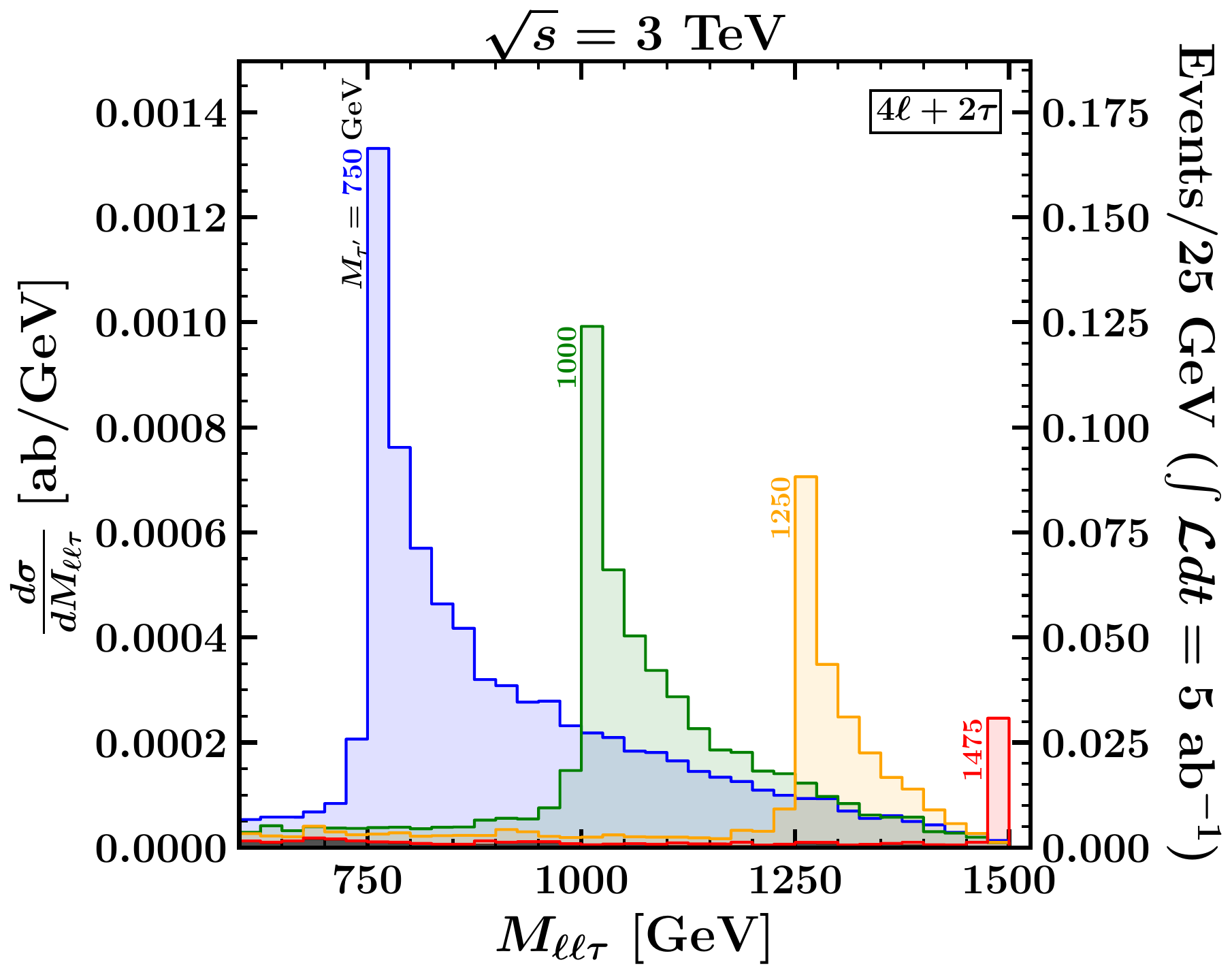}
  \end{minipage}
  \end{center}
  \vspace{0.2cm}
  \begin{minipage}[]{0.495\linewidth}
    \includegraphics[width=8cm]{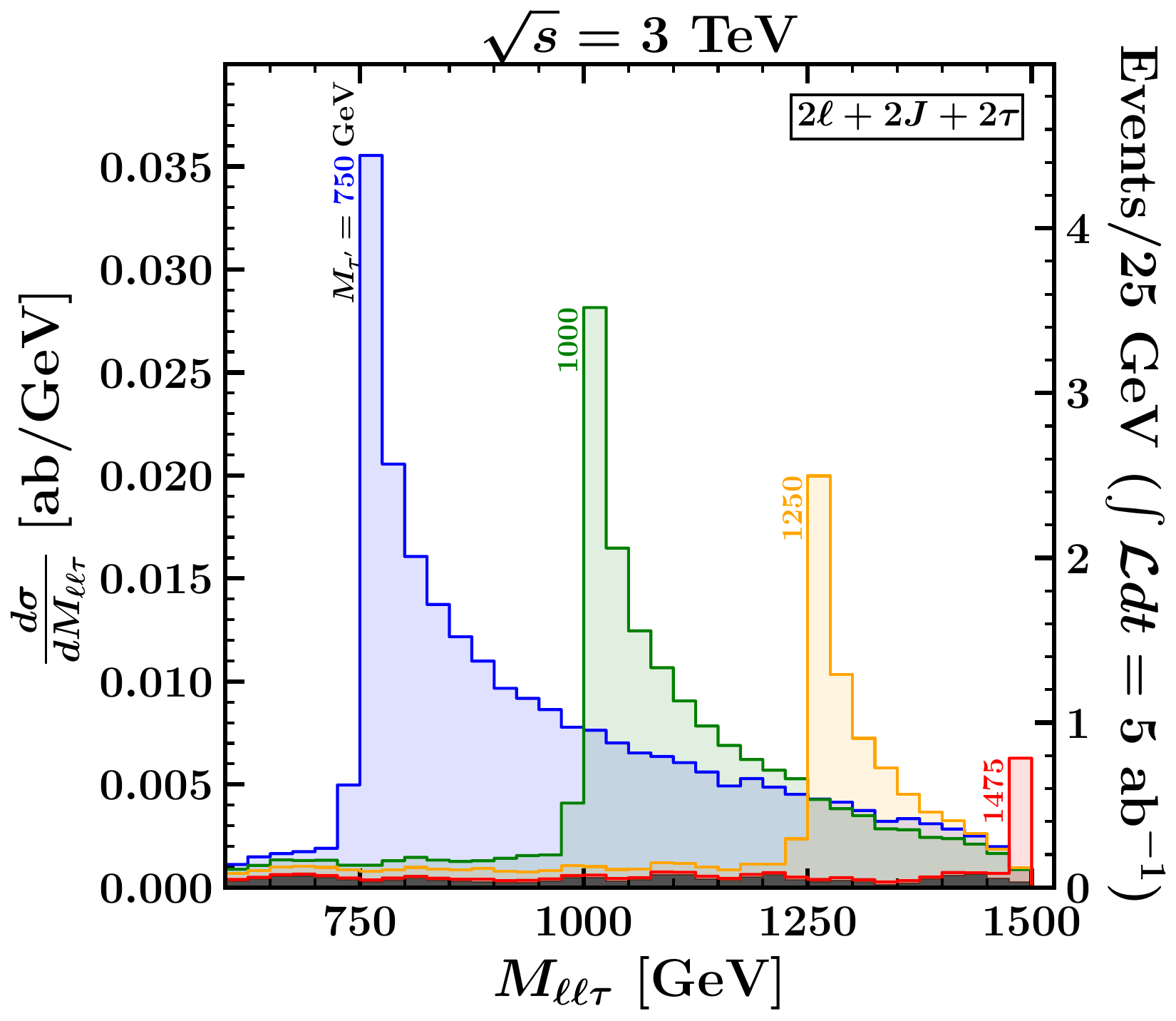}
  \end{minipage}
  \begin{minipage}[]{0.495\linewidth}
    \includegraphics[width=8cm]{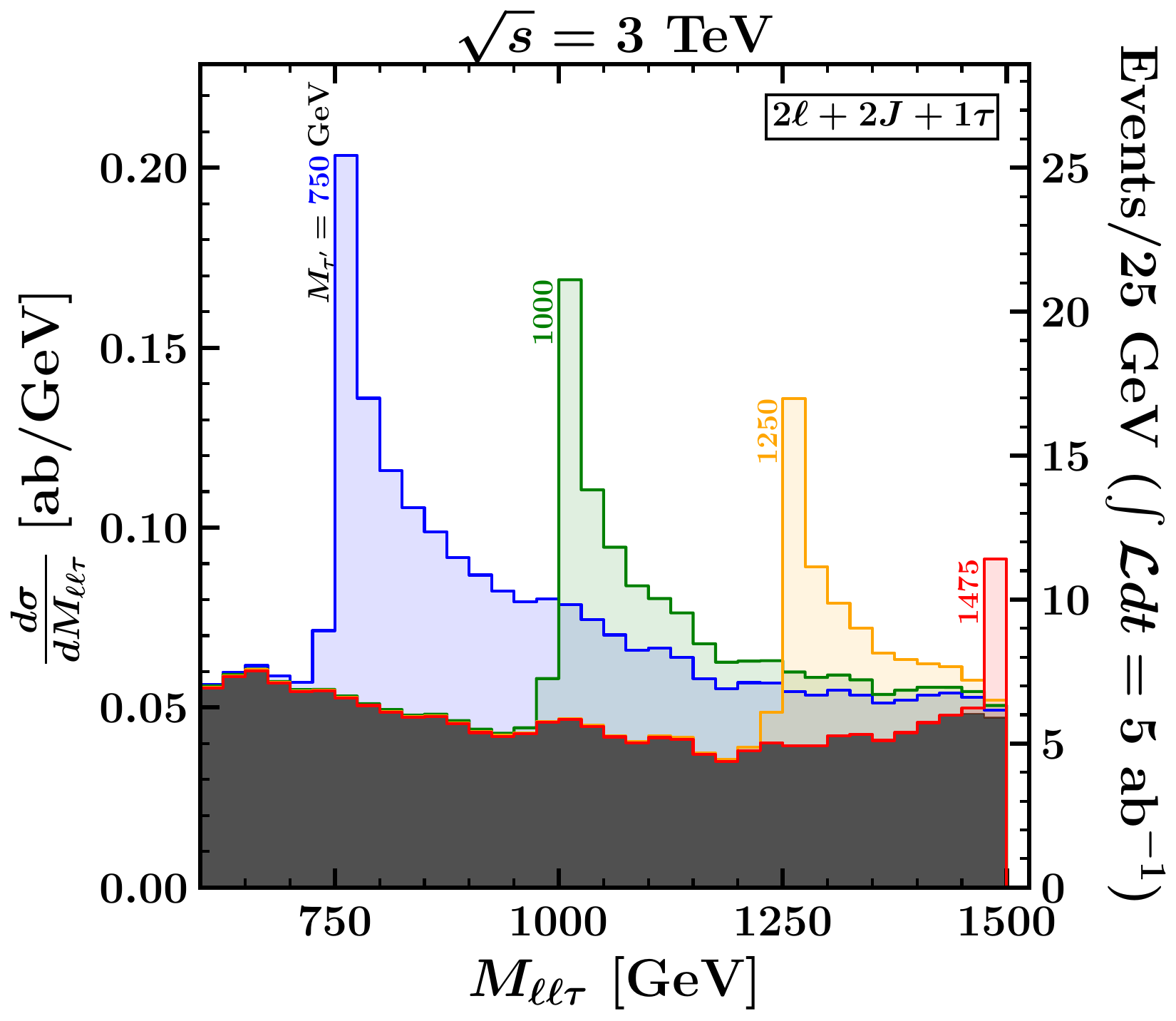}
  \end{minipage}
  \vspace{0.2cm}
  \begin{minipage}[]{0.495\linewidth}
    \includegraphics[width=8cm]{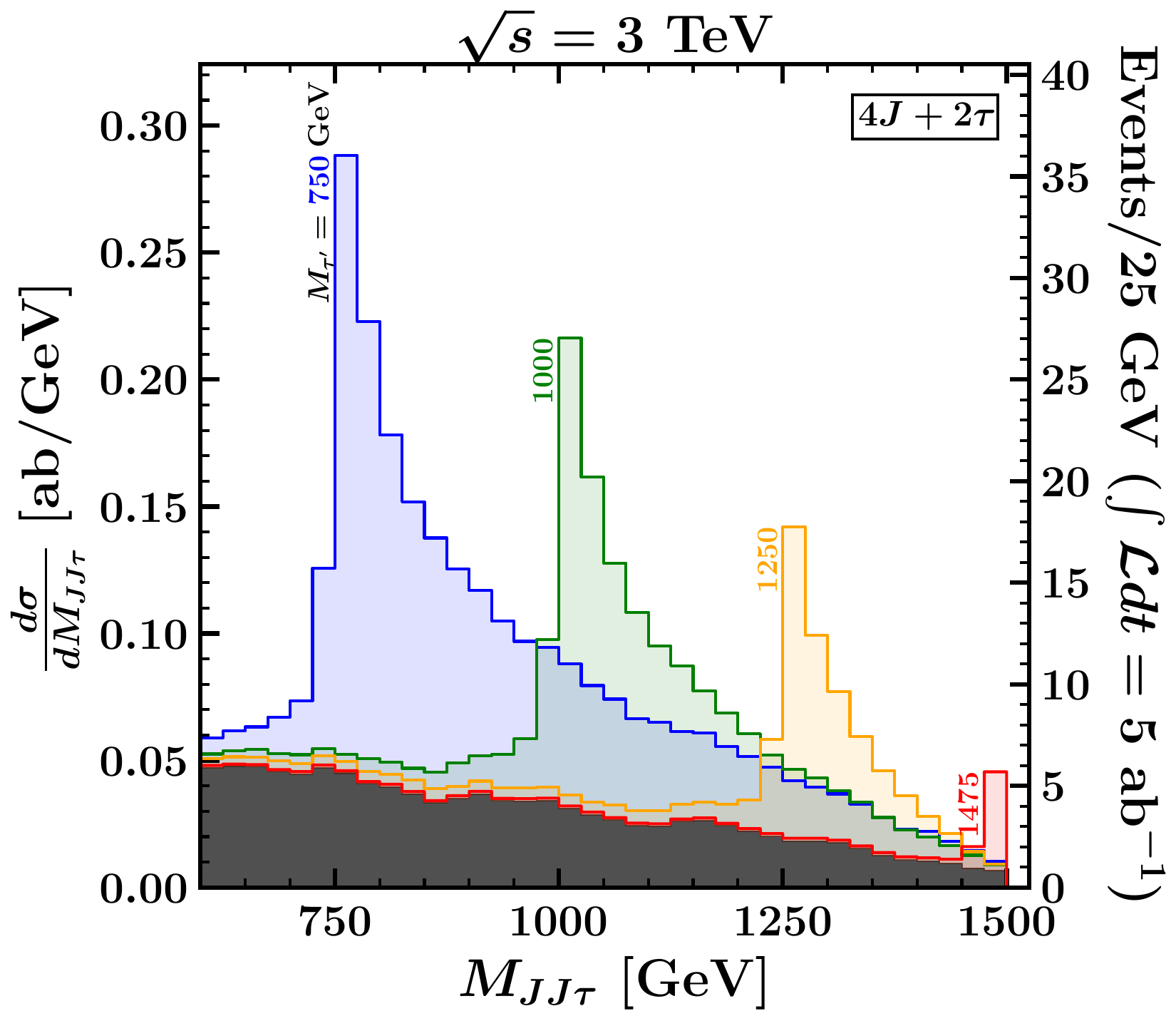}
  \end{minipage}
  \begin{minipage}[]{0.495\linewidth}
     \includegraphics[width=8cm]{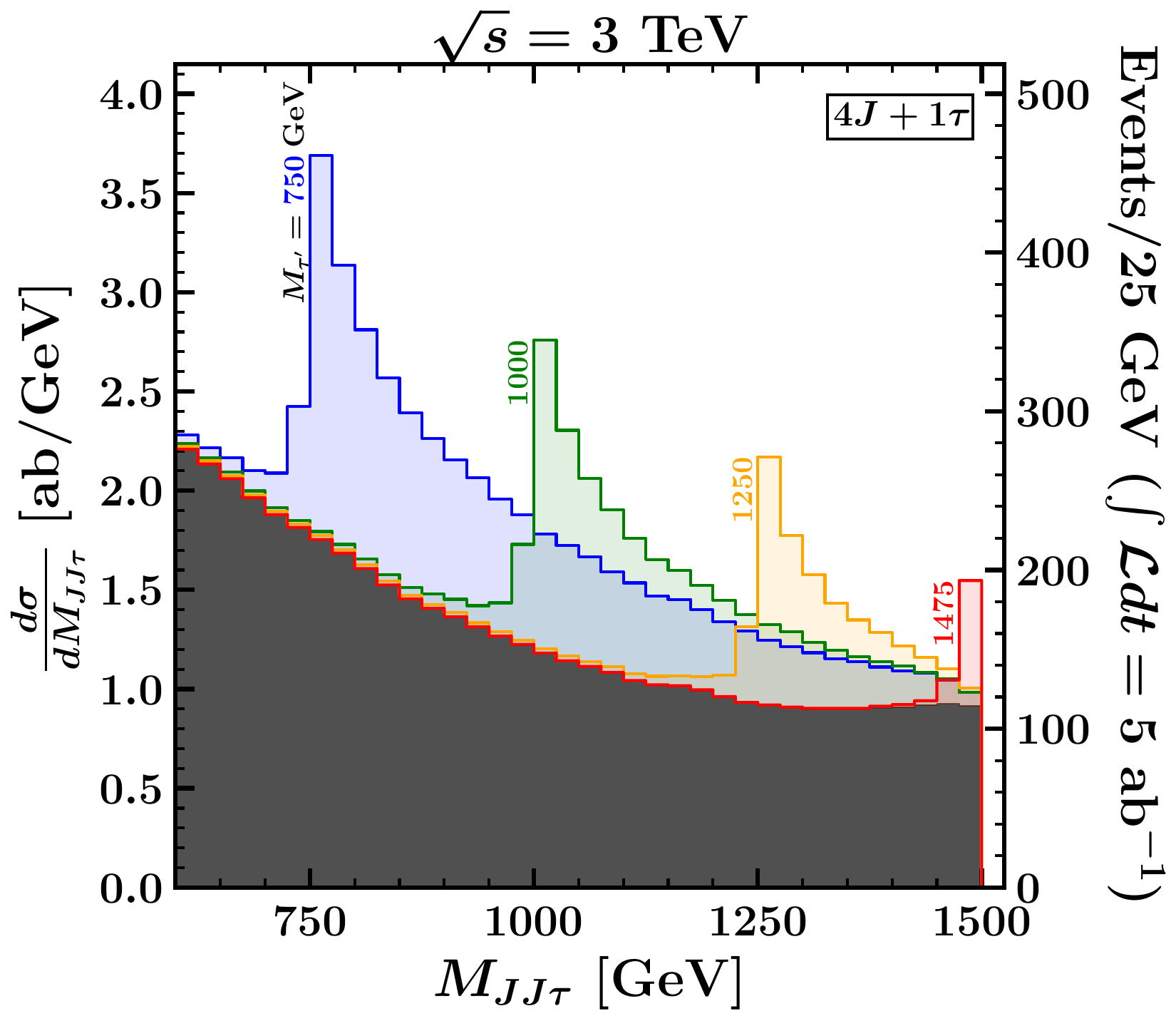}
   \end{minipage}
 \caption{Reconstructed $\tau^\prime$ mass peaks in five different signal 
 regions, for $M_{\tau^\prime} = $ 750, 1000, 1250, and 1475 GeV as labeled, for $e^+ e^-$ collisions at $\sqrt{s} = 3$ TeV. Backgrounds are shown as gray-shaded histograms stacked together with the signal histograms.
\label{fig:Mtaup_sqrts3000}}
\end{figure}

The results for the $\tau^{\prime}$ mass peaks are shown in Figs.~\ref{fig:Mtaup_sqrts1000}, \ref{fig:Mtaup_sqrts1500}, and \ref{fig:Mtaup_sqrts3000} for $\sqrt{s} = 1$, 1.5, and 3 TeV, respectively.  Again, we have shown numbers of events corresponding to benchmark luminosities taken from
Ref.~\cite{Narain:2022qud} on the right-hand side of the plots.   Because the production cross section falls with $\sqrt{s}$, a lack of adequate statistics can be an issue. This is especially true in the $4 \ell + 2 \tau$ signal region, which is clearly not viable at $\sqrt{s} = 1.5$ TeV with $2.5$ ab$^{-1}$ and at 3 TeV with $5$ ab$^{-1}$, and is expected to give a few events at most even for $\sqrt{s}= 1$ TeV. In the case of $\sqrt{s} = 1$ TeV, the other four signal regions are expected to give an observable peak. 
However, for the $\sqrt{s} = 1.5$ TeV and 3 TeV cases, the small numbers of events expected with the assumed integrated luminosities becomes problematic in some of the signal regions, especially if the $\tau^{\prime}$ mass is close to the kinematic limit. Also, for the 3 TeV case, backgrounds become relatively more significant (see the last three panels of Fig.~\ref{fig:Mtaup_sqrts3000}), but with a smooth mass distribution that should be under good theoretical control. The most significant mass peak in those cases will be from the $4J + 1\tau$ signal region, and estimations of branching ratios may be statistics-limited.
As can be seen from Fig.~\ref{fig:Sigma} and the discussion surrounding Eqs.~(\ref{eq:sigmahat})-(\ref{eq:aR}), running CLIC in the mode with $P_{e-} = +0.8$ and $P_{e+}=0$ would enhance the signal cross-section, but only by about 50\%.

\FloatBarrier

\section{Outlook\label{sec:outlook}}
\setcounter{equation}{0}
\setcounter{figure}{0}
\setcounter{table}{0}
\setcounter{footnote}{1}

We have investigated the prospects for the discovery and study of an isosinglet $\tau^{\prime}$ at future lepton colliders.  We have shown that discovery of such a particle should be possible up to close to the kinematic limit by an effective reconstruction of the $\tau^{\prime}$ mass peak. We have not attempted a precise numerical characterization of the discovery reach very near threshold, as this is likely to depend on detailed collider and detector characteristics that are not reliably estimated so far in advance.

We have also shown that examination of different signal regions gives access to different decays of the $\tau^{\prime}$, with nearly pure samples of 5 of the 6  final states for $\tau^{\prime+}\tau^{\prime-}$, which were listed in Eqs.~(\ref{eq:finalstats_2tau})-(\ref{eq:finalstats_0tau}).  Should the discovery of a $\tau^{\prime}$ be made, our results show that the examination of these different signal regions will therefore allow a determination of the $\tau^{\prime}$ branching ratios.  Comparison with the predictions should allow for a definitive verification of the identity of the new particle as an isosinglet $\tau^{\prime}$.

{\it Acknowledgments:}
PNB thanks Daniel Jeans, Wolfgang Kilian, Aakaash Narayanan, Evan Petrosky, J\"{u}rgen Reuter, Keping Xie, and Aleksander Filip \.{Z}arnecki for helpful discussions. This research was supported in part through computational resources and services provided by Advanced Research Computing (ARC), a division of Information and Technology Services (ITS) at the University of Michigan, Ann Arbor. This research was also supported in part through computational resources and services provided by NICADD compute cluster at Northern Illinois University. This work is supported in part by the National Science Foundation under grant number 2013340,
and by the Department of Energy under grant number DE-SC0007859.


\end{document}